\documentclass[%
reprint,
nofootinbib,
amsmath,amssymb,
aps,
prx,
]{revtex4-2}

\usepackage{graphicx}% Include figure files
\usepackage{dcolumn}% Align table columns on decimal point
\usepackage{bm}% bold math
\usepackage{subfigure}
\usepackage{floatrow}
\newfloatcommand{capbtabbox}{table}[][0.45\textwidth]
\usepackage{comment}
\usepackage{xcolor}
\usepackage{bbold}

\newcommand{\be}{\begin{equation}}
\newcommand{\ee}{\end{equation}}
\newcommand{\ba}{\begin{eqnarray}}
\newcommand{\ea}{\end{eqnarray}}

\newcommand{\snn}{\sigma_{nn}}
\newcommand{\Enn}{E_{12}}
\newcommand{\pdf}{\operatorname{PDF}}

\newcommand{\ave}[1]{\left\langle#1\right\rangle}

\newcommand{\tou}[1]{{\boldsymbol{#1}}}

\def\TM{\textcolor{black}}

\begin{document}

%\preprint{}

%\title{Intergranular stresses in elastic materials}
\title{Grain boundary stresses in elastic materials}
%\title{Elasticity of grain boundaries}




\author{S.~El~Shawish}
\email{samir.elshawish@ijs.si}
\affiliation{Jo$\check{z}$ef Stefan Institute, SI-1000, Ljubljana, Slovenia}

\author{T.~Mede}
\email{timon.mede@ijs.si}
\affiliation{Jo$\check{z}$ef Stefan Institute, SI-1000, Ljubljana, Slovenia}

%\author{J.~Hure}
%\email{jeremy.hure@cea.fr}
%\affiliation{Universit\'{e} Paris-Saclay, CEA, Service d'\'Etudes des Mat\'{e}riaux Irradi\'{e}s, 91191 Gif-sur-Yvette cedex, France}

\date{\today}

% ==========================================================================
\begin{abstract}

A \TM{simple analytical model} of \TM{intergranular} normal stresses is \TM{proposed} for a general elastic polycrystalline material with arbitrary shaped and randomly oriented grains under uniform loading. The \TM{model} provides algebraic expressions for the local \TM{grain-boundary-normal} stress \TM{and the} corresponding uncertainties, \TM{as a function of} the grain-boundary type, its inclination with respect to the \TM{direction of} external loading and material-elasticity parameters. The knowledge of intergranular normal stresses is a necessary prerequisite in any local damage modeling approach, \TM{\textit{e.g.}}, to predict the intergranular stress-corrosion cracking, grain-boundary sliding or fatigue-crack-initiation sites in structural materials.%A theory of grain-boundary-normal stresses is established for a general elastic polycrystalline material with randomly shaped and oriented crystalline grains under arbitrary applied uniform loading. The theory provides an algebraic expression for the local grain boundary (also intergranular) normal stress, with corresponding uncertainties, given the information about the grain boundary type, its inclination with respect to the external loading and material elasticity parameters. The knowledge of intergranular normal stresses is a necessary prerequisite in any local damage modeling approach, for example, to predict the intergranular stress-corrosion cracking, grain boundary sliding or fatigue crack initiation sites in structural materials.

%%%For a particular solid, grain boundary stresses may be accurately estimated using computationally demanding simulations, however, such an approach deems impractical for a general case and provides little or no insight into the principles involved.

The \TM{model} is derived in a perturbative manner, starting with the exact solution of a simple setup and later successively refining it to account for higher order complexities of realistic polycrystalline materials. In the simplest scenario, a bicrystal model is embedded in an isotropic elastic medium and solved for uniaxial loading conditions, assuming 1D Reuss and Voigt approximations on different length scales. In the final iteration, \TM{the grain boundary becomes a part of} a 3D \TM{structure consisting} of five 1D chains \TM{with arbitrary number of grains and surrounded by an anisotropic elastic medium. Constitutive equations can be solved for arbitrary uniform loading, for any grain-boundary type and choice of elastic polycrystalline material.} At each iteration, the algebraic expressions for the local grain-boundary-normal stress, \TM{along with the corresponding statistical distributions, are derived and their accuracy} systematically verified and validated against the finite element simulation results of different Voronoi microstructures. %The theory is derived in a perturbative manner, starting with the exact solution of a simple setup and later successively refining that solution to account for higher-order complexities of a realistic polycrystalline material. In the simplest scenario, a bicrystal model is embedded in an isotropic elastic medium and solved for uniaxial loading conditions, assuming 1D Reuss and Voigt approximations on different length scales. In the final iteration, a 3D pattern of five 1D grain chains of variable lengths is introduced around the grain boundary and solved approximately for the anisotropic elastic medium and arbitrary uniform loading. At each iteration, the derived algebraic expressions for the grain-boundary-normal stress are verified and validated systematically against the finite element simulation results of different Voronoi microstructures. An accurate semi-analytical expression for the local grain-boundary-normal stress is derived, along with the corresponding statistical distributions, for arbitrary grain boundary types, elastic polycrystalline materials and loading conditions, providing a useful tool in local damage modeling and characterization.

\end{abstract}
% ==========================================================================

%\keywords{}

%\keywords{
%  IGSCC, grain boundary type, coincidence site lattice, intergranular
%  normal stress fluctuations, finite element method, grain boundary
%  engineering
%}

\maketitle

\section{\label{intro}Introduction}
% ==========================================================================

% from ARRS project proposal
Predicting damage initiation and its progression in structural materials relies heavily on the knowledge of local mechanical stresses present in the material. For particular aging processes, the material damage is initiated at the grain boundaries (GBs), where intergranular microcracks form. With time, these microcracks may grow along the GBs and combine into larger macroscopic cracks, which can eventually compromise the structural integrity of the entire component under load. %It is well known {\red\cite{}} that macroscopic cracks, visible to detection instruments, only exist for the final $\sim$10\% of the component’s life. For the majority of its lifetime, however, the microcracks are invisible to non-destructive inspection techniques.
\TM{Since microcracks are invisible to non-destructive inspection techniques, the detection instruments can only reveal the existence of macroscopic cracks, which roughly appear in the final $10\%$ of the component’s lifetime.} 
Having accurate models for predicting the component’s susceptibility to microcracking in its earlier stages is therefore of uttermost importance in many different applications, as this could reduce the costs needed for frequent inspections and replacements.

InterGranular Stress-Corrosion Cracking (IGSCC) is one of the most significant ageing-degradation mechanisms. It corresponds to the initiation and propagation of microcracks along the GBs and is common in alloys, that are otherwise typically corrosion-resistant (austenitic stainless steels~\cite{nishioka2008,lemillier,stephenson2014,gupta,fujii2019}, zirconium alloys~\cite{cox,cox1990}, nickel based alloys~\cite{rooyen1975,shen1990,panter2006,IASCC_IAEA}, high strength aluminum alloys~\cite{speidel,burleigh} and ferritic steels~\cite{wang,arafin}). The IGSCC is a multi-level process that includes electro-chemical, micro-mechanical and thermo-mechanical mechanisms. The activation of these mechanisms depends on material properties, corrosive environment and local stress state. It is believed that GB stresses are the driving force of intergranular cracking, therefore they need to be accurately determined, in order to make quantitative predictions about IGSCC initiation. 

Various approaches \TM{to IGSCC-initiation modeling are being considered}. One such approach is to treat IGSCC phenomenon on a local GB scale, where GB-normal stresses $\snn$ \TM{can be} studied separately (decoupled) from the environmental effects that degrade the GB strength $\sigma_c$;
${\rm IGSCC}\approx\mathcal{F}(\snn)\cdot\mathcal{F}(\sigma_c)$.
A GB-normal stress $\snn$ is defined here as a \TM{component of local} stress tensor \TM{along} the GB-normal direction, \textit{i.e.}, perpendicular to the GB plane.%
\footnote{In the terminology of fracture mechanics, $\snn$ corresponds to the opening-mode stress (Mode~$1$).}
%
%Because of such a local approach, a single stress-based criterion for a local IGSCC initiation can be assumed on every GB: IGSCC initiates locally when $\snn>\sigma_c$.
\TM{Hence}, a single stress-based criterion for a local IGSCC initiation can be assumed on every GB: IGSCC gets initiated wherever $\snn>\sigma_c$, \TM{with both these quantities being local in a sense, that they in principle depend on the position of the GB within the aggregate}. %\TM{Note, that in principle $\sigma_c$ is also a local quantity, depending on the position of the GB within the aggregate (including the exact configuration of all the grains in its neighborhood) --- the dominant effect, however, can be attributed to the GB type it belongs to.} 

The introduced local criterion can be used to evaluate the probability, that a randomly selected GB on a component’s surface, where it is in contact with the corrosive environment, is overloaded (or soon-to-be cracked). This probability can be estimated by calculating a fraction $\eta$ of GBs with $\snn>\sigma_c$ as
$\eta=\int_{\sigma_c}^{\infty}\pdf(\snn)d\snn$,
for the assumed probability-density function $\pdf(\snn)$. If a fraction of overloaded GBs exceeds a threshold value, $\eta>\eta_f$, a specimen-sized crack may develop, possibly resulting in a catastrophic failure of the component.
%
%Such an approach is needed due to the stochastic nature of the process, where microcracks should be dense enough in some region for the macrocrack to form.
%
This approach, based on the accurate knowledge of GB-normal stresses, seems feasible when a GB strength $\sigma_c$ is known and approximately constant within the examined surface section. Unfortunately, this is not the case in real materials.

Measurements have shown that different GBs show different IGSCC sensitivities~\cite{rahimi2011,fujii2019,rahimi2009,liu2019}, implying that GB strength $\sigma_c$ depends not only on the material and environmental properties but also strongly on a GB type;
$\sigma_c=\mathcal{F}({\rm GB\ type, material, environment})$.
Here, a GB type denotes a GB microstructure (inter-atomic arrangements in the vicinity of the GB), which affects the GB energy and, eventually, its strength $\sigma_c$. In the continuum limit, five parameters are \TM{needed to} define a GB neighborhood: four parameters are \TM{required} to specify a GB plane in \TM{crystallographic systems of} the two adjacent grains and one parameter defines a twist rotation between the associated crystal lattices about the plane normal. %However, not all five parameters need to be pre-defined in a particular GB type. For example, in the $[abc]$-$[def]$ GB type, where a GB plane is set equal to the $[abc]$ plane in one grain and $[def]$ plane in the other grain, the twist angle is assumed random (undefined).
\TM{In principle, the term ``GB type'' thus refers to GBs with the same GB strength. Sometimes it is convenient to specify a GB type by less than five parameters (\textit{e.g.}, when the values of skipped parameters do not affect the $\snn$ distribution). 
For instance, in the $[abc]$-$[def]$ GB type, with a GB-plane normal along the $[abc]$ direction in one grain and $[def]$ direction in the other grain, the twist angle can be assumed random (and thus remains unspecified).
%However, the same $\snn$ distribution might correspond to different GB types. In such case, we sometimes speak ``loosely'' of a common GB type, for whose specification less than five parameters are needed, even though the term ``GB type'' is in principle reserved for GBs with the same GB strength.         
%However, the same GB strength and $\snn$ distribution might correspond to different GB types, which can thus be treated as essentially the same type. In such case, even less than five parameters might be needed to specify it. For instance, in the $[abc]$-$[def]$ GB type, with a GB-plane normal along the $[abc]$ direction in one grain and $[def]$ direction in the other grain, the twist angle can be assumed random (and can remain unspecified).
} 

In \TM{addition} to $\sigma_c$ being a function of GB type, also the distributions of GB-normal stresses should be evaluated for different GB types in order to later perform a meaningful calculation of fraction $\eta$. Hence,
$\pdf(\snn)=\mathcal{F}({\rm GB\ type, applied\ stress, material})$.
Since exact general solutions for both the local $\snn$ and statistical $\pdf(\snn)$ are too complex to be derived analytically, researchers have restricted themselves to numerical simulations limited to few selected materials and specific (usually \TM{uniaxial}) loading conditions.

% from previous paper (rearranged)

% numerical simulations - specific cases
Crystal-plasticity finite element (FE) simulations~\cite{diard2002,diard2005,kanjarla,gonzalez2014,hure2016,elshawish2018} and crystal-plasticity fast Fourier transform simulations~\cite{lebensohn2012,disen2020} have been used to obtain intergranular stresses on random GBs in either synthetic or realistic polycrystalline aggregates, providing valuable information for IGSCC initiation in those specific cases. In particular, the fluctuations of intergranular normal stresses (the widths of $\pdf(\snn)$) have been found to depend primarily on the elasto-plastic anisotropy of the grains with either cubic~\cite{gonzalez2014,hure2016,elshawish2018} or hexagonal lattice symmetries~\cite{elshawish2018}.

% numerical simulations - common statistical behavior
Although the computationally demanding simulations can provide accurate results, such an approach deems impractical for a general case and provides little insights into involved physics. Thus, efforts have been made to identify most influential parameters affecting the GB-normal stresses on \TM{any single} GB type~\cite{west2011,elshawish2021}. In the elastic regime of grains with cubic lattice symmetry, Zener elastic anisotropy index $A$~\cite{zener} and effective GB stiffness $\Enn$, measuring the average stiffness of GB neighborhood along the GB-normal direction, have been identified and demonstrated to be sufficient for quantifying normal-stress fluctuations on any GB type in a given material under uniaxial external loading~\cite{elshawish2021}. The empirical relation (still lacking a satisfactory explanation) has been established for the standard deviation of $\snn$ \TM{distribution} evaluated on $[abc]$-$[def]$ GBs, which is a function of $A$ and $\Enn$. On the contrary, the mean value of the same $\snn$ \TM{distribution} has been shown to be independent of the chosen material and/or the GB type on which it is calculated.

% analytical model derived for flow stress=yield stress
To account for \TM{elastic--perfectly plastic} grains at applied tensile yield stress, a simple Schmid-Modified Grain-Boundary-Stress model has been proposed~\cite{west2011} to investigate the initiation of an intergranular crack, based on a normal stress acting at GB. The model considers combined effects of GB-plane orientation and grain orientations through their Schmid factors. It has been pointed out, that intergranular cracks occur most likely at highly stressed GBs. In other similar studies~\cite{stratulat2014,zhang2019,fujii2019}, the same model has been used to discuss crack initiation in austenitic stainless steel, concluding that initiation sites coincide with the most highly stressed GBs.

% goal
Building upon partial results, limited to either specific loading conditions~\cite{west2011,elshawish2021} and/or specific grain-lattice symmetries~\cite{elshawish2021}, the goal of this study is to develop a \TM{model} of GB-normal stresses, that would provide accurate analytic or semi-analytic expressions for $\snn$, with the corresponding statistical measure $\pdf(\snn)$, depending on a general GB type, general applied stress and general elastic polycrystalline material. Once the knowledge of GB strength $\sigma_c$ becomes available, the resulting expressions will be directly useful in the mechanistic modeling of GB-damage initiation (such as IGSCC%
\footnote{Since material and mechanical aspects of IGSCC are decoupled from the environmental factors hidden in the GB strength, this study may also be relevant for other degradation mechanisms, where GB-normal stresses are the driving force for crack initiation, such as GB sliding or fatigue.
}%
) and should therefore become a quick and reliable tool to all the experts dealing with local damage modeling and characterization.

The paper is structured as follows: in Sec.~\ref{sec:2} typical material and GB-type effects on GB-normal stresses are introduced. In Sec.~\ref{sec:3} the \TM{perturbative framework for predicting} GB-normal stresses is developed, providing  analytical and semi-analytical models along with their solutions. In Sec.~\ref{sec:4} the upgraded models are verified with FE-simulation results. Practical implications are discussed in Sec.~\ref{sec:5} and in Sec.~\ref{sec:6} some concluding remarks are given. All technical details are deferred to the set of Appendices.

% ==========================================================================
\section{Material and grain boundary type effects on intergranular normal stresses}
\label{sec:2}
% ==========================================================================

The anisotropic elasticity of crystals is governed by the generalized Hooke's law, $\sigma_{ij}=C_{ijkl} \, \epsilon_{kl}$, where $C_{ijkl}$ is a 3D fourth-order stiffness tensor. Depending on the symmetry of the underlying grain lattice, $C_{ijkl}$ can be expressed in terms of two (isotropic), three (cubic), or more (up to $21$ for triclinic) independent elastic parameters. All grains in a polycrystalline aggregate are assigned the same elastic material properties, but different, \TM{random} crystallographic orientations \TM{(no texture)}. %While the majority of grains is assigned a random orientation (no texture), a relatively small fraction of grains is assigned a specific orientation which is characteristic to a particular GB type.
\TM{%These should in principle be random for all the grains (no texture). 
However, for practical purposes, we artificially increase the share of GBs of a certain type in our %numerical simulations
finite element aggregate models, by imposing specific orientations to a relatively small fraction of grains.} 

In the continuum limit%
\footnote{
	On the atomistic scale, more parameters would be required to characterize a GB \TM{by describing} the arrangement of atoms on both sides of the GB plane (\textit{e.g.}, coherent vs. non-coherent GBs).
},
a general GB type is defined by five independent parameters, which specify the orientations of two nearest grains relative to the GB plane. It can be expressed in the form of $[abc]$-$[def]$-$\Delta\omega$ GBs, where their GB plane is the $[abc]$ plane in one grain and $[def]$ plane in the other grain, with $\Delta\omega$ denoting a relative twist of the two grain orientations \TM{about} the GB normal.%
\footnote{GB type can also be defined by specifying less than five parameters. In such cases, \TM{the value of certain} parameters \TM{can be} assumed random. For example, misorientation GBs have only one fixed parameter and coincidence-site-lattice GBs, such as $\Sigma$3, $\Sigma$5, $\Sigma$7, have three fixed parameters~\cite{elshawish2021}.}

\begin{figure}
	\includegraphics[width=\columnwidth]{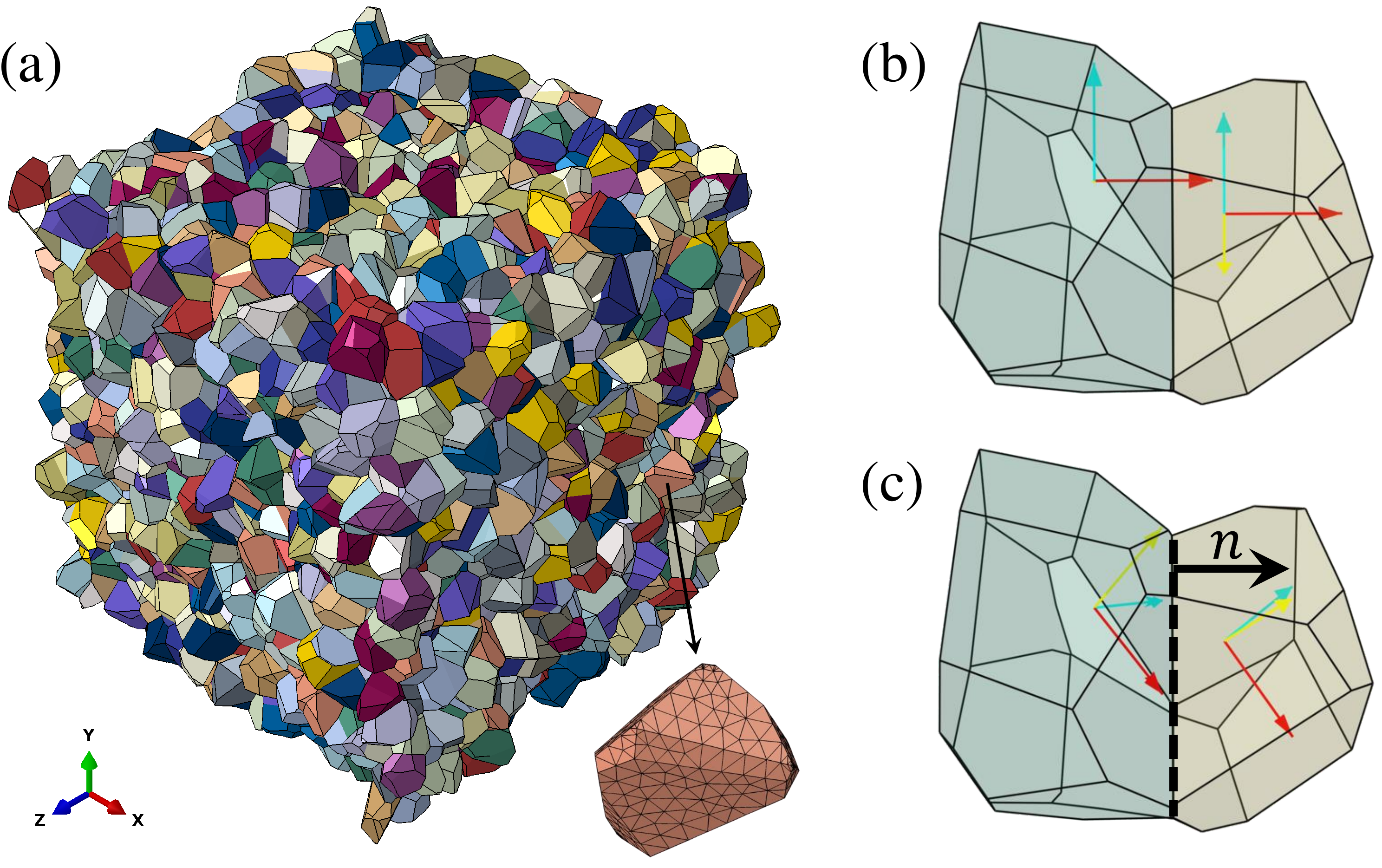}
	\caption{(a) 3D periodic Voronoi aggregate with $4000$ grains used in this study. Different grains are denoted by different colors. Finite element mesh is shown for one selected grain. Visualization of two different GBs with fixed GB plane (with normal $n$) but different crystallographic orientations: (b) $[001]$-$[001]$-$30^{\circ}$ GB and (c) $[111]$-$[111]$-$30^{\circ}$ GB.}
	\label{fig:geom}
\end{figure}

Due to topological constraints, not all GBs can be assigned the same GB character. In practice, a particular $[abc]$-$[def]$-$\Delta\omega$ GB type can be ascribed to at most $\sim$17\% of the GBs in a given aggregate, with the remaining GBs being of random type (\textit{i.e.}, defined by two randomly oriented neighboring grains). A polycrystalline aggregate and two particular GB types are visualized in Fig.~\ref{fig:geom}.

The constitutive equations of the generalized Hooke's law \TM{are solved numerically for a chosen uniform loading with FE solver Abaqus~\cite{abaqus}% for all finite elements (in all the grains) of the aggregate model
. The obtained stresses $\sigma$, corresponding to the nearest integration points of a particular GB $k$, are then used to produce a single value $\snn(k)$ as their weighted average. Besides local stresses $\snn(k)$, first two statistical moments of $\pdf(\snn)$, the mean value and standard deviation, are calculated for the distribution of stresses on GBs of a chosen, overrepresented $[abc]$-$[def]$-$\Delta\omega$ GB type, whose density was artificially boosted} (see Appendix~\ref{app:fem} for further details).

\begin{figure}
	\includegraphics[width=0.9\columnwidth]{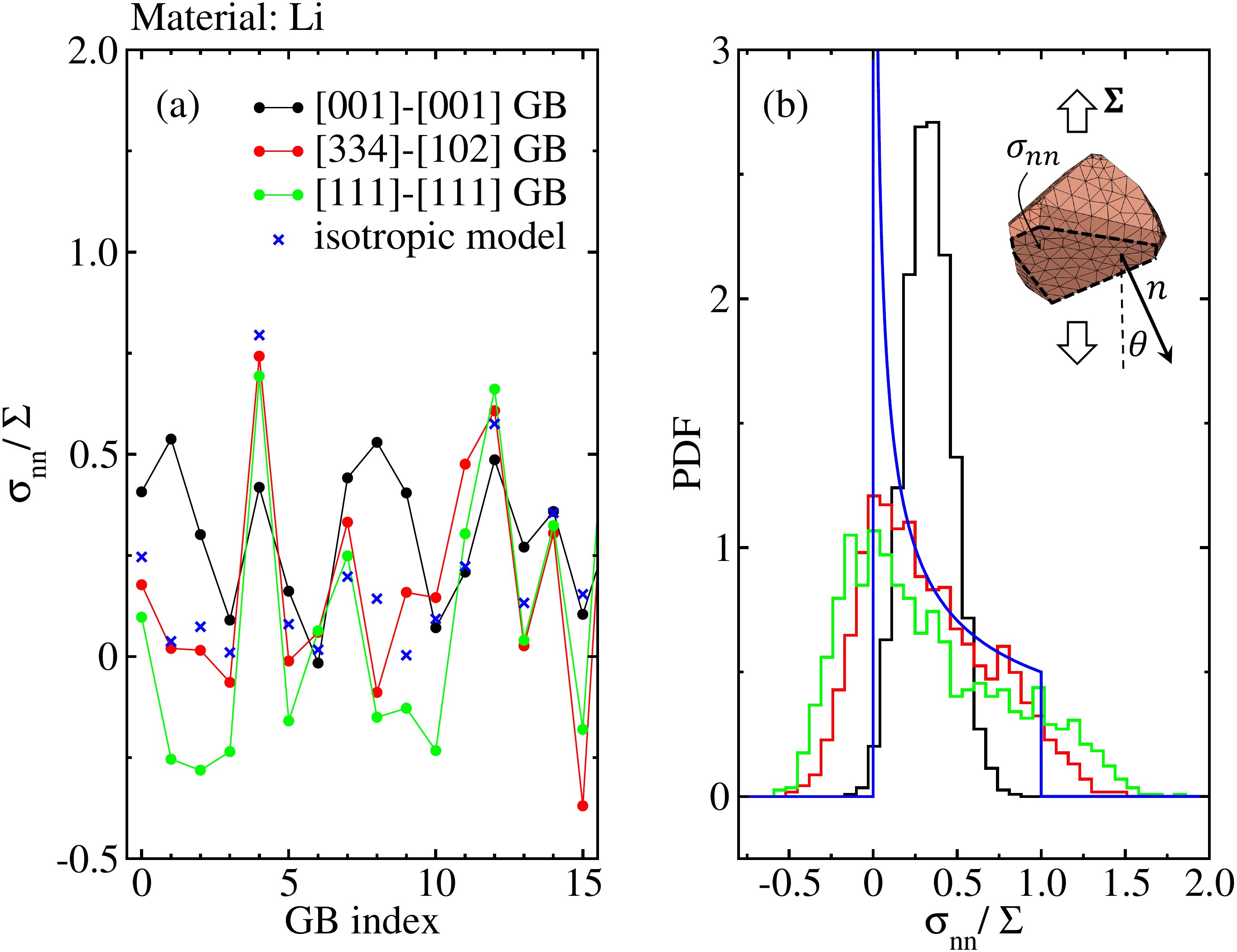}
	\caption{(a) \TM{Normalized} local stress responses $\snn/\Sigma$ and (b) \TM{their statistical distributions} $\pdf(\snn/\Sigma)$ in a polycrystalline lithium under macroscopic tensile loading $\Sigma$ \TM{for $3$ different GB types. Large influence of a chosen GB type and poor prediction capability of the isotropic model ($\snn/\Sigma=\cos^2\theta$) are clearly visible. In panel (a) the results are shown for just $15$ randomly selected GBs of each type (GB index).}}
	\label{fig:effectGB}
\end{figure}
\begin{figure}
	\includegraphics[width=0.9\columnwidth]{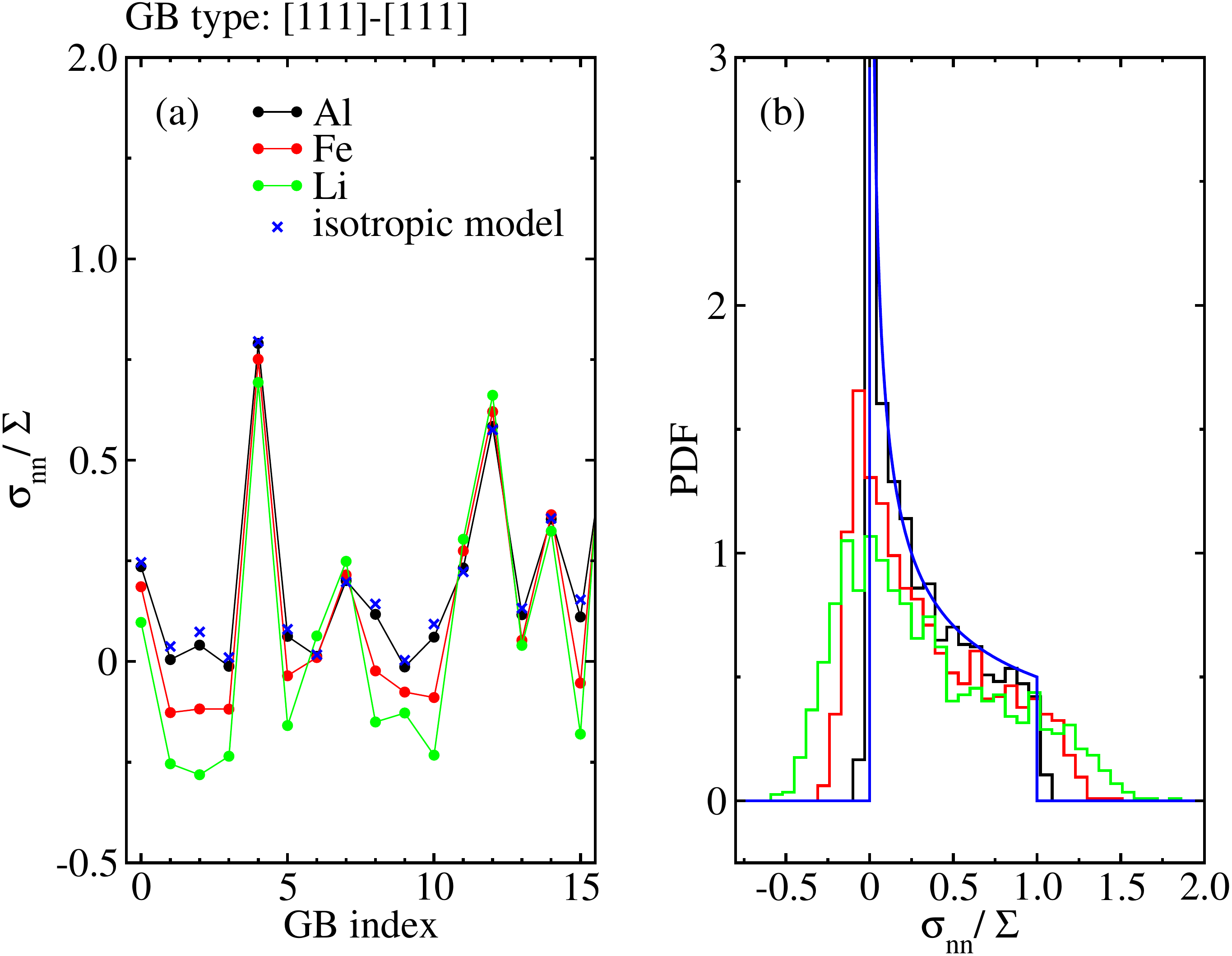}
	\caption{%(a) Local $\snn/\Sigma$ and (b) statistical stress response $\pdf(\snn/\Sigma)$ evaluated on $[111]$-$[111]$ type of GBs under macroscopic tensile loading $\Sigma$, showing significant influence of elastic material properties and poor prediction capability of the isotropic model in case of anisotropic grains (Fe, Li). In panel (a) the results are shown for only 15 GBs.
	\TM{Similarly as in Fig.~\ref{fig:effectGB}, but evaluated on $[111]$-$[111]$ GB type in different materials to demonstrate the effect of their elastic properties. Panel (b) shows how isotropic model begins to fail with the growing anisotropy of the grains.}}
	\label{fig:effectMat}
\end{figure}

Figs.~\ref{fig:effectGB} and~\ref{fig:effectMat} show typical (strong) effects of different GB types and different materials on both, local stresses $\snn$ and the corresponding stress distributions $\pdf(\snn)$, for the assumed macroscopic uniaxial tensile loading $\Sigma$. \TM{In Fig.~\ref{fig:effectGB}, a comparison of different $[abc]$-$[def]$ GB types is made, with $\Delta\omega$ assumed random. Each value of GB index refers to a particular %position of the 
GB within the %aggregate. It implies a specific inclination of the GB plane and fixed configuration and geometry of all the grains, while their crystallographic orientations are allowed to change. 
aggregate of fixed grain topology, shown in Fig.~\ref{fig:geom}(a).
In this way, the effect of the GB type can be isolated from %the contribution of a particular GB neighborhood.
other contributions.%GBs with the same GB index always correspond to the same position within the aggregate, regardless of their GB type, \textit{i.e.}, the GB-plane inclination and grain geometries remain the same, and only crystallographic orientations of all the grains might change to isolate the effect of the GB type.
} 
While the mean stress is independent of the GB type (with $\ave{\snn}=\Sigma/3$ for all types), the stress fluctuations are much larger on the (stiffest) $[111]$-$[111]$ GBs than on the (softest) $[001]$-$[001]$ GBs~\cite{elshawish2021}.

A similar behavior is observed in Fig.~\ref{fig:effectMat}, where the effect of different material properties is isolated from other contributions by comparing $\snn$ on \TM{identical} GBs. All stress distributions $\pdf(\snn)$ are again centered around $\ave{\snn}=\Sigma/3$, \TM{while they are at the same time getting} considerably wider with increasing grain anisotropy~\cite{elshawish2021}.

In most cases depicted in Figs.~\ref{fig:effectGB} and~\ref{fig:effectMat}, a poor prediction capability of the isotropic model%
\footnote{Isotropic model assumes isotropic material properties of the grains, resulting in local stresses that are equal to the applied stress.}
is observed, implying that local GB stresses are non-trivially dependent on the GB type, material properties and loading conditions.

% ==========================================================================
%\section{\label{model}Perturbation theory of grain boundary normal stresses}
\section{\label{model}Perturbative model of grain boundary normal stresses}
\label{sec:3}
% ==========================================================================

\subsection{Assumptions}
%\begin{enumerate}
%\item
%Matrika R, 3 koti, opisi funkcijo vsakega Eul. kota
%
%\item
%Slika pert. iteracij. Zacni z isotropnim modelom, potem bikristal+aksialno vpetje. Podrobno opisi vse korake do koncnega rezultata za snn (in povprecje in std). Upostevaj splosno simetrijo!
%
%\item
%buffer axial, buffer Lt, hkrati vodi se povarianto z povprecenimi twisting koti, ce je rezultat preklobasast, zapisi le funkcijsko odvisnost.
%
%\item
%Na koncu tabela rezultatov za snn, <snn> in std(snn), za nekaj iteracij.
%
%\end{enumerate}

\begin{figure*}
	\includegraphics[width=\columnwidth]{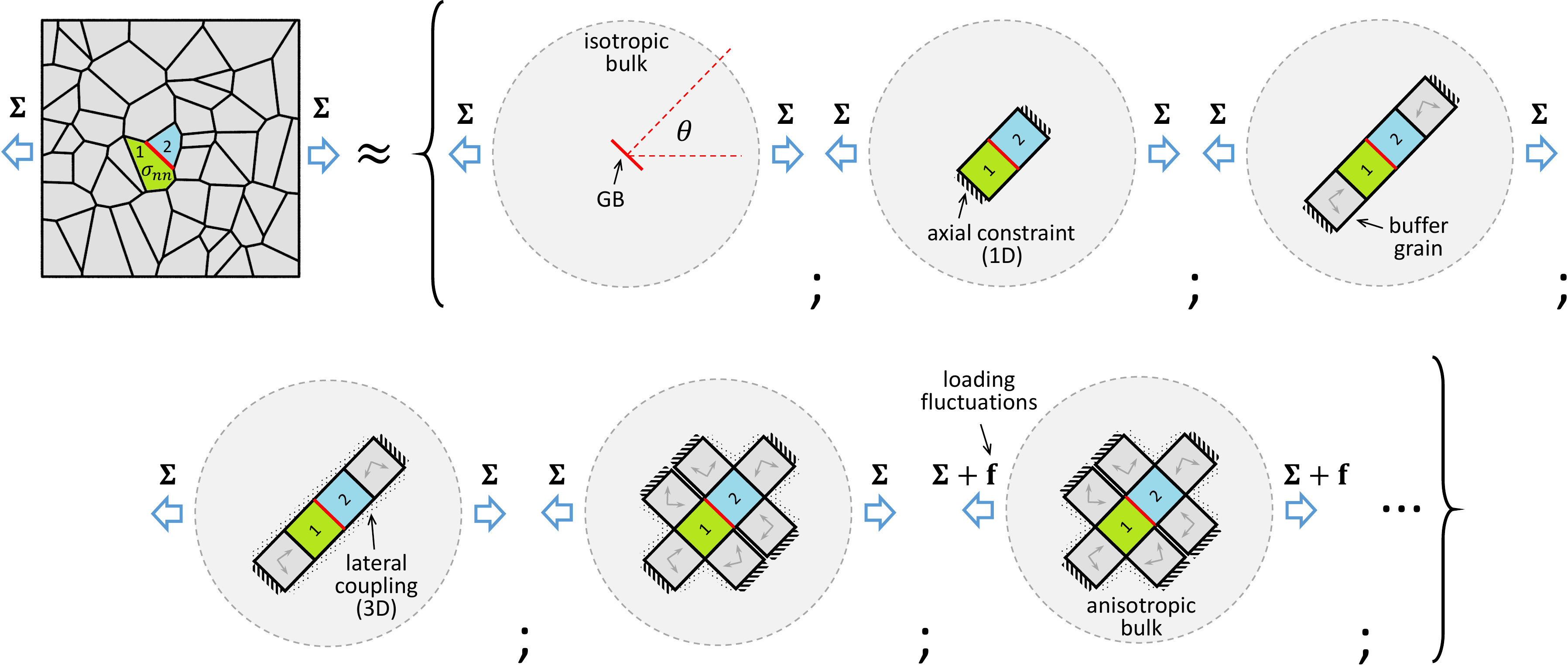}
	\caption{\TM{Perturbation-theory based} strategy for finding GB-normal stress $\snn^{(k)}$. \TM{In each successive step $k$, a more complex GB neighborhood is taken into account. For simplicity, the scheme presented here is only 2D and subjected to tensile loading $\mathbf{\Sigma}$, but in practise a 3D case for a general uniform loading is considered.}}
	\label{fig:pert}
\end{figure*}

To develop an accurate \TM{prediction} for $\snn$ (and the corresponding $\pdf(\snn)$), a %incremental/gradual/stepwise
\TM{step-by-step} approach is taken, \TM{inspired by} perturbation theory. In this sense, the solution for $\snn^{(k)}$, starting from the trivial isotropic-grain solution $\snn^{(0)}$, is \TM{refined} in each successive step $k$ by considering \TM{the contribution of more distant grains% (and thus larger GB neighborhood) to the INS
.} To provide an analytic solution, \TM{sensible} approximations and assumptions are used. For example, following Saint Venant’s principle, the effects of more distant neighborhood on a GB are described in less detail, using only average quantities such as elastic grain anisotropy $A^u$~\cite{ranganathan} or isotropic bulk stiffness $\ave{E}$. The strategy for building a perturbative model is shown schematically in Fig.~\ref{fig:pert}. 
\TM{In the simplest approximation ($k=0$), the neighborhood of a chosen GB can be %neglected entirely
modeled as isotropic, in which case the only relevant degree of freedom is the orientation of the GB plane. %This corresponds to isotropic case. 
In the next order iteration ($k=1$), the two (anisotropic) grains enclosing the GB are %also 
considered% (including their specific structure and orientation), while the remaining grains are modeled by bulk properties. ``Buffer'' grains can then be added to soften the response of the GB stress to the incorrect ``outer'' boundary conditions being used. In the next-to-next order approximation (not considered here), even more grains surrounding the GB could be taken into account.
, while their combined (axial) strain is assumed equal as if both grains were made from (isotropic) bulk material.}

\TM{%Simplest such example is the bicrystal model, where we %prescribe
%postulate, that each pair of grains deforms the same as bulk material of equal dimensions would, if subject to the same external loading. 
This assumption works well for average grains, but the stiffer or softer the grains in the pair are, the more it starts to fail. %For a very stiff GB the amplitude of induced stress is smaller than predicted (for both tensile and compressive loadings) and for a very soft GB the opposite, meaning the estimated standard-deviation curve as a function of GB stiffness is unrealistically steep, see Fig.~\ref{fig:effectE12}. The problem lies in the assumption, that bicrystal pair deforms as bulk material. 
To relax that condition, in the next order iteration ($k=2$), ``buffer'' grains are introduced along the GB-normal (axial) direction. Then not only bicrystal pair, but the whole axial chain (containing also the buffer grains) is supposed to deform as if it was made from bulk material.} In a similar manner, buffer grains are added also along the transverse directions, forming transverse chains of grains whose axial strain is constrained by the bulk ($k=3$).

In the isotropic-grain solution ($k=0$), the GB-normal stress is equal to the externally applied stress projected onto a GB plane, $\snn^{(0)}=\Sigma_{zz}$, which for \TM{uniaxial} loading $\Sigma$ translates to $\snn^{(0)}=\Sigma \cos^2\theta$, where $\theta$ is the angle between the GB normal and loading direction. The isotropic-grain solution may be a good \TM{initial approximation}, but \TM{it turns to be} a poor solution for moderate and highly anisotropic materials, see Figs.~\ref{fig:effectGB} and~\ref{fig:effectMat}. 

\begin{figure}
	\includegraphics[width=0.8\columnwidth]{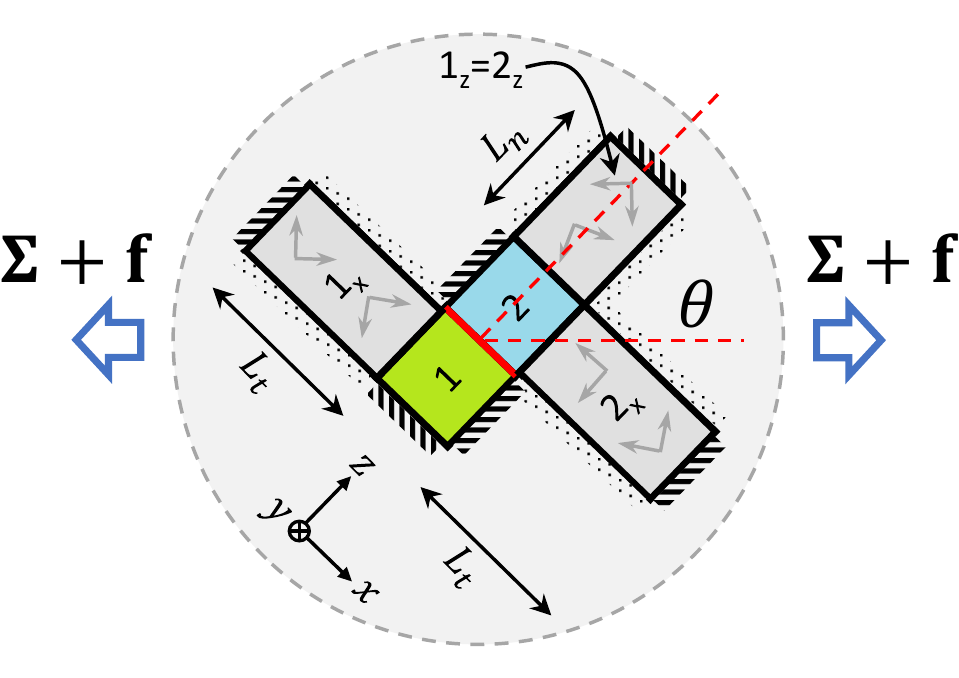}
	\caption{A 2D sketch of \TM{perturbative model for GB stresses}, \TM{consisting} of two anisotropic grains of unit size, \TM{enclosing} the GB, and several isotropic buffer grains of variable length, composing one axial chain of length $L_n+2$ and two (four in 3D) transverse chains of length $L_t+1$. Stresses and strains are assumed constant within the grains. Total strain of each chain is \TM{prescribed} to \TM{match} that of isotropic bulk \TM{of the same length and under the same external loading} (Voigt-like assumption on a chain\TM{-length} scale). 3D coupling of the chains with the surrounding bulk is modeled by assuming variable chain stiffness. External loading $\mathbf{\Sigma}$ is dressed by fluctuations $\mathbf{f}$.}
	\label{fig:chains}
\end{figure}

To obtain higher-order ($k>0$) solutions $\snn^{(k)}$, the effect of two nearest grains \TM{enclosing} the GB is considered in more detail, while the effect of \TM{more distant, buffer} grains is accounted for \TM{less rigorously}. Instead of a full 3D solution, several partial 1D solutions are obtained simultaneously and properly combined to accurately approximate $\snn^{(k)}$. Schematically, the corresponding general model can be viewed in Fig.~\ref{fig:chains}, as composed of one axial chain of length $L_n+2$ and four lateral chains of length $L_t+1$ crossing the two grains, \TM{that are adjacent to GB}.

The %Lateral 
chains are assumed decoupled from each other, \TM{but} they interact with the \TM{surrounding} bulk. \TM{The bulk is taken as} isotropic, with \TM{average (bulk) properties, such as} elastic stiffness $\ave{E}$ and Poisson's ratio $\ave{\nu}$. The chain-bulk interaction is, in the first approximation (\textit{i.e.}, without lateral 3D coupling), assumed to be along the chain direction. \TM{It constrains} the total strain of the chain %\TM{(and thus its length)} 
to that of the isotropic bulk. This boundary condition corresponds to the Voigt-like assumption, but on a chain\TM{-length} scale.

Buffer grains are assumed isotropic \TM{as well, but} with elastic stiffness $E_b$ and Poisson's ratio $\nu_b$, both corresponding to the average response of a chain \TM{with} $L_n$ (or $L_t$) randomly oriented grains. However, when accounting for the lateral 3D effects (cf.~Sec.~\ref{sec:3Deffects}), the chains are \TM{allowed} to interact also laterally with the bulk \TM{and in the limit of long chains} both parameters \TM{approach} those of the bulk, $E_b\sim\ave{E}$ and $\nu_b\sim\ave{\nu}$.

The two grains \TM{on either side of} the GB are assumed anisotropic, with \TM{their crystallographic orientations determining the $[abc]$-$[def]$-$\Delta\omega$ type of the corresponding GB}.

Finally, the stresses and strains are considered homogeneous within all the grains. In addition, a general analytical expression for $\snn^{(k)}$ \TM{is derived by applying a reduced set of %``inner'' 
boundary conditions.
To facilitate a simple, closed-form solution for $\sigma_{nn}^{(k)}$, only the conditions for stresses are imposed at the GB, while those for strains are neglected.
%It is complicated to write down the solution of the Hooke's law for a chosen set of variables (either all the stress or all the strain components) in a simple, closed-form expression. The main problem lies in ``inner'' boundary conditions, since some apply to stresses and some to strains.
%Specifically, we used here only conditions for stresses, \textit{i.e.}, we prescribed only the relation between the stress components on both sides of the GB. 
Hence}, the stress equilibrium is fulfilled everywhere in the model, while the strain compatibility at the GB is not guaranteed.
% 
%\footnote{\TM{Reducing the number of conditions must be compensated by reduction of the number of variables. In our case, we assign the values to the shear-stress modes in both grains %(mind, that shear-strain components of the bicrystal pair's deformation are also not prescribed; cf.~extended (3D) model for $L_n = 0$ and $L_t = 0$)
%as being identical to the corresponding external-stress local components, \textit{i.e.}, $\sigma_{ij}^{(1)} = \sigma_{ij}^{(2)} := \Sigma_{ij}$ ($i\neq j$). This is exactly true only in isotropic case, which thus correctly fulfils all the boundary conditions.}}
These assumptions \TM{will be justified \textit{a posteriori} by comparing the model %of five decoupled 1D chains with known axial strains, whose response is based on them, with the numerical simulations using a complete set of boundary conditions.
results with those from numerical simulations.}

\subsection{Analytical models}
\subsubsection{General setup}

\begin{figure}
	\includegraphics[width=0.8\columnwidth]{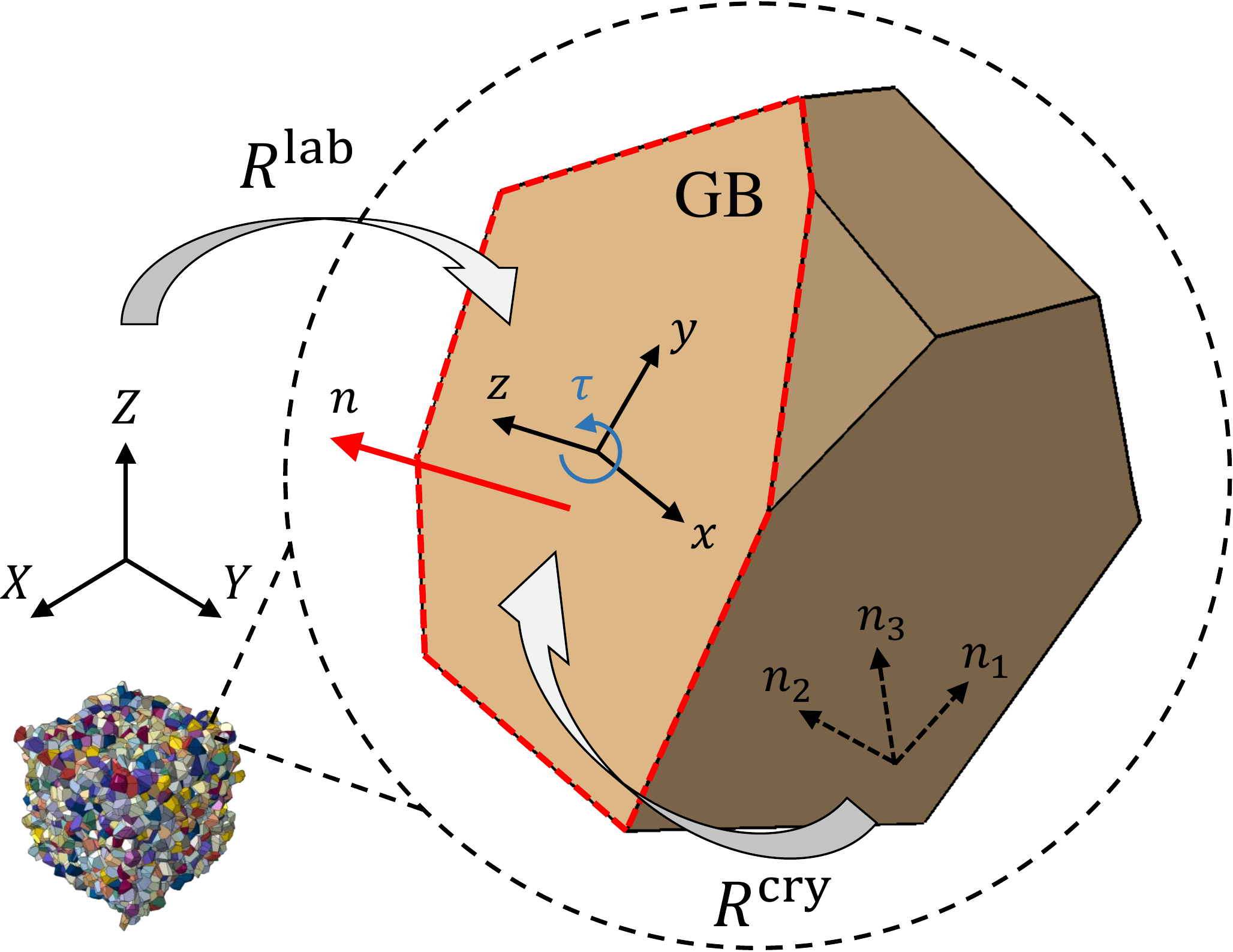}
	\caption{Definition of three coordinate systems: laboratory coordinate system $(X,Y,Z)$, local-grain coordinate system $(n_1,n_2,n_3)$, and GB coordinate system $(x,y,z)$. The latter \TM{can be arbitrarily chosen with respect to the} twist angle $\tau$ \TM{about} the GB normal $\hat{n}||\hat{z}$. Passive rotations $\mathbf{R}^{\text{lab}}$ and $\mathbf{R}^{\TM{\text{cry}}}$ \TM{transform} external and local\TM{-grain} quantities, respectively, to the GB coordinate system. \TM{Since in the following, two grains will be considered, four coordinate systems will be in use, namely one crystallographic system for each grain, together with associated rotations $\mathbf{R}^{\TM{\text{cry}, abc}}$ and $\mathbf{R}^{\TM{\text{cry}, def}}$.}}
	\label{fig:cs}
\end{figure}

Analytical expressions for $\snn^{(k)}$ are \TM{presented in} the GB coordinate system $(x,y,z)$ \TM{with $z$-axis along the GB normal}. All quantities expressed in the local-grain coordinate system $(n_1,n_2,n_3)$, \TM{aligned with crystallographic (eigen-)axes}, and the laboratory coordinate system $(X,Y,Z)$, therefore need to be appropriately \TM{transformed} using the following (passive) rotations $\mathbf{R}^{\TM{\text{cry}}}$ and $\mathbf{R}^{\text{lab}}$, respectively (see also Fig.~\ref{fig:cs}),
\begin{widetext}
\ba
\label{eq:Rloc}
\mathbf{R}^{\text{cry}}&=&\left(
\begin{array}{ccc}
	\phantom{-}\frac{h l \cos \omega }{\sqrt{h^2+k^2} \sqrt{h^2+k^2+l^2}} - \frac{k \sin \omega}{\sqrt{h^2+k^2}} & \phantom{-}\frac{k l \cos \omega}{\sqrt{h^2+k^2} \sqrt{h^2+k^2+l^2}} + \frac{h \sin \omega}{\sqrt{h^2+k^2}} & -\frac{\sqrt{h^2+k^2} \cos \omega
		}{\sqrt{h^2+k^2+l^2}} \\
	-\frac{h l \sin \omega}{\sqrt{h^2+k^2} \sqrt{h^2+k^2+l^2}} - \frac{k \cos \omega}{\sqrt{h^2+k^2}} & -\frac{k l \sin \omega}{\sqrt{h^2+k^2} \sqrt{h^2+k^2+l^2}} + \frac{h \cos \omega}{\sqrt{h^2+k^2}} & \phantom{-}\frac{\sqrt{h^2+k^2} \sin \omega
		}{\sqrt{h^2+k^2+l^2}} \\
	\frac{h}{\sqrt{h^2+k^2+l^2}} & \frac{k}{\sqrt{h^2+k^2+l^2}} & \frac{l}{\sqrt{h^2+k^2+l^2}} \\
\end{array}
\right),\\
\label{eq:Rlab}
\mathbf{R}^{\text{lab}}&=&\left(
\begin{array}{ccc}
	\phantom{-}\cos \theta \cos \psi \cos \phi - \sin \psi \sin \phi & \phantom{-}\cos \theta \sin \psi \cos \phi + \cos \psi \sin \phi & -\sin \theta \cos \phi \\
	-\cos \theta \cos \psi \sin \phi - \sin \psi \cos \phi & -\cos \theta \sin \psi \sin \phi + \cos \psi \cos \phi & \phantom{-}\sin \theta \sin \phi \\
	\sin \theta \cos \psi & \sin \theta \sin \psi & \cos \theta \\
\end{array}
\right).
\ea
\end{widetext}
While standard notation with three Euler angles $(\psi,\theta,\phi)$, \TM{corresponding to a sequence of rotations $\mathbf{R}_1$ about $\hat{n}_3$ (angle $\psi$), $\mathbf{R}_2$ about $\mathbf{R}_1 \hat{n}_2$ (angle $\theta$) and $\mathbf{R}_3$ about $\mathbf{R}_2 \mathbf{R}_1 \hat{n}_3 = \hat{z}$ (angle $\phi$)}, is used for matrix $\mathbf{R}^{\text{lab}}$, the rotation $\mathbf{R}^{\text{cry}}$ is expressed in terms of $(h,k,l,\omega)$, where the GB 
%plane 
normal corresponds to the $[h k l]$ direction%
\footnote{The $[h k l]$ direction is determined by two (not three) independent parameters.} 
in the local-grain coordinate system, and $\omega$ denotes a twist angle \TM{about} the GB normal. This notation is particularly useful for analyzing the response of $[abc]$-$[def]$-$\Delta\omega$ GBs. In the following, we shall always use $(x,y,z)$ to refer to the axes of the GB coordinate system and $(X,Y,Z)$ for laboratory system associated with the external loading $\mathbf{\Sigma}$. In this respect,
\be
\label{eq:sigLAB}
\mathbf{\Sigma}^{\text{lab}}=\left(
\begin{array}{ccc}
	\Sigma _{XX} & \Sigma _{XY} & \Sigma _{XZ} \\
	\Sigma _{XY} & \Sigma _{YY} & \Sigma _{YZ} \\
	\Sigma _{XZ} & \Sigma _{YZ} & \Sigma _{ZZ} \\
\end{array}
\right),
\ee
and
\be
\label{eq:sigGB}
	\mathbf{\Sigma}^{\text{GB}}=\mathbf{R}^{\text{lab}}\mathbf{\Sigma}^{\text{lab}}(\mathbf{R}^{\text{lab}})^T=
	\left(
	\begin{array}{ccc}
		\Sigma _{xx} & \Sigma _{xy} & \Sigma _{xz} \\
		\Sigma _{xy} & \Sigma _{yy} & \Sigma _{yz} \\
		\Sigma _{xz} & \Sigma _{yz} & \Sigma _{zz} \\
	\end{array}
	\right).
\ee

\TM{To find the solution of perturbative models in Fig.~\ref{fig:pert}, the number of variables needs to match the number of boundary conditions.
In isotropic limit, $\sigma_{ij}^{(0)} = \Sigma_{ij}$, and thus there are no constraints and no degrees of freedom. %Nonetheless, it is the exact solution in a sense that it respects/fulfils all the boundary conditions, while in higher order iterations the actual shear-stress modes in either grain are not properly considered. Instead they are approximated by $\Sigma_{ij}$, meaning the correct BCs for strains at the GB have been given up. 
In a bicrystal model with (1D) axial constraint, there is only a single unknown ($\sigma_{zz}^{(1)} = \sigma_{zz}^{(2)}$), and also a single %``outer'' 
constraint on the axial strain ($\epsilon_{zz}^{(1)} + \epsilon_{zz}^{(2)} = 2 \epsilon_{zz}^{\text{bulk}}$). The situation does not change even when buffer grains are added. 
%For the extended (3D) model, there are five variables ($\sigma_{zz}^{(1)} = \sigma_{zz}^{(2)}$, $\sigma_{xx}^{(1)}$, $\sigma_{xx}^{(2)}$, $\sigma_{yy}^{(1)}$ and $\sigma_{yy}^{(2)}$), corresponding to $5$ ``outer'' boundary conditions, while all other stress components are approximated with suitable $\Sigma^{\text{GB}}_{ij}$ components. Thus in reference to Fig.~\ref{fig:chains}}, strains of the five 1D chains are constrained by the following five equations,
For a (3D) model in Fig.~\ref{fig:chains}}, the following set of conditions is used, constraining the axial strains of all five 1D chains:
\ba
\label{eq:constr}
\begin{split}
	L_n \epsilon_{zz}^{(\TM{1_z = 2_z})} +\epsilon_{zz}^{(1)} + \epsilon_{zz}^{(2)}&=(L_n+2)\epsilon_{zz}^{\text{bulk}} , \\
	L_t \epsilon_{xx}^{(1_x)} +\epsilon_{xx}^{(1)}&=(L_t+1)\epsilon_{xx}^{\text{bulk}} , 
	\\
	L_t \epsilon_{xx}^{(2_x)} +\epsilon_{xx}^{(2)}&=(L_t+1)\epsilon_{xx}^{\text{bulk}} , 
	\\
	L_t \epsilon_{yy}^{(1_y)} +\epsilon_{yy}^{(1)}&=(L_t+1)\epsilon_{yy}^{\text{bulk}} , 
	\\
	L_t \epsilon_{yy}^{(2_y)} +\epsilon_{yy}^{(2)}&=(L_t+1)\epsilon_{yy}^{\text{bulk}}.
\end{split}	
\ea
\TM{Strain of each grain is weighted by its length, \textit{i.e.}, either $L_n, L_t\ge 0$ for buffer grains, or $1$ for unit-size GB grains.
Superscript label of each strain-tensor component (and similarly for stresses) indicates to which particular grain it corresponds; $N=1,2$ for GB grains or $N_i$ for buffer grains in $i=x,y,z$ directions.}
%The corresponding grain, to which a particular strain-tensor component belongs to, is indicated by its superscript label; $N=1,2$ for GB grains or $N_i$ for buffer grains in $i=x,y,z$ directions.
%For any strain-tensor component, the corresponding grain is indicated by a superscript label, $N=1,2$ for GB grains or $N_i$ for buffer grains in $i=x,y,z$ directions.
%Superscript labels indicate the corresponding grains, to which particular strain-tensor components belong to; $N=1,2$ for GB grains and $N_i$ for buffer grains in $i=x,y,z$ directions, see Fig.~\ref{fig:chains}.
%$\epsilon_{ii}^{(N)}$ and $\epsilon_{ii}^{(N_i)}$, with $i=x,y,z$ and $N=1,2$, is used to denote normal strain component $i$ in GB grain $N$ and buffer grain $N_i$, where $N$ and $i$ indicate, respectively, the corresponding GB grain and chain direction to which the buffer grain belongs to. 
%Same labeling scheme applies also to stresses.

\TM{Applying} the generalized Hooke's law to GB grain $N$, the $ii$ component of its strain tensor can be written as%
\footnote{\TM{The summation indices $1,2,3$ correspond to $x,y,z$, respectively.}}
\ba
    \epsilon_{ii}^{(N)}&=& \sum_{k,l=1}^{3} s_{iikl}^{\text{GB},N} \, \TM{\sigma_{kl}^{(N)}}\\
    &=& \hspace{-.5cm} \sum_{k,l,m,n,o,p=1}^{3} \hspace{-.6cm} R_{im}^{\text{cry},N} R_{in}^{\text{cry},N} R_{ko}^{\text{cry},N} R_{lp}^{\text{cry},N} \, s_{mnop}^{\text{cry}} \, \TM{\sigma_{kl}^{(N)}} \nonumber,
\ea
%
\begin{comment}
\ba
	\epsilon_{ii}^{(N)}&=& \TM{\sum_{k=1}^{3} s_{iikk}^{\text{GB},N} \, \sigma_{kk}^{(N)}} + \sum_{k=1}^{3}\sum_{l\neq k} s_{iikl}^{\text{GB},N} \, \Sigma_{kl}^{\text{GB}}\\
%
	&=& \TM{\hspace{-.5cm} \sum_{k,m,n,o,p=1}^3 \hspace{-.5cm} R_{im}^{\text{cry},N} R_{in}^{\text{cry},N} R_{ko}^{\text{cry},N} R_{kp}^{\text{cry},N} \, s_{mnop}^{\text{cry}} \, \sigma_{kk}^{(N)}} + \nonumber \\
%	
	& & + \hspace{-.8cm} \sum_{k,l\neq k,m,n,o,p=1}^3 \hspace{-.8cm} R_{im}^{\text{cry},N} R_{in}^{\text{cry},N} R_{ko}^{\text{cry},N} R_{lp}^{\text{cry},N} \, s_{mnop}^{\text{cry}} \, \Sigma_{kl}^{\text{GB}}\nonumber,
\ea
\end{comment}
%
\TM{with all stress-tensor components $\sigma_{kl}^{(N)}$ listed in Table~\ref{tab:load}. Note that shear stresses do not appear as variables in either grain, but have their values assigned ($\sigma_{ij}^{(N)}=\Sigma_{ij}$ for $i\ne j$), \textit{i.e.}, they are set equal to the components of external-stress tensor, rotated to a local GB system; cf.~Eq.~\eqref{eq:sigGB}. Out of the $6$ remaining stress components in both grains, two are set equal due to stress-continuity condition ($\sigma_{zz}^{(1)}=\sigma_{zz}^{(2)}:=\sigma_{zz}$), hence the number of unknowns (five) matches the number of constraints in Eq.~\eqref{eq:constr}.
%For buffer grains $N_i$ similar expression applies, only there all stress components correspond to $\Sigma^{\text{GB}}_{ij}$, except for the stress along the chain length.
}

Compliance tensor $s^{\text{cry}}$, expressed in the \TM{local (crystallographic) coordinate system of the grain, is readily transformed to the GB system, where rotation matrices $\mathbf{R}^{\text{cry}}$ can be different for both grains}. Depending on the symmetry of the grain lattice, $s^{\text{cry}}$ can be expressed as a function of minimum two (isotropic) and maximum $21$ (triclinic) independent elastic parameters. Here, no preference \TM{for} the underlying symmetry is assumed, thus keeping the approach as general as possible.

To maintain the clarity \TM{of the manuscript}, only functional dependence of $\epsilon_{ii}^{(N)}$ is retained here%
\footnote{Parameters in $\mathcal{F}$ are grouped into three sets, \TM{separated by semicolons. They are related either to} material properties, GB orientation or loading.}
(with full analytic expressions for cubic lattice symmetry presented in Appendix~\ref{app:epsnn}),
\ba
\label{eq:eps12}
\begin{split}
	\epsilon_{ii}^{(1)}=\mathcal{F}(s^{\text{cry}};a,b,c,\omega_1;\sigma_{xx}^{(1)},\sigma_{yy}^{(1)},\sigma_{zz},\Sigma_{xy},\Sigma_{xz},\Sigma_{yz}) , \\
	\epsilon_{ii}^{(2)}=\mathcal{F}(s^{\text{cry}};d,e,f,\omega_2;\sigma_{xx}^{(2)},\sigma_{yy}^{(2)},\sigma_{zz},\Sigma_{xy},\Sigma_{xz},\Sigma_{yz}) .
\end{split}
\ea
\TM{Generic $(h,k,l,\omega)$ parameters} in $\mathbf{R}^{\text{cry}}$ have been replaced by specific values $(a,b,c,\omega_1)$ and $(d,e,f,\omega_2)$ in GB grains $1$ and $2$, respectively. This \TM{setting} corresponds to a well-defined GB type $[abc]$-$[def]$-$\Delta\omega$ with $\Delta\omega:=\omega_2-\omega_1$. 

\TM{Similar expressions apply also to buffer grains $N_i$. The only difference is, that there all stress components correspond to projected external loading $\mathbf{\Sigma}^{\text{GB}}$. The only exception is the axial stress $\sigma_{ii}^{(N_i)}$, which matches $\sigma_{ii}^{(N)}$ in GB grain $N$ due to stress-continuity along the chain length. Stress components in each of the grains are summarized in Table~\ref{tab:load}.}

\begin{table}[h]
	\caption{\label{tab:load}%
		Assumed stress components in different grains of the model. Buffer grain \TM{label $N_i$ denotes the corresponding GB grain ($N=1,2$) and the direction of the chain, to which} it belongs ($i=x,y,z$).
	}
	\begin{ruledtabular}
		\begin{tabular}{lll}
			Grain& Assigned stresses&	Unknown stresses\\
			\colrule
			GB grain $1$ & $\sigma_{ij}^{(1)}=\Sigma_{ij}$, $i\ne j$ & $\sigma_{xx}^{(1)}$, $\sigma_{yy}^{(1)}$, $\sigma_{zz}^{(1)}$\\
			GB grain $2$ & $\sigma_{ij}^{(2)}=\Sigma_{ij}$, $i\ne j$ & $\sigma_{xx}^{(2)}$, $\sigma_{yy}^{(2)}$, $\sigma_{zz}^{(2)}=\sigma_{zz}^{(1)}$\\
			buffer $1_x$ & $\sigma_{ij}^{(1_x)}=\Sigma_{ij}$, $\TM{ij\ne xx}$ & $\sigma_{xx}^{(1_x)}=\sigma_{xx}^{(1)}$\\
			buffer $1_y$ & $\sigma_{ij}^{(1_y)}=\Sigma_{ij}$, $\TM{ij\ne yy}$ & $\sigma_{yy}^{(1_y)}=\sigma_{yy}^{(1)}$\\
			buffer $2_x$ & $\sigma_{ij}^{(2_x)}=\Sigma_{ij}$, $\TM{ij\ne xx}$ & $\sigma_{xx}^{(2_x)}=\sigma_{xx}^{(2)}$\\
			buffer $2_y$ & $\sigma_{ij}^{(2_y)}=\Sigma_{ij}$, $\TM{ij\ne yy}$ & $\sigma_{yy}^{(2_y)}=\sigma_{yy}^{(2)}$\\
			buffer $1_z(=2_z)$ & $\sigma_{ij}^{(1_z)}=\Sigma_{ij}$, $\TM{ij\ne zz}$  & $\sigma_{zz}^{(1_z)}=\sigma_{zz}^{(1)}$	
		\end{tabular}
	\end{ruledtabular}
\end{table}

\TM{Sufficiently far from the GB, the grains can be treated as isotropic. This allows for much simpler expressions for strain components $\epsilon_{ii}$ (with $i=x,y,z$) in both, buffer grains and the bulk material,}
\ba
\label{eq:epsbuf}
	\epsilon_{ii}^{(N_i)}&=&\frac{1}{E_b}\left(\sigma_{ii}^{(N)}-\nu_b(\operatorname{tr}(\mathbf{\Sigma}^{\text{GB}})-\Sigma_{ii})\right),\\
\label{eq:epsbul}
	\epsilon_{ii}^{\text{bulk}}&=&\frac{1}{\ave{E}}\left(\Sigma_{ii}-\ave{\nu}(\operatorname{tr}(\mathbf{\Sigma}^{\text{GB}})-\Sigma_{ii})\right).
\ea

\TM{With relevant strain components in individual grains defined in Eqs.~\eqref{eq:eps12}, \eqref{eq:epsbuf} and~\eqref{eq:epsbul}, the set of conditions in Eq.~\eqref{eq:constr}, constraining the axial strains of all five chains, can be solved analytically} for all five unknown stresses $\sigma_{ii}^{(N)}$, including the \TM{GB-normal stress} $\snn:=\sigma_{zz}$.

However, the resulting $\snn$ has a significant deficiency. It depends on the choice of the \TM{local} GB coordinate system (the value of twist angle $\tau$ in Fig.~\ref{fig:cs}). This dependence originates in the prescribed directions of the four lateral chains, which are directed along the \TM{local} $x$ and $y$ axes. A different choice of $x$ and $y$ axes would produce different lateral constraints, \TM{which would result in a} different $\snn$. To avoid this ambiguity, a symmetrized lateral boundary condition is derived below.

\subsubsection{Symmetrized model}

\TM{The model is symmetrized by averaging the lateral boundary condition over all possible GB coordinate systems. The twist of the local GB system for an arbitrary angle $\tau$ about the GB normal ($z$-axis) changes how $\mathbf{\Sigma}^{\text{GB}}$ and $s^{\text{GB},N}$ are expressed in that system. Specifically, the rotation changes the Euler angles $\omega_N$ and $\phi$ in transformation matrices~\eqref{eq:Rloc} and~\eqref{eq:Rlab}, respectively,
\ba
\begin{split}
\omega_N & \to \omega_N+\tau \, ; \quad (N=1,2), \\
\phi & \to \phi + \tau, 
\end{split}
\ea
which in turn affect Eqs.~\eqref{eq:eps12}, \eqref{eq:epsbuf} and~\eqref{eq:epsbul}, and make them $\tau$ dependent.
Since all twist rotations should be equivalent}, averaging over $\tau$ \TM{replaces Eq.~\eqref{eq:constr} with} new, symmetrized boundary conditions
\ba
\begin{split}
\label{eq:constrSym}
	\TM{\tfrac{1}{2\pi}}\hspace{-.1cm}\int_0^{2\pi}\hspace{-.2cm}\left(L_n \epsilon_{zz}^{(\TM{1_z = 2_z})} +\epsilon_{zz}^{(1)} +	\epsilon_{zz}^{(2)}\right) d\tau&=\TM{\tfrac{1}{2\pi}}\hspace{-.1cm}\int_0^{2\pi}\hspace{-.2cm}(L_n+2)\epsilon_{zz}^{\text{bulk}} d\tau , \\
	\TM{\tfrac{1}{2\pi}}\hspace{-.1cm}\int_0^{2\pi}\hspace{-.2cm}\left(L_t \epsilon_{xx}^{(1_x)} +\epsilon_{xx}^{(1)}\right) d\tau&=\TM{\tfrac{1}{2\pi}}\hspace{-.1cm}\int_0^{2\pi}\hspace{-.2cm}(L_t+1)\epsilon_{xx}^{\text{bulk}}d\tau , \\
	\TM{\tfrac{1}{2\pi}}\hspace{-.1cm}\int_0^{2\pi}\hspace{-.2cm}\left(L_t \epsilon_{xx}^{(2_x)} +\epsilon_{xx}^{(2)}\right) d\tau&=\TM{\tfrac{1}{2\pi}}\hspace{-.1cm}\int_0^{2\pi}\hspace{-.2cm}(L_t+1)\epsilon_{xx}^{\text{bulk}}d\tau , \\
	\TM{\tfrac{1}{2\pi}}\hspace{-.1cm}\int_0^{2\pi}\hspace{-.2cm}\left(L_t \epsilon_{yy}^{(1_y)} +\epsilon_{yy}^{(1)}\right) d\tau&=\TM{\tfrac{1}{2\pi}}\hspace{-.1cm}\int_0^{2\pi}\hspace{-.2cm}(L_t+1)\epsilon_{yy}^{\text{bulk}}d\tau , \\
	\TM{\tfrac{1}{2\pi}}\hspace{-.1cm}\int_0^{2\pi}\hspace{-.2cm}\left(L_t \epsilon_{yy}^{(2_y)} +\epsilon_{yy}^{(2)}\right) d\tau&=\TM{\tfrac{1}{2\pi}}\hspace{-.1cm}\int_0^{2\pi}\hspace{-.2cm}(L_t+1)\epsilon_{yy}^{\text{bulk}}d\tau .
\end{split}	
\ea
Solving the above set of symmetrized equations for five unknowns $\sigma_{ii}^{(N)}$ (with $i=x,y,z$ and $N=1,2$), provides analytical $\snn:=\sigma_{zz}$, independent of $\tau$. However, \TM{for the most general case} the resulting expression is too cumbersome to be \TM{presented here}. \TM{Hence, we again resort to its} functional dependence
\begin{widetext}
\ba
\begin{split}
\label{eq:snnGeneral}
	\snn&={\mathcal F}(s^{\text{cry}},\ave{E},\ave{\nu},E_b,\nu_b;a,b,c,\omega_1,d,e,f,\omega_2;L_n,L_t;\mathbf{\Sigma}^{\text{\TM{GB}}}) \\
		&={\mathcal F}(s^{\text{cry}},\ave{E},\ave{\nu},E_b,\nu_b;a,b,c,\omega_1-\phi,d,e,f,\omega_2-\phi;L_n,L_t;\mathbf{\Sigma}^{\text{lab}},\theta,\psi) ,
\end{split}
\ea
\end{widetext}
from which it is clear, that $\snn$ does not depend on the choice of the GB coordinate system due to observed $\omega_1-\phi$ and $\omega_2-\phi$ dependence. The normal stress $\snn$ is a \TM{complicated} function of many parameters%
\footnote{To account for loading fluctuations due to anisotropy of the bulk, a universal elastic anisotropy index $A^u$ should be added to the list of influencing parameters (see Sec.~\ref{sec:gauss}). \TM{On the other hand, $A^u$, $\ave{E}$ and $\ave{\nu}$ are all only functions of $s^{\text{cry}}$.}}
(\textit{e.g.}, up to \TM{$39$} independent parameters in a material with triclinic lattice symmetry and for a most general external loading). However, not all parameters are of same significance, as shown is Sec.~\ref{sec:comp}, where $\snn$ is tested against numerical results. In order to derive a compact, but still meaningful expression, further approximations are needed.

So far, the strategy \TM{was based on} adding more complexity to the model when getting closer to the GB. In this respect, grains closest to it have been modeled in greater detail (\textit{e.g.}, employing anisotropic elasticity and mostly unknown loading conditions), while the grains further away required less modeling (\textit{e.g.}, employing isotropic elasticity and mostly known loading conditions).

With the goal to provide a compact and accurate analytical expression for $\snn$ (and the corresponding $\pdf(\snn)$), few selected limits of the general result, Eq.~$\eqref{eq:snnGeneral}$, are investigated and discussed in more detail. Some of these limits will become very useful later, when a comparison with the numerical results is done in Sec.~\ref{sec:comp}.

\subsubsection{Isotropic limit ($k=0$)}
\label{sec:iso}

\TM{The initial (zeroth order) approximation $\snn^{(0)}$, representing the exact solution in the isotropic material limit, can be reproduced from Eq.~$\eqref{eq:snnGeneral}$ in two ways, either by assuming isotropic properties of the grains (\textit{i.e.,} by taking the appropriate $s^{\text{cry}}$) or taking the limit of very long chains ($L_n,L_t\to\infty$) with average properties ($E_b=\ave{E}$, $\nu_b=\ave{\nu}$)}, in which the chain-strain constraints become ineffective, resulting in stresses equal to external loading,
\ba
	\snn^{(0)}&=&\Sigma_{zz}\\
	 	&=&\Sigma_{XX}\sin^2\theta\cos^2\psi+\Sigma_{YY}\sin^2\theta\sin^2\psi \nonumber\\
	 	&+&\Sigma_{ZZ}\cos^2\theta + \Sigma_{XY}\sin^2\theta\sin2\psi \nonumber\\
		&+&\Sigma_{XZ}\sin2\theta\cos\psi + \Sigma_{YZ}\sin2\theta\sin\psi\nonumber.
\ea
Having a sufficient number of GBs with normals uniformly distributed on a sphere, the corresponding first two statistical moments of $\pdf(\snn^{(0)})$, the mean \TM{value} and standard deviation, can be straightforwardly expressed as
\ba
\begin{split}
\label{eq:isopdf}
	\ave{\snn^{(0)}}&=&\frac{1}{3} \operatorname{tr}(\mathbf{\Sigma}^{\text{lab}}) , \\
	s(\snn^{(0)})&=&\frac{2}{3\sqrt{5}}\ \Sigma^{\text{lab}}_{\text{mis}} ,
\end{split}
\ea
where \TM{$\tfrac{1}{3} \operatorname{tr}(\mathbf{\Sigma}^{\text{lab}})$ is a hydrostatic pressure, related to volume change of the aggregate,} and $\Sigma^{\text{lab}}_{\text{mis}}$ corresponds to von Mises external stress,
\TM{traditionally associated with the yielding of ductile materials% and the transition from elastic regime to plasticity
. Von Mises stress is related to deviatoric tensor (responsible for volume preserving shape changes of the aggregate),
\ba
\begin{split}
	\label{eq:mises}
	\Sigma^{\text{lab}}_{\text{mis}}&:=\frac{\sqrt{3}}{\sqrt{2}} \sqrt{\operatorname{tr}\left ((\mathbf{\Sigma}^{\text{lab}}_{\text{dev}})^2 \right )} , \\
    \label{eq:deviatoric}
    \mathbf{\Sigma}^{\text{lab}}_{\text{dev}} &:= \mathbf{\Sigma}^{\text{lab}} - \frac{1}{3}  \operatorname{tr}(\mathbf{\Sigma}^{\text{lab}}) \mathbb{1}_{3\times 3}.
\end{split}
\ea
Both, $\operatorname{tr}(\mathbf{\Sigma}^{\text{lab}})$ and $\Sigma^{\text{lab}}_{\text{mis}}$, are rotational invariants and thus assume identical form in all coordinate systems.% Hence, Eq.~\eqref{eq:isopdf} %does not change under coordinate transformations and remains 
%looks exactly the same when expressed with laboratory stress components $\mathbf{\Sigma}^{\text{lab}}$ or local GB components $\mathbf{\Sigma}^{\text{GB}}$; cf.~Eqs.~\eqref{eq:sigLAB} and~\eqref{eq:sigGB}, respectively.
}

\TM{Even though Eq.~\eqref{eq:isopdf} is derived for isotropic case ($k=0$), the same functional dependence of the first two statistical moments on $\mathbf{\Sigma}^{\text{lab}}$ is retained for all orders $k$, suggesting that the loading part can be trivially decoupled from the material and GB-type contributions. In a specific case, when the external stress is of hydrostatic form (\textit{i.e.}, proportional to identity matrix; $\mathbf{\Sigma}^{\text{lab}} := \Sigma_0 \, \mathbb{1}_{3\times 3}$), this can be easily confirmed. In that case, there is no effect of grain orientations, since stress tensor is invariant to rotations. Hence, the trivial (hydrostatic) solution applies to the whole aggregate ($\snn^{(k)} = \Sigma_0$), resulting in an infinitely narrow stress (and strain) distribution. On the other hand $\Sigma^{\text{lab}}_{\text{mis}} = 0$, therefore $s(\snn)\sim\Sigma^{\text{lab}}_{\text{mis}}$ applies for any, not just isotropic material.}

\subsubsection{Axially constrained bicrystal ($k=1$)}
\label{sec:bi}

The first non-trivial solution $\snn^{(1)}$ corresponds to a bicrystal, embedded axially in the isotropic bulk ($L_n\to 0$). As there are no lateral constraints imposed \TM{on} the two GB grains, this model corresponds to the $L_t\to\infty$ limit of the general model shown in Fig.~\ref{fig:chains}. However, to obtain a compact expression for $\snn^{(1)}$, another \TM{simplification is required}, which will be justified in Sec.~\ref{sec:comp}. Since we are interested in the response of $[abc]$-$[def]$-$\Delta\omega$ GBs, which have a well-defined difference of the two twist angles, $\snn^{(1)}$ is obtained by \TM{replacing $\omega_2$ in Eq.~\eqref{eq:snnGeneral} with $\omega_1+\Delta\omega$, and averaging it over $\omega_1$:}  
\ba
\begin{split}
\label{eq:bi}
	\snn^{(1)}&:=\frac{1}{2\pi}\int_{0}^{2\pi}\left.\left( \lim_{\substack{L_n\to0\\L_t\to\infty}}\snn\right)\right\vert_{\omega_2=\omega_1+\Delta\omega} \hspace{-1.5cm} d\omega_1\\
	&= \Enn\Sigma_{zz} + \Enn\left(\nu_{12}-\ave{\nu}\right)\left(\Sigma_{xx}+\Sigma_{yy}\right) ,
\end{split}
\ea
where
\ba
\label{eq:e12nu12}
\begin{split}
	\Enn&=\frac{2\ave{E}^{-1}}{E_{abc}^{-1}+E_{def}^{-1}}=\frac{2\ave{E}^{-1}}{s^{\text{GB},abc}_{3333}+s^{\text{GB},def}_{3333}} , \\
	\nu_{12}&=-\frac{\ave{E}}{4}\left(s^{\text{GB},abc}_{3311}+s^{\text{GB},abc}_{3322}+s^{\text{GB},def}_{3311}+s^{\text{GB},def}_{3322}\right) ,
\end{split}
\ea
and
\ba
	s^{\text{GB},hkl}_{33jj}&=&\hspace{-.4cm}\sum_{m,n,o,p=1}^3 \hspace{-.4cm} R_{3m}^{\text{cry},hkl} R_{3n}^{\text{cry},hkl} R_{jo}^{\text{cry},hkl} R_{jp}^{\text{cry},hkl} s_{mnop}^{\text{cry}} , \phantom{xxx}
\ea
for $j=1,2,3$ and $hkl = abc$ or $def$.
\TM{This approximation removes (averages out) all the twist-angle degrees of freedom. We will refer to it as the \textit{reduced} version of the model, intended to mimic the behavior observed in numerical studies.}
The derived compact expression for $\snn^{(1)}$ is the first main result of this study. It suggests that GB-normal stress is a simple function of the loading part, \TM{contained in} $\Sigma_{xx}$, $\Sigma_{yy}$ and $\Sigma_{zz}$, and the GB\TM{-type} (and material) part, which is represented compactly by only two (composite) parameters $\Enn$ and $\nu_{12}$. While $\Enn$ has already been introduced in Ref.~\cite{elshawish2021} as an effective GB stiffness, measuring the average stiffness of GB neighborhood along the GB-normal direction, the newly introduced $\nu_{12}$ can be seen as an effective GB Poisson's \TM{ratio}, measuring the average ratio of \TM{transverse and} axial responses \TM{(strains)} in both GB grains. Both $\Enn$ and $\nu_{12}$ are unitless and characterize the $[abc]$-$[def]$-$\Delta\omega$ GB neighborhood in terms of local material and GB-type parameters%
\footnote{\TM{With the exception of $\Delta\omega$, whose influence is implicitly removed from Eq.~\eqref{eq:bi} by integration over $\omega_1$.}},
\ba
	\Enn&=&\mathcal{F}(s^{\text{cry}},\ave{E},a,b,c,d,e,f) , \\
	\nu_{12}&=&\mathcal{F}(s^{\text{cry}},\ave{E},a,b,c,d,e,f) .
\ea
Full analytic expressions for $\Enn$ and $\nu_{12}$ \TM{(as well as $\ave{E}$ and $\ave{\nu}$)} depend on the choice of the grain lattice symmetry (expressions for cubic lattice symmetry are given in Appendix~\ref{app:epsnn}). Note that expressions simplify considerably with more symmetric lattices. In cubic lattices, for example, a GB is fully characterized by $\Enn$ \TM{alone}, since $\nu_{12}=\ave{\nu}+\tfrac{1}{2}(\Enn^{-1}-1)$. In isotropic grains, $\Enn=1$ and $\nu_{12}=\ave{\nu}$, which recovers the $\snn^{(0)}$ solution.

Switching to a statistical behavior of \TM{infinitely} many $[abc]$-$[def]$-$\Delta\omega$ GBs with randomly \TM{oriented} GB planes, the first two statistical moments of $\pdf(\snn^{(1)})$, the mean \TM{value} and standard deviation, become
\ba
\begin{split}
\label{eq:bipdf}
	\ave{\snn^{(1)}}&=\frac{\operatorname{tr}(\mathbf{\Sigma}^{\text{lab}})}{3}\Enn\left(1+2(\nu_{12}-\ave{\nu})\right),\\
	s(\snn^{(1)})&=\frac{2\ \Sigma^{\text{lab}}_{\text{mis}}}{3\sqrt{5}} \Enn\sqrt{\left(1-\nu_{12}+\ave{\nu}\right)^2}.
\end{split}
\ea
For cubic lattices they simplify to $\ave{\snn^{(1)}}=\operatorname{tr}(\mathbf{\Sigma}^{\text{lab}})/3$ and $s(\snn^{(1)})=\Sigma^{\text{lab}}_{\text{mis}}/(3\sqrt{5}) \left|1-3\Enn\right|$%, \TM{and for isotropic grains to Eq.~\eqref{eq:isopdf}}
. 
The fact that the mean stress is equal to $\Sigma/3$ for the uniaxial loading $\Sigma$, \TM{while the fluctuation of GB-normal stress in cubic grains is a monotonic function of a single GB parameter $\Enn$ (although the functional dependence differs from that of Eq.~\eqref{eq:bipdf}),
%with $\Enn$ being the only GB parameter determining the fluctuations of GB-normal stresses} in cubic grains, 
has already been identified in (realistic) FE simulations~\cite{elshawish2021}.} However, the observed behavior can now be easily extended to other non-cubic lattices and for general external loading.
The accuracy of the derived expressions for local $\snn^{(1)}$, Eqs.~\eqref{eq:bi}--\eqref{eq:e12nu12}, and statistical $\pdf(\snn^{(1)})$, Eq.~\eqref{eq:bipdf}, is investigated in more detail in Sec.~\ref{sec:comp}.

\subsubsection{Axially constrained chain \TM{with} $L_n+2$ grains ($k=2$)}
\label{sec:axial}

The next-order solution $\snn^{(2)}$ corresponds to a single chain with $L_n+2$ grains, axially constrained by the isotropic bulk. The reason for adding a buffer grain of length $L_n>0$ \TM{to the bicrystal is to relax the axial strain constraint. In the previous ($k=1$) iteration, this constraint applies directly to the bicrystal, which produces too large (resp. small) stresses $\snn^{(1)}$ on very stiff (resp. soft) GBs, see Sec.~\ref{sec:comp}.}

Following the same reasoning and steps as in the bicrystal model, the resulting \textit{reduced} version is derived for a general grain-lattice symmetry and arbitrary external loading
\begin{widetext}
\ba
\begin{split}
\label{eq:chain}
	\snn^{(2)}&:=\frac{1}{2\pi}\int_{0}^{2\pi}\left.\left( \lim_{L_t\to\infty}\snn\right)\right\vert_{\omega_2=\omega_1+\Delta\omega} \hspace{-1.5cm} d\omega_1 \\
	&= \frac{2+L_n}{2\Enn^{-1}+L_n E_3^{-1}}\Sigma_{zz}+\frac{2}{2\Enn^{-1}+L_n E_3^{-1}}  \left(\nu_{12}-\ave{\nu}-\frac{1}{2} L_n\left(\ave{\nu}-\nu_b E_3^{-1}\right)\right) \left(\Sigma_{xx}+\Sigma_{yy}\right) \\
	&\approx\frac{2+L_n}{2\Enn^{-1}+L_n}\Sigma_{zz}+\frac{2 \left(\nu_{12}-\ave{\nu}\right)}{2\Enn^{-1}+L_n}\left(\Sigma_{xx}+\Sigma_{yy}\right) .
\end{split}
\ea
\end{widetext}
Same definitions for $\Enn$ and $\nu_{12}$ apply as in Eq.~\eqref{eq:e12nu12}, while $E_3:= E_b/\ave{E}$ and $\nu_b$ denote, respectively, the normalized elastic stiffness and Poisson's \TM{ratio} of the (isotropic) buffer grain. \TM{Its response corresponds to the average response of a chain with $L_n$ randomly oriented grains} 
\ba
\begin{split}
	\label{eq:buffer}
	E_b&:= E_{L_n}^{\text{rnd}}=\ave{\frac{L_n}{\sum_{i}s^{\text{GB},i}_{3333}}}_{L_n} , \\
	\nu_b&:=\nu_{L_n}^{\text{rnd}}=-\ave{\frac{\sum_{i} s^{\text{GB},i}_{1133}}{\sum_{i} s^{\text{GB},i}_{3333}}}_{L_n} .
\end{split}
\ea
The averaging $\ave{\ldots}_{L_n}$ is assumed over all \TM{possible linear configurations of $L_n$ grains with random orientations}, and the summation \TM{index $i$ runs} over the grains in each chain. %Using the averaging $\ave{\ldots}_{L_n}$, an isotropic response is produced even for finite (small) $L_n$.

The elastic response of a buffer grain, \TM{calculated in this way, is usually} softer than that of the bulk ($E_3<1$). Nevertheless, it is convenient to assume $E_3\approx 1$ and $\nu_b\approx\ave{\nu}$. In fact, this assumption becomes realistic, when the 3D effects are considered, \textit{e.g.,} the lateral coupling of buffer grain \TM{to} the neighboring bulk (see Sec.~\ref{sec:3Deffects}).

Assuming $E_3=1$ and $\nu_b=\ave{\nu}$, the mean \TM{value} and standard deviation of $\pdf(\snn^{(2)})$ become
\ba
\begin{split}
\label{eq:chainpdf}
	\ave{\snn^{(2)}}&=&\frac{\operatorname{tr}(\mathbf{\Sigma}^{\text{lab}})}{3}\frac{2+L_n+4(\nu_{12}-\ave{\nu})}{2\Enn^{-1}+L_n} , \\
	s(\snn^{(2)})&=&\frac{2\ \Sigma^{\text{lab}}_{\text{mis}}}{3\sqrt{5}}\frac{\sqrt{\left(2+L_n-2(\nu_{12}-\ave{\nu})\right)^2}}{2\Enn^{-1}+L_n},
\end{split}
\ea
which simplify for cubic lattices to $\ave{\snn^{(2)}}=\operatorname{tr}(\mathbf{\Sigma}^{\text{lab}})/3$, $s(\snn^{(2)})=2\Sigma^{\text{lab}}_{\text{mis}}/(3\sqrt{5}) \sqrt{\left ((3+L_n)-\Enn^{-1}\right )^2}/(2\Enn^{-1}+L_n)$. 

In contrast to the bicrystal model, the $\snn^{(2)}$ expression depends also on the parameter $L_n$, which \TM{makes it} a mixture of bicrystal \TM{solution} $\snn^{(1)}$ (reproduced for $L_n\to 0$) and isotropic \TM{solution} $\snn^{(0)}$ (reproduced for $L_n\to\infty$). However, the effect of $L_n$ is negligible for GBs with $\Enn\sim 1$ and $\nu_{12}\sim\ave{\nu}$. As shown in Sec.~\ref{sec:comp}, the value $L_n\sim 2$ best replicates the numerical results.\\

\subsubsection{Axially constrained chains \TM{with} $L_n+2$ and $L_t+1$ grains ($k=3$)}
\label{sec:extended}

The highest-order solution considered in this study is  $\snn^{(3)}$. It corresponds to the complex configuration of chains, shown in Fig.~\ref{fig:chains}. The axial chain consists of $L_n+2$ grains and the four transverse chains of $L_t+1$ grains. All the chains are assumed to be axially constrained to the \TM{strain of} isotropic bulk \TM{of equal length}. \TM{In a similar fashion to previous iterations}, the \textit{reduced} version can be derived for a general grain-lattice symmetry and \TM{arbitrary} external loading
\ba
\begin{split}
\label{eq:INS_general}
	\snn^{(3)}&:= \frac{1}{2\pi}\int_{0}^{2\pi}\snn\bigg\vert_{\omega_2=\omega_1+\Delta\omega} \hspace{-1.4cm} d\omega_1 \\
	&=A^{(3)}\Sigma_{zz} + B^{(3)} (\Sigma_{xx}+\Sigma_{yy}) ,
\end{split}
\ea
where, assuming $E_3=1$ and $\nu_b=\ave{\nu}$,
\begin{widetext}
\ba
\begin{split}
	A^{(3)} &= \frac{(2+L_n) (s^{abc}_{tt}+\ave{E}^{-1} L_t)  (s^{def}_{tt}+\ave{E}^{-1} L_t) + \ave{\nu}  ((s^{abc}_{tt}+\ave{E}^{-1} L_t) s^{def}_{tl} + (s^{def}_{tt}+\ave{E}^{-1} L_t) s^{abc}_{tl})}{(2 E_{12}^{-1}+L_n) (s^{abc}_{tt}+\ave{E}^{-1} L_t)  (s^{def}_{tt}+\ave{E}^{-1} L_t) - \tfrac{1}{2} \ave{E} ((s^{abc}_{tt}+\ave{E}^{-1} L_t) (s^{def}_{tl})^2 + (s^{def}_{tt}+\ave{E}^{-1} L_t) (s^{abc}_{tl})^2)} , \\
	B^{(3)} &= -\frac{2 \, \ave{\nu} (s^{abc}_{tt}+\ave{E}^{-1} L_t)  (s^{def}_{tt}+\ave{E}^{-1} L_t) + \tfrac{1}{2} (1+L_t-\ave{\nu}) ((s^{abc}_{tt}+\ave{E}^{-1} L_t) s^{def}_{tl} + (s^{def}_{tt}+\ave{E}^{-1} L_t) s^{abc}_{tl})}{(2 E_{12}^{-1}+L_n) (s^{abc}_{tt}+\ave{E}^{-1} L_t)  (s^{def}_{tt}+\ave{E}^{-1} L_t) - \tfrac{1}{2} \ave{E} ((s^{abc}_{tt}+\ave{E}^{-1} L_t) (s^{def}_{tl})^2 + (s^{def}_{tt}+\ave{E}^{-1} L_t) (s^{abc}_{tl})^2)} , \\ \label{eq:C_trans_general} 
\end{split}
\ea
\end{widetext}

for\\

\ba
\begin{split}
	s^{hkl}_{tt} &:= \frac{1}{2} \left(s_{1111}^{\text{GB}, hkl} + s_{2222}^{\text{GB}, hkl}\right) + s_{1122}^{\text{GB}, hkl} , \\ 
	s^{hkl}_{tl} &:= s_{1133}^{\text{GB}, hkl} + s_{2233}^{\text{GB}, hkl} , \\
	s^{hkl}_{ll} &:= s_{3333}^{\text{GB}, hkl} := E_{hkl}^{-1} .  \label{eq:combination}
\end{split}
\ea
The combinations of compliance-tensor components%
\footnote{\TM{The compliance-tensor components $s_{ijkl}^{\text{GB}, hkl}$ depend on the twist angle $\omega$, but their linear combinations, defined in Eq.~\eqref{eq:combination}, do not. Hence, the reduced model solution $\snn^{(3)}$ in Eq.~\eqref{eq:INS_general} is indeed independent of $\Delta\omega$.}
},
introduced in Eq.~\eqref{eq:combination}, are related through a material dependent (but GB type independent) linear combination
\ba
\begin{split}
	2 s^{hkl}_{tt} + 2 s^{hkl}_{tl} + s^{hkl}_{ll} = & (s_{1111}^{\text{cry}} + s_{2222}^{\text{cry}} + s_{3333}^{\text{cry}})+ \\
	&+ 2 (s_{1122}^{\text{cry}} + s_{1133}^{\text{cry}} + s_{2233}^{\text{cry}}),
\end{split}	
\ea
which suggests that $\snn^{(3)}$ is a function of (at most) four local GB parameters (in addition to bulk properties $\ave{E}$, $\ave{\nu}$ and chain parameters $L_n$, $L_t$). \TM{In the $L_t\to\infty$ limit, $\snn^{(3)}$ reduces to $\snn^{(2)}$, see Eq.~\eqref{eq:chain}.}

The corresponding first two statistical moments can also be expressed analytically (but they are not shown here for brevity). They have the already familiar loading dependence,
\ba
\begin{split}
	\ave{\snn^{(3)}}&\sim\frac{\operatorname{tr}(\mathbf{\Sigma}^{\text{lab}})}{3} , \\
	s(\snn^{(3)})&\sim\frac{2\Sigma^{\text{lab}}_{\text{mis}}}{3\sqrt{5}} .
\end{split}
\ea

Expressions simplify further for higher lattice symmetries. For cubic lattices, for example, $A^{(3)}$ and $B^{(3)}$ become (again) only functions of Young's moduli $E_{abc}$ and $E_{def}$ along the GB-normal direction (see Appendix~\ref{app:cubic})%
\footnote{\TM{For cubic lattices, the $E_{abc}$ and $E_{def}$ parameters appear in a single combination ($\Enn$) in $\snn^{(1)}$ and $\snn^{(2)}$, while in $\snn^{(3)}$ there are two such combinations ($\Enn$ and $\Delta_{12}$), see Appendix~\ref{app:cubic}.}}.

All \TM{compact-form} solutions $\snn^{(k)}$, representing the special limits of the general solution, Eq.~\eqref{eq:snnGeneral}, are summarized in Table~\ref{tab:models}.

\begin{table*}
	\caption{\label{tab:models}
		A summary of derived models. \TM{Analytical solutions can be written in a compact form only in certain limits.}
	}
	\begin{ruledtabular}
		\begin{tabular}{llllll}
			\TM{$k$} & Model & Version\footnote{In contrast to the full version, the reduced version of the model \TM{eliminates} the twist-angle degrees of freedom, which makes the solution only approximate, but significantly more condensed. Note that both versions provide the same mean \TM{value} $\ave{\snn^{(k)}}$.} & Assumptions\footnote{Assumptions are taken with respect to the general solution; cf.~Eq.~\eqref{eq:snnGeneral}.}&	Fitting parameters & Compact solution\footnote{Compact solutions are derived for a general grain-lattice symmetry.}\\
			\colrule
			$0$ & isotropic & full & $A^u=0$ or $L_n,L_t\to\infty$& - & $\snn^{(0)}=\mathcal{F}(\mathbf{\Sigma}^{\text{GB}})$\\
%
%			\colrule
			$1$ & bicrystal & full & $L_n\to 0,L_t\to\infty$ & - & - \\
					 & & reduced        & $L_n\to 0,L_t\to\infty, \int d\omega_1\vert_{\omega_2=\omega_1+\Delta\omega}$  & - &  $\snn^{(1)}=\mathcal{F}(\mathbf{\Sigma}^{\text{GB}},\Enn,\nu_{12},\ave{E},\ave{\nu})$\\
%
%			\colrule
			$2$ & $L_n$-chain & full & $L_t\to\infty$ & $L_n\ge 0$ & -\\
				        & & reduced        & $L_t\to\infty, \int d\omega_1\vert_{\omega_2=\omega_1+\Delta\omega}$ & $L_n\ge 0$ & $\snn^{(2)}=\mathcal{F}(\mathbf{\Sigma}^{\text{GB}},\Enn,\nu_{12},\ave{E},\ave{\nu},L_n)$\\
%			
%			\colrule
			$3$ & $L_n$-$L_t$-chain & full & - & $L_n, L_t\ge 0$ & -\\
							  & & reduced        & $\int d\omega_1\vert_{\omega_2=\omega_1+\Delta\omega}$ & $L_n, L_t\ge 0$ & $\snn^{(3)}=\mathcal{F}(\mathbf{\Sigma}^{\text{GB}},s^{abc}_{tt},s^{abc}_{ll},s^{def}_{tt},s^{def}_{ll},\ave{E},\ave{\nu},L_n,L_t)$
		\end{tabular}
	\end{ruledtabular}
\end{table*}

\subsection{Models validation}
\label{sec:comp}

\begin{figure}
	\includegraphics[width=0.9\columnwidth]{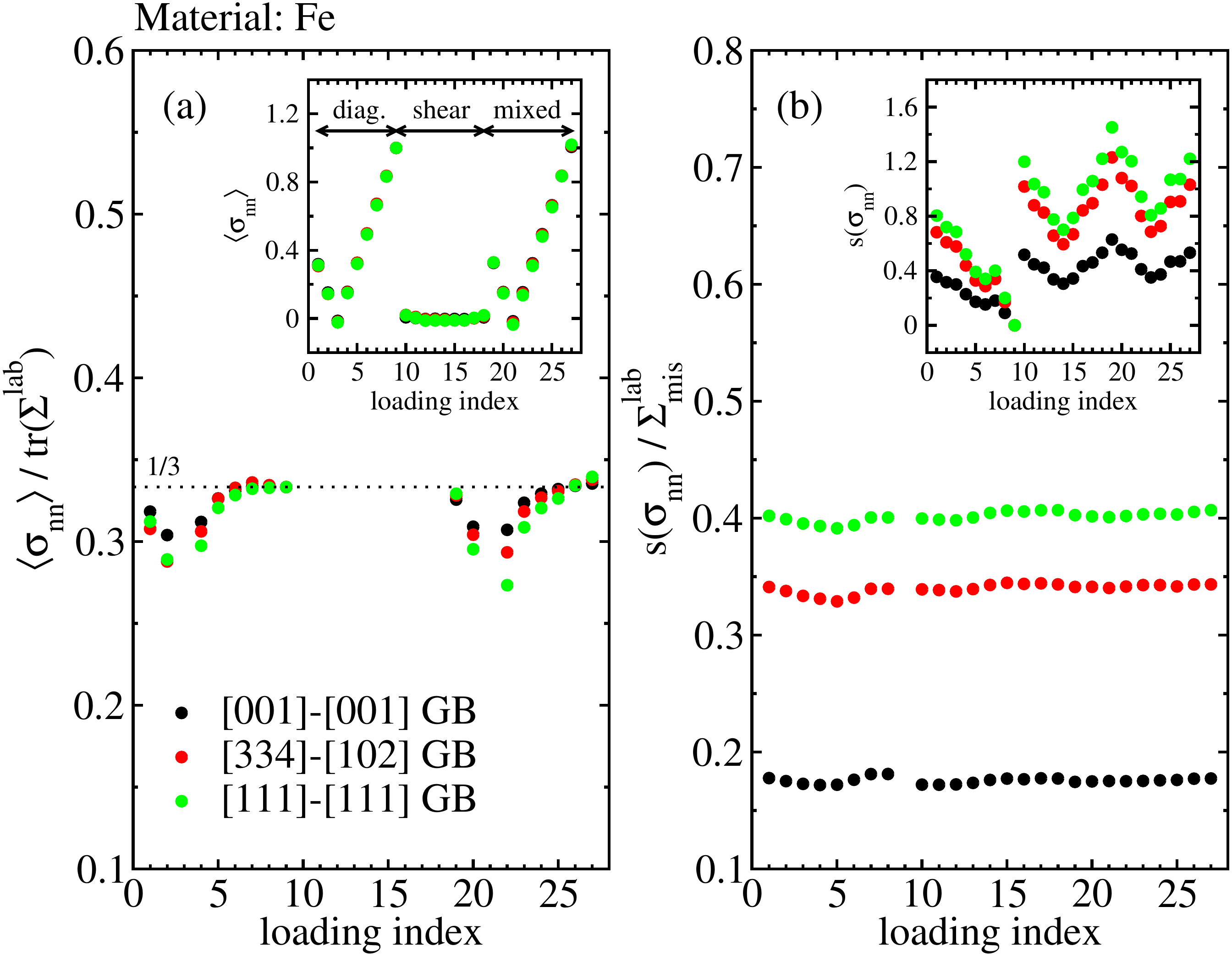}
	\vskip 0.5cm
	\includegraphics[width=0.9\columnwidth]{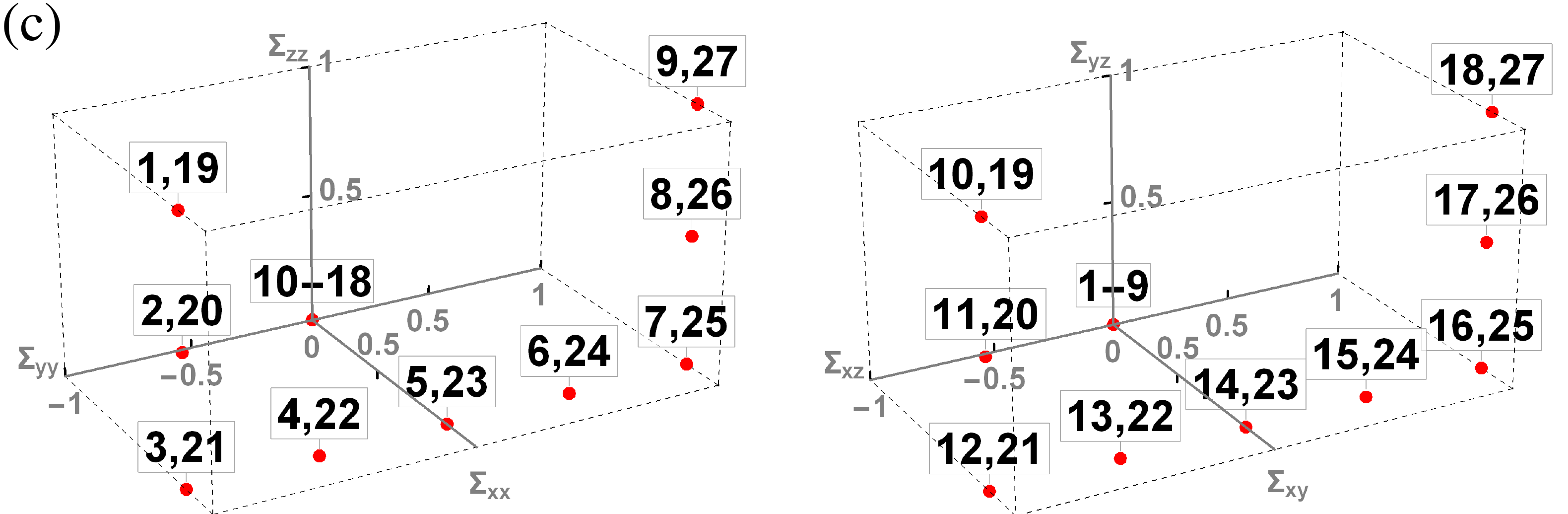}
	\caption{Effect of external loading can be decoupled from other influences by a suitable choice of normalization factor (a) $\operatorname{tr}(\mathbf{\Sigma}^{\text{lab}})$ for mean value $\ave{\snn}$ and (b) $\Sigma^{\text{lab}}_{\text{mis}}$ for standard deviation $s(\snn)$. Simulation results are shown for Fe, $27$ different external loadings $\mathbf{\Sigma}^{\text{lab}}$ and three GB types. Non-normalized values are shown in the insets~(a) and~(b). Panel~(c) shows correspondence between $\mathbf{\Sigma}^{\text{lab}}$ and loading index ($1$-$27$).}
	\label{fig:effectLoadNorm}
\end{figure}

In this section, the solutions $\snn^{(k)}$ of derived models are tested against numerical results.%
\footnote{Having the exact constitutive (Hooke's) law, there are practically no physical uncertainties in numerical simulations besides finite size effects, which can be diminished by using sufficiently large aggregates and sufficiently dense finite element meshes.}
For demonstration purposes, only cubic elastic materials are chosen for comparison (see Appendix~\ref{app:mat} for the corresponding elastic properties).

Following the derived expressions, Eqs.~\eqref{eq:isopdf}, \eqref{eq:bipdf} and~\eqref{eq:chainpdf}, the mean \TM{value} $\ave{\snn}$ and standard deviation $s(\snn)$ of $\pdf(\snn)$ should depend trivially on the external loading $\mathbf{\Sigma}^{\text{lab}}$. Using suggested normalization, $\ave{\snn}/\operatorname{tr}(\mathbf{\Sigma}^{\text{lab}})$ and $s(\snn)/\Sigma^{\text{lab}}_{\text{mis}}$, the first two statistical moments become independent of $\mathbf{\Sigma}^{\text{lab}}$, which is demonstrated in Fig.~\ref{fig:effectLoadNorm} for $27$ different loading configurations. Very good agreement%
\footnote{Observed deviations from $1/3$ in Fig.~\ref{fig:effectLoadNorm}(a) are due to numerical artifacts which result from the division of two small numbers, $\ave{\snn}/\operatorname{tr}(\mathbf{\Sigma}^{\text{lab}})$, and the fact that $\ave{\snn}$ is approximate.}
between the prediction and numerical results confirms the validity of the derived expressions being of the form $\snn^{(k)}=A^{(k)} \Sigma_{zz}+B^{(k)} (\Sigma_{xx}+\Sigma_{yy})$ for any $k$. Hence, a tensile loading $\Sigma$ will be used hereafter without the loss of generality.

\begin{figure}
	\includegraphics[width=0.9\columnwidth]{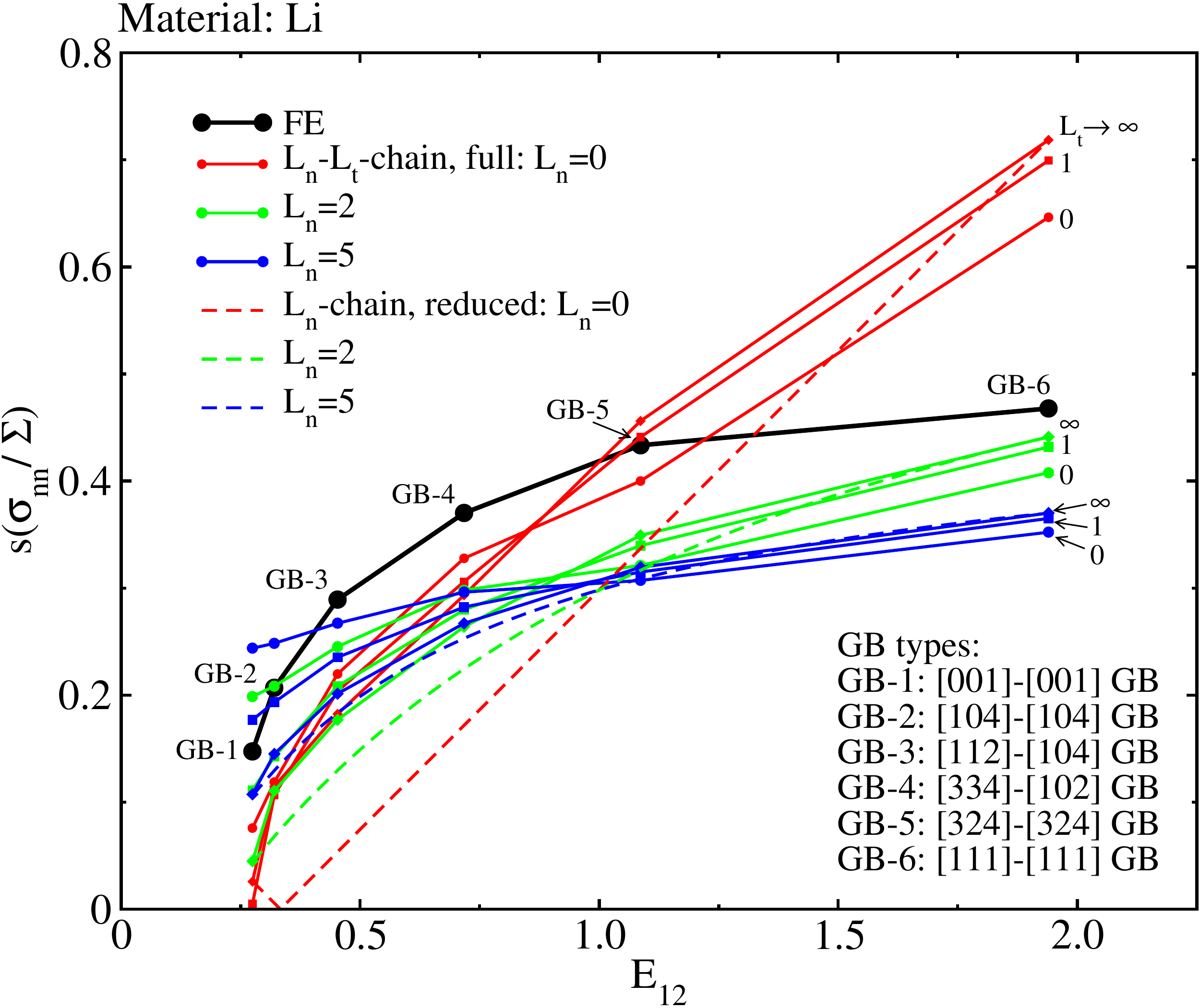}
	\caption{Standard deviation $s(\snn/\Sigma)$ as a function of effective GB-stiffness parameter $\Enn$. A comparison is shown between numerical results (FE) and different model predictions from Table~\ref{tab:models}. The results are evaluated for Li on all GBs of a specific type, and thus corresponding to a certain $\Enn$ value. Six GB types are used in total (here, solid lines are unphysical and are meant only to indicate the trend).}
	\label{fig:effectE12}
\end{figure}

In Fig.~\ref{fig:effectE12} the normalized standard deviation $s(\snn/\Sigma)$ is shown for polycrystalline Li (cubic symmetry) as a function of effective GB stiffness parameter $\Enn$, which is a single characteristic parameter of the $[abc]$-$[def]$ GB. Results of different models from Table~\ref{tab:models} are compared with the results of finite element simulations. The Li is chosen because of very high elastic anisotropy ($A^u=7.97$), which makes the comparison more challenging. 

Although none of model predictions for $s(\snn/\Sigma)$ are very accurate, some of the models are more appropriate than others. The $L_n$-$L_t$-chain (full version) model results are grouped into three families (with a given color) with a common axial chain length $L_n=0,2$ or 5. While the response of the $L_n=0$ (red) family is too steep for all transverse chain lengths $L_t$, overestimating the $s(\snn/\Sigma)$ at large $\Enn$, the response of $L_n=2$ (green) and $L_n=5$ (blue) families is too gradual for $L_t\lesssim 2$, overestimating the $s(\snn/\Sigma)$ at small $\Enn$. These models are recognized as inappropriate. In addition, all three model families show, for $L_t=0$, a sudden change in the slope of $s(\snn/\Sigma)$, which is not observed numerically, suggesting that $L_t=0$ models are also unsuitable. Most favorable are therefore $2\lesssim L_n\lesssim 5, L_t\gtrsim 1$ models, which predict $s(\snn/\Sigma)$ consistently below the numerical curve. This systematic underestimation of fluctuations is compensated later in Sec.~\ref{sec:gauss} by accounting for loading fluctuations, which are generated by external loading mediated through the anisotropic bulk surrounding the $L_n$-$L_t$-chain model (see last stage in Fig.~\ref{fig:pert}).

In Fig.~\ref{fig:effectE12} also the results of the $L_n$-chain (reduced version) model are shown for comparison. The advantage of the latter is the compact formulation of the $\snn^{(2)}$ and its statistical moments. The resulting $s(\snn/\Sigma)$ curves show similar dependence of $\Enn$ as the corresponding $L_n$-$L_t\to\infty$ (full version) models, however, with slightly reduced fluctuations in the mid-$\Enn$ range%
\footnote{Since the responses on $[001]$-$[001]$ GB type (with corresponding $E_{12,\text{min}}$) and $[111]$-$[111]$ GB type (with corresponding $E_{12,\text{max}}$) are independent of twist angles $\omega_1, \omega_2$, the predictions of the $L_n$-chain model (reduced version) and $L_n$-$L_t$-chain model (full version, with $L_t\to\infty$) are the same.}.
Since in either case additional fluctuations need to be added to fit the numerical results, also the validity of the $L_n$-chain model, with $2\lesssim L_n\lesssim 5$, is considered appropriate.

\begin{figure}
	\includegraphics[width=0.9\columnwidth]{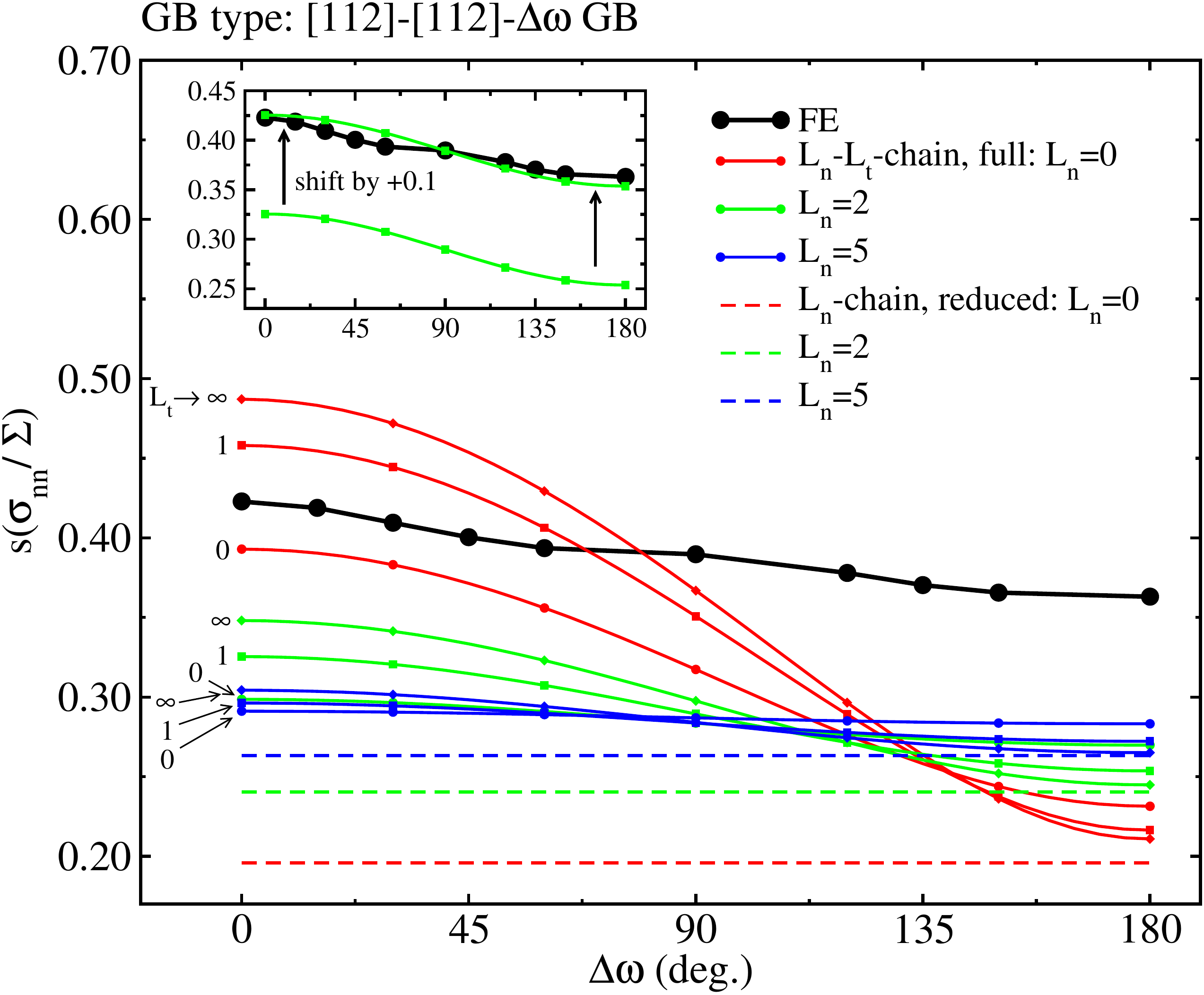}
	\caption{Standard deviation $s(\snn/\Sigma)$ as a function of twist-angle difference $\Delta\omega$ associated with the $[112]$-$[112]$-$\Delta\omega$ GB type. A comparison is shown between numerical results (FE) and different model predictions from Table~\ref{tab:models}. The  properties of Li are used. Note that $[112]$-$[112]$-$\Delta\omega$ GBs correspond to $\Enn=0.77$, irrespective of the value  of $\Delta\omega$. Inset shows the agreement between the FE result and the response of $L_n$-$L_t$-chain model for $L_n=2$ and $L_t=1$ (note the artificial shift accounting for the missing fluctuations).}
	\label{fig:effectdom}
\end{figure}

In Fig.~\ref{fig:effectdom} a response of a single $[112]$-$[112]$-$\Delta\omega$ GB type is shown in terms of $s(\snn/\Sigma)$ as a function of $\Delta\omega$, using numerical simulations and model predictions from Table~\ref{tab:models} for a polycrystalline Li to associate with results of Fig.~\ref{fig:effectE12}. According to the numerical curve, very small variations in $s(\snn/\Sigma)$ are observed across the whole $\Delta\omega$ range, which is consistent with previous results~\cite{elshawish2021}. Using the same coloring and labeling scheme as in Fig.~\ref{fig:effectE12}, a very good agreement with simulations is achieved for the $L_n$-$L_t$-chain model for $L_n=2$ and $L_t=1$ (see the inset of Fig.~\ref{fig:effectdom}). The other two families of curves produce either too big ($L_n=0$, in red) or too small ($L_n=5$, in blue) variations across the $\Delta\omega$ range. Since the twist angle degrees of freedom are integrated out, the response of the $L_n$-chain (reduced version) model is independent of $\Delta\omega$, which is, by design, also in good agreement with numerical results.

Results of Figs.~\ref{fig:effectE12} and~\ref{fig:effectdom} seem to \TM{favor} the $L_n$-$L_t$-chain model with $L_n\sim 2$ and $L_t\sim 1$. This is further corroborated by noting that the overall shift in $s(\snn/\Sigma)$ (by $\sim$0.1), used to fit the simulation results in the inset of Fig.~\ref{fig:effectdom}, is matching very well the gap at $\Enn=0.77$ (corresponding to $[112]$-$[112]$-$\Delta\omega$ GB) between the two corresponding curves in Fig.~\ref{fig:effectE12}.

\begin{figure}
	\includegraphics[width=0.9\columnwidth]{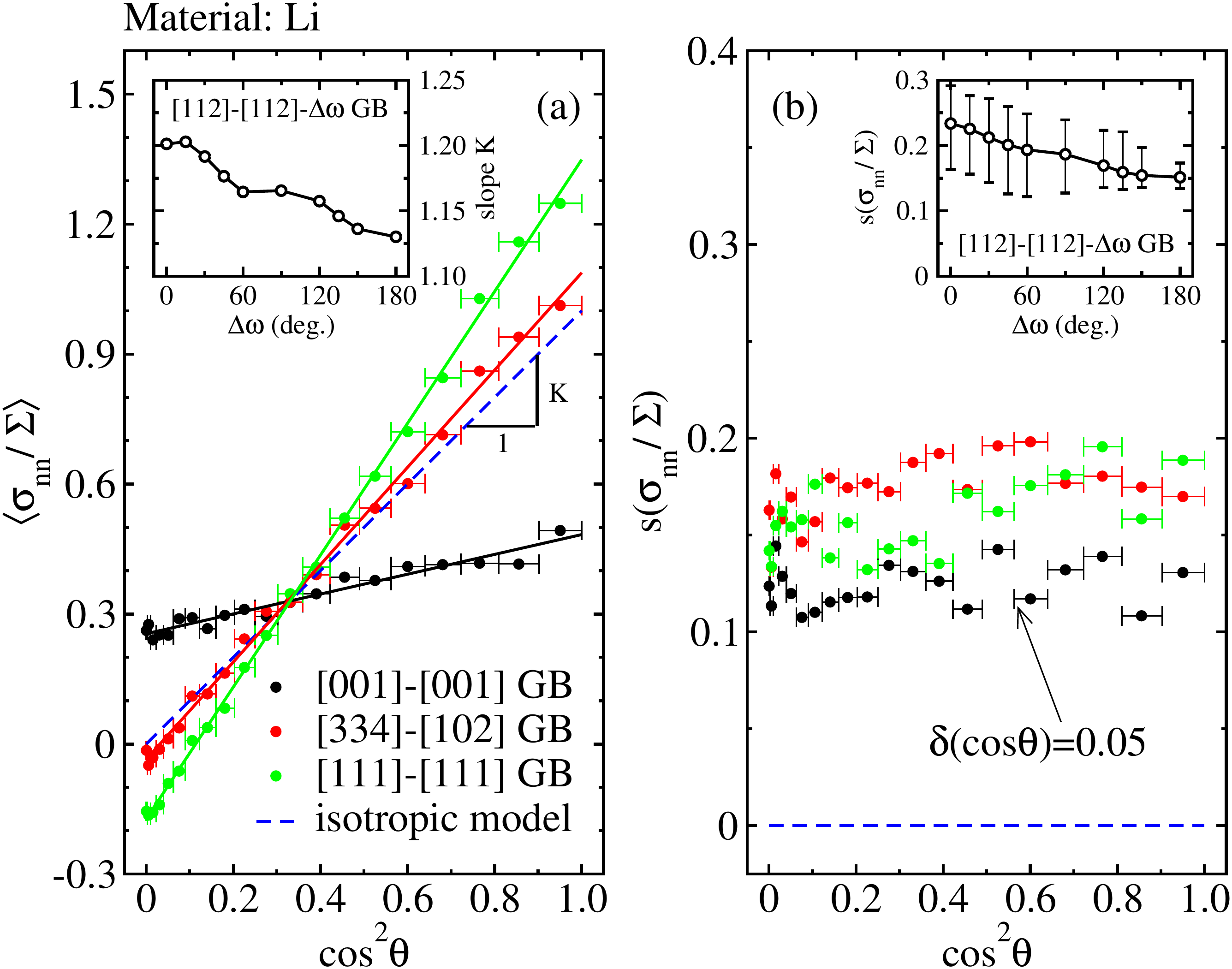}
	\caption{(a) Mean local stress $\ave{\snn/\Sigma}$ and (b) corresponding standard deviation $s(\snn/\Sigma)$, both evaluated numerically as a function of $\cos^2\theta$ for a finite range of GB tilt angles, $\delta(\cos\theta)=0.05$. The averaging range is denoted by horizontal error bars and the averaged values by dots. Lines in panel (a) are linear fits with slope $K$. Twist angle difference $\Delta\omega$ in $[112]$-$[112]$-$\Delta\omega$ GB has a negligible influence on slope $K$ (inset (a)) but a significant effect on the standard deviation (inset (b)).}
	\label{fig:effectCos2}
\end{figure}

\begin{figure}
	\includegraphics[width=0.9\columnwidth]{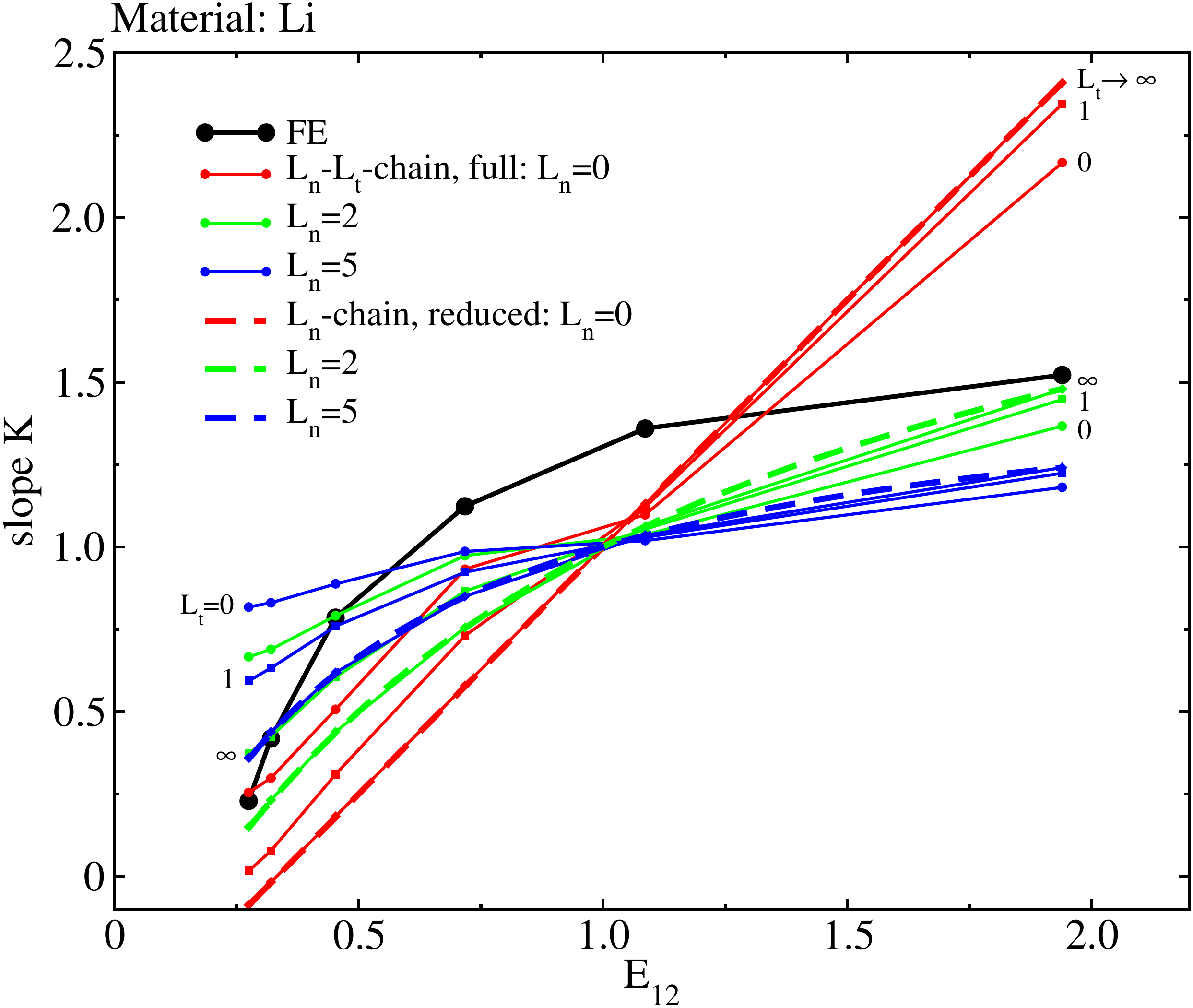}
	\caption{Slope $K$, obtained from Fig.~\ref{fig:effectCos2}(a), versus effective GB-stiffness parameter $\Enn$. A comparison is shown between numerical results (FE) and different model predictions from Table~\ref{tab:models}. The  properties of Li are used. Note that $L_n$-chain model (reduced version) and $L_n$-$L_t$-chain model (full version with $L_t\to\infty$) provide identical slopes $K=1+3(\Enn-1)/(2+L_n\Enn)$, assuming $E_3=1$. There is no effect of $L_n$ when $\Enn=E_3= 1$.}
	\label{fig:effectK}
\end{figure}

In the following, the evaluation of derived models is shifted from macro- to mesoscale using a linear correlation property, $\snn^{(k)}/\Sigma=A^{(k)}+B^{(k)}\cos^2\theta$, derived for the external uniaxial loading $\Sigma$%
\footnote{For cubic lattices and $E_3=1$, $A^{(2)}=(1-\Enn)/(2+L_n\Enn)$ and $B^{(2)}=1+3(\Enn-1)/(2+L_n\Enn)$.}.
Statistical analysis employed on a subset of GBs with a fixed angle $\theta$ (or $\cos^2\theta$) between the GB normal and uniaxial loading direction is useful because it allows one to test the validity of individual parts of expressions in $\snn^{(k)}$ (\textit{e.g.}, $A^{(k)}$ and $B^{(k)}$). Such analyses are demonstrated in Figs.~\ref{fig:effectCos2} and~\ref{fig:effectK} for polycrystalline Li.

The local mean $\ave{\snn/\Sigma}$ and standard deviation%
\footnote{The $s(\snn/\Sigma)$ results from Fig.~\ref{fig:effectCos2}(b) are discussed latter in Sec.~\ref{sec:gauss}.}
$s(\snn/\Sigma)$ are shown in Fig.~\ref{fig:effectCos2} as a function of $\cos^2\theta$. Due to finite aggregate size, the mean and standard deviation are obtained at given $\cos^2\theta$ by averaging over Euler angles $\psi$ and $\phi$ on a finite (but small) range of GB tilt angles $\delta(\cos\theta)=0.05$. The proposed linear trend is nicely reproduced, showing a clear effect of different GB types on the corresponding slopes $K$ of fitted lines. In general, slope $K$ increases with increasing GB stiffness (parameter $\Enn$, see also Fig.~\ref{fig:effectK}). However, there is a very weak effect of $\Delta\omega$ on the corresponding slope $K$ when evaluated on the $[112]$-$[112]$-$\Delta\omega$ GB. This suggests that, on average, the GB stiffness (which is independent of $\Delta\omega$, see Eq.~\eqref{eq:E12cubic}) is the main contributor to $\snn$ at given $\cos^2\theta$.
	
It is interesting to note a crossing point in Fig.~\ref{fig:effectCos2}(a) at $\cos^2\theta=1/3$ at which $\snn$ becomes independent of both material and GB type properties. This point is exactly reproduced by all non-trivial models ($\snn^{(k)}, k>0$). The value of $\snn$ at this point is $\Sigma/3$ (actually $\operatorname{tr}(\mathbf{\Sigma})/3$ for arbitrary loading). 

The simulation results from Fig.~\ref{fig:effectCos2}(a) are analyzed further in Fig.~\ref{fig:effectK} where the actual dependence of slope $K$ with $\Enn$ is presented and compared with predictions of the models from Table~\ref{tab:models}. While the increasing trend is captured well by all the models, none of the presented curves fit the numerical result very accurately for all $\Enn$. Actually, this holds true for any combination of $L_n, L_t$ values in the $L_n$-$L_t$-chain model. The most suitable solution is chosen to be that of the $L_n$-chain model (reduced version) for $L_n=2$, which matches the true $K$ values at the two extreme $\Enn$ points. The same agreement is observed also for other cubic materials (not shown).

The $L_n$-$L_t$-chain model with $L_n\sim 2$ and $L_t\sim 1$, which has been selected as the most suitable model at the macroscale (see Figs.~\ref{fig:effectE12} and~\ref{fig:effectdom}), provides in Fig.~\ref{fig:effectK} a very similar $K(\Enn)$ response as the $L_n$-chain model for $L_n=2$. In this sense, both models seem equally well acceptable, however, the latter one will be preferred due to much more compact formulation.

In summary, although the qualitative behavior of $\snn$ is well reproduced by the selected model ($\snn^{(2)}$ for $L_n\sim 2$) on a wide range of parameters (associated with external loading, material properties and GB type), two ingredients still seem to be missing. The first one is related to the systematic shortage of stress fluctuations observed on the macroscale and the second one is linked to the insufficient agreement of mean stresses on the mesoscale. Both issues are addressed in the next section.

\subsection{Model upgrades}
\subsubsection{Variable axial strain constraint and 3D effects}
\label{sec:3Deffects}

\begin{figure}
	\includegraphics[width=\columnwidth]{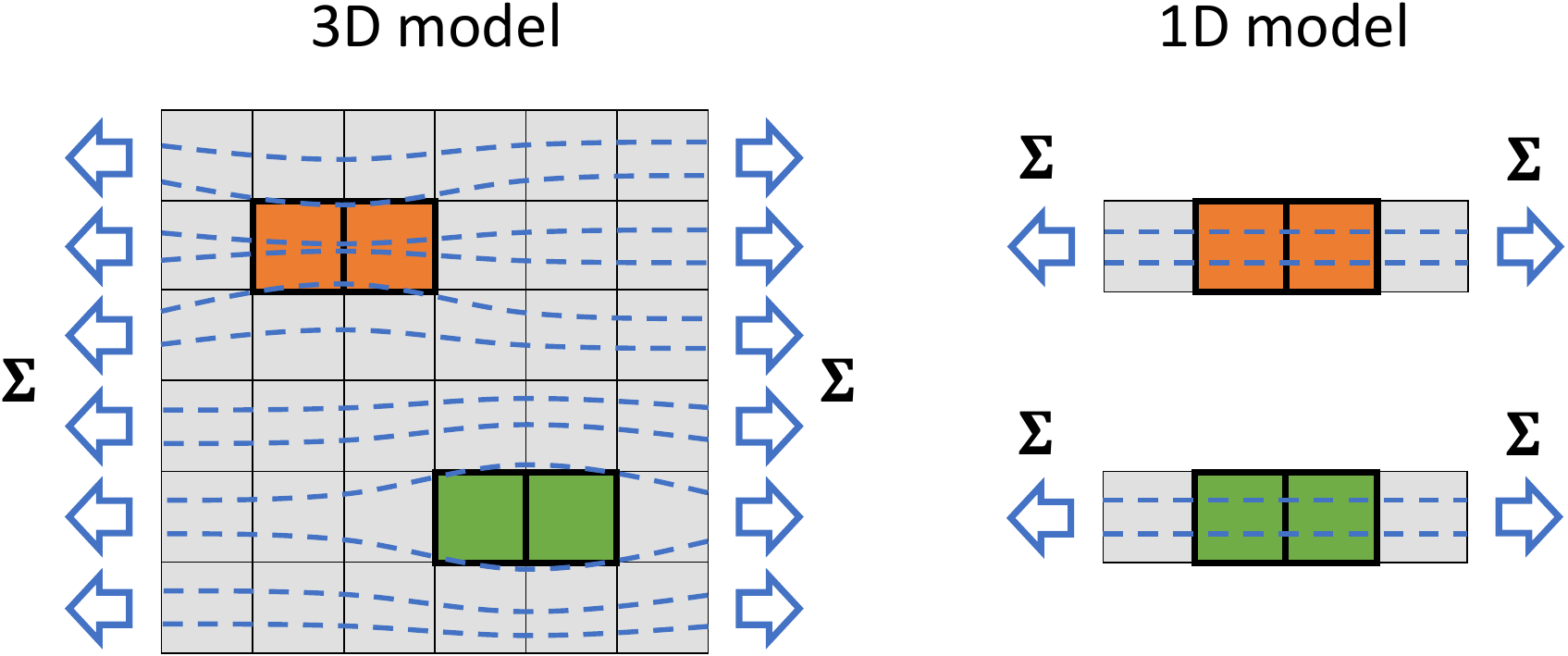}
	\caption{Schematic view of field lines crossing through the stiff (orange) and soft (green) GB grains in 3D and 1D models.}
	\label{fig:dE12}
\end{figure}

The observed inconsistency in Fig.~\ref{fig:effectK} is attributed to the (i) imposed axial strain constraint of the $L_n$-chain model and (ii) 3D effects which have been omitted in the model derivation. The 3D effects include primarily a non-zero lateral coupling of the axial grain chain with the elastic bulk. Depending on the relative axial stiffness of the chain with respect to the bulk, this coupling may effectively either increase or decrease the chain stiffness, resulting in larger or lower $\snn$, respectively. To model this in 1D framework, the elastic properties of both GB grains and buffer grain need to be amended%
\footnote{Alternatively, one could try to resolve the observed inconsistency by simply assuming a variable buffer length $L_n=\mathcal{F}(\Enn,A^u)$. However, it becomes clear from Fig.~\ref{fig:effectK} that such an approach fails to produce correct slopes $K$ for $\Enn\sim 1$ (as there is no effect of $L_n$ for $\Enn=E_3= 1$). This confirms that the observed mismatch cannot be resolved solely by assuming a variable axial strain constraint, controlled by $L_n$ in Eq.~\eqref{eq:constrSym}, and that 3D effects need to be employed on $\Enn$ and $\nu_{12}$, too.}.
Regarding the GB grains, therefore
\ba
\begin{split}
\label{eq:dE12}
	\Enn&\to \Enn+\delta\Enn , \\
	\delta\Enn&=\mathcal{F}(\Enn,\nu_{12},L_n,A^u) ,
\end{split}
\ea
and similarly 
\ba
\begin{split}
\label{eq:dnu12}
	\nu_{12}&\to \nu_{12}+\delta\nu_{12} , \\
	\delta\nu_{12}&= \mathcal{F}(\Enn,\nu_{12},L_n,A^u) .
\end{split}
\ea
The assumed functional dependence of $\delta\Enn$ in Eq.~\eqref{eq:dE12} can be explained with the help of Fig.~\ref{fig:dE12} where the field lines are used to visualize schematically the force field around the two GB grains under tensile loading. While force lines are always parallel in the 1D model (no lateral coupling with the bulk), they concentrate within/outside the stiffer/softer (larger/smaller $\Enn$) GB grains in the 3D model. Obviously, the effect gets stronger for $\Enn\to E_{12,\text{max}}$ or $\Enn\to E_{12,\text{min}}$ and for increasing material anisotropy $A^u$. To account for more (less) field lines in stiffer (softer) GB grains, $\delta\Enn>0$ ($\delta\Enn<0$) should be used in 1D modeling. However, using a non-zero $\delta\Enn$ (or $\delta\nu_{12}$) affects also the boundary condition applied on the chain scale in Eq.~\eqref{eq:constrSym}. Since the latter is regulated also by the length of the buffer grain $L_n$, both $\delta\Enn$ and $L_n$ are coupled as indicated in Eq.~\eqref{eq:dE12}.

In a similar way, the properties of the buffer grain (of length $L_n$) are modified due to lateral coupling with the bulk
\ba
\begin{split}
\label{eq:dbuffer}
	E_3&\to E_3+\delta E_3\approx 1 , \\
	\nu_b&\to \nu_b+\delta\nu_b\approx\ave{\nu} .
\end{split}
\ea
The above mapping follows from the fact that a chain of randomly oriented grains, when coupled laterally to the bulk, should, on average, behave similarly as the bulk itself. An equality in Eq.~\eqref{eq:dbuffer} is achieved for $L_n\to\infty$, while very small deviations are observed at $L_n=2$ (see footnote~\ref{f1}). As already mentioned, this modification has been already implemented in Eq.~\eqref{eq:chain} by setting $E_3= 1$ and $\nu_b=\ave{\nu}$.

Since $E_b$ and $\nu_b$ can be evaluated (\textit{e.g.}, numerically) by Eqs.~\eqref{eq:buffer} for a given material and $L_n$, the corresponding increments can be estimated directly from Eqs.~\eqref{eq:dbuffer}. By design, the same increments should also apply to GB grains, $(\delta\Enn,\delta\nu_{12})=(\delta E_b,\delta\nu_b)$, if $(\Enn,\nu_{12})=(E_b,\nu_b)$. Unfortunately, there seems to be no analytical approach to identify the increments for a general pair $(\Enn,\nu_{12})$. In the following, the functional dependence of $\delta\Enn$ (and $\delta E_3$) is therefore derived empirically for materials with cubic lattice symmetry, where further simplification is used due to mutual dependence of $\Enn$ and $\nu_{12}$ ($\nu_{12}=\ave{\nu}+(\Enn^{-1}-1)/2$).

The expression for the reduced version of the $L_n$-chain model, Eq.~\eqref{eq:chain}, simplifies for cubic materials and general macroscopic loading to
\ba
\begin{split}
\label{eq:cubic}
	\snn^{(2)}&=\frac{2+L_n}{2\Enn^{-1}+L_n E_3^{-1}}\Sigma_{zz} + \\
	&+\frac{1}{2}\left(1-\frac{2+L_n}{2\Enn^{-1}+L_n E_3^{-1}}\right) \left(\Sigma_{xx}+\Sigma_{yy}\right) ,
\end{split}
\ea
which reduces further for uniaxial macroscopic loading $\Sigma$ as
\ba
\begin{split}
	\snn^{(2)}/\Sigma&=\frac{1}{2}\left(1-\frac{2+L_n}{2\Enn^{-1}+L_n E_3^{-1}}\right) + \\
	&+\frac{3}{2}\left(-\frac{1}{3}+\frac{2+L_n}{2\Enn^{-1}+L_n E_3^{-1}}\right) \cos^2\theta ,
\end{split}
\ea
where $\theta$ is an angle between the GB normal and uniaxial loading direction. As discussed before, the model can be upgraded by assuming $\delta\Enn=\mathcal{F}(\Enn,L_n,A^u)$ for the two GB grains and $\delta E_3=\mathcal{F}(E_3,L_n,A^u)$ for the buffer grain. Both increments can be calculated numerically from the requirement that the resulting modified slope (a factor in front of $\cos^2\theta$),
\be
\label{eq:slopeK}
	K=\frac{3}{2}\left(-\frac{1}{3}+\frac{2+L_n}{2(\Enn+\delta\Enn)^{-1}+L_n (E_3+\delta E_3)^{-1}}\right),
\ee
is matching the corresponding $K^{\text{FE}}$ slope obtained from the FE simulations for different $\Enn$ values and materials (see Fig.~\ref{fig:effectK} where the results for Li are shown)%
\footnote{The corresponding increments are deduced in two steps. First, $\delta E_3$ is identified from $K(\Enn= E_3,E_3)=K^{\text{FE}}(E_3)$ for the assumed $\Enn= E_3$ and $\delta\Enn=\delta E_3$ in Eq.~\eqref{eq:slopeK}, where $E_3$ is evaluated numerically using Eq.~\eqref{eq:buffer} for a given material $A^u$ and buffer length $L_n$. In practice, the $K^{\text{FE}}(E_3)$ value is estimated by interpolating from several $K^{\text{FE}}(\Enn)$ values. Once $\delta E_3$ is known, $\delta\Enn$ is obtained from $K(\Enn,E_3)=K^{\text{FE}}(\Enn)$.}.

\begin{figure}
	\includegraphics[width=0.9\columnwidth]{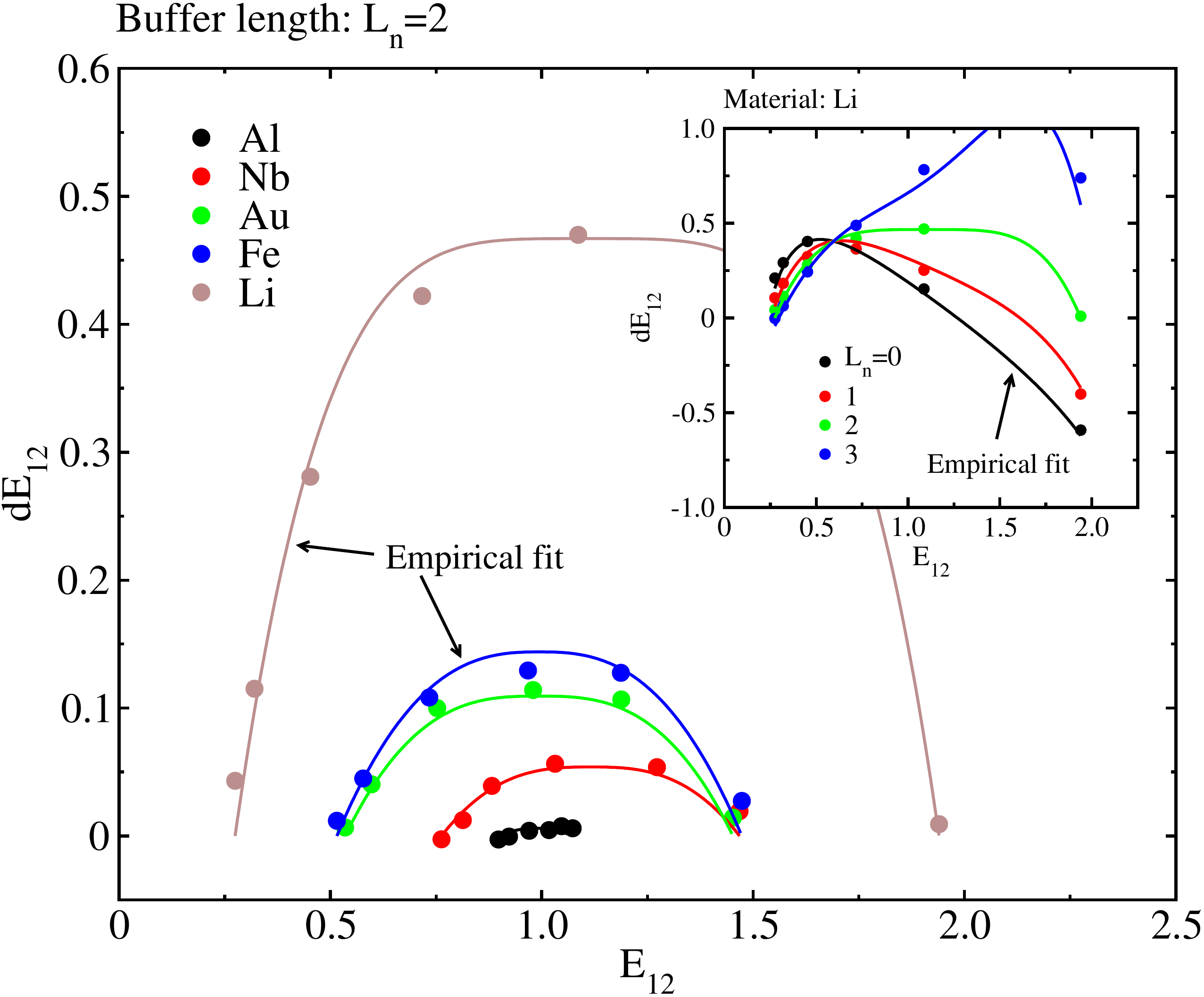}
	\caption{Calculated (dots) and fitted (lines) increments $\delta\Enn$ as a function of $\Enn$ for the chosen reduced $L_n$-chain model (with $L_n=2$) applied to various materials with cubic lattice symmetry. Fitting function, Eq.~\eqref{eq:fit}, has been \TM{chosen} based on the observed symmetry $\delta\Enn(\Enn)$ for $L_n=2$. Inset shows that very good agreement is preserved also when the symmetric $L_n=2$ fitting function is used for other $L_n$ (shown for Li).}
	\label{fig:dE12cubic}
\end{figure}

The results for $\delta\Enn$ are shown in Fig.~\ref{fig:dE12cubic} for $L_n=2$ and various materials%
\footnote{\label{f1}Results also show that the relative buffer stiffness, when coupled to the bulk, is bounded by $1\le E_3+\delta E_3\le 1.03$ for $L_n\ge 2$ and all the materials shown in Fig.~\ref{fig:dE12cubic}.}.
As anticipated, $\delta\Enn$ depends strongly on the GB stiffness $\Enn$, elastic grain anisotropy $A^u$ and buffer length $L_n$ (inset of Fig.~\ref{fig:dE12cubic}). Interestingly, for $L_n=2$ a (quasi) symmetry is recognized in $\delta\Enn(\Enn)$ curves for all investigated materials%
\footnote{The authors haven't resolved yet whether the observed symmetry is a coincidence or an intrinsic property of the (reduced) $L_n$-chain model.}. 
The symmetry is lost when $L_n\ne2$.

Based on the observed symmetry in Fig.~\ref{fig:dE12cubic} for $L_n=2$, the following empirical fit is proposed for all cubic materials with corresponding elastic anisotropy index $A^u$
\ba
\begin{split}
\label{eq:fit}
	\delta\Enn&=C_1-\left|\Enn-\bar{E}_{12}\right|^{C_2} , \\
	\bar{E}_{12}&=\frac{1}{2} \left(E_{12,\text{min}}+E_{12,\text{max}}\right) , \\
	\delta\Enn(E_{12,\text{min}})&= 0 , \\
	\delta\Enn(E_{12,\text{max}})&= 0 .
\end{split}
\ea
The best agreement with FE results is obtained for
\ba
\begin{split}
\label{eq:fit2}
	C_1&=0.08 (A^u)^{0.85} , \\
	C_2&=\frac{\log C_1}{\log (E_{12,\text{max}}-E_{12,\text{min}})-\log 2} .
\end{split}
\ea
It seems quite surprising that the proposed fitting function, Eq.~\eqref{eq:fit}, with only two adjustable parameters (0.08 and 0.85) in Eq.~\eqref{eq:fit2} provides such a good agreement for a wide range of (cubic) materials shown in Fig.~\ref{fig:dE12cubic}. Good agreement remains also when the $L_n=2$ fitting function is used for $L_n\ne2$ models (assuming $E_3+\delta E_3= 1$ in Eq.~\eqref{eq:slopeK}) as shown in the inset of Fig.~\ref{fig:dE12cubic}. The identified empirical relation represents the second main result of this study.

\subsubsection{Stochastic loading fluctuations}
\label{sec:gauss}

So far, the original external loading $\Sigma^{\text{lab}}$ (also $\Sigma$) has been assigned to all GB models from Table~\ref{tab:models}. However, in reality, this assumption is true only on average. In fact, a GB and its immediate neighborhood far away from the external surfaces feel an external loading modified by fluctuations, $\Sigma+f$, where $f$ stands for the fluctuation stress tensor. The fluctuations $f$ arise as a consequence of bringing far-away loading $\Sigma$ onto a GB neighborhood through the elastic bulk of anisotropic grains (see the last stage in Fig.~\ref{fig:pert}).

%To account for loading fluctuations $f$ in the estimation of $\snn$ and $\pdf(\snn)$, it is assumed for simplicity that $f$, at a position of a GB, depends on external loading and on grain anisotropy, $f\approx\mathcal{F}(\Sigma,A^u)$, and that $f$ is a stochastic variable with a known distribution. The dependence of $f$ on internal GB degrees of freedom (\textit{e.g.}, $\theta$, $\Enn$, $\omega_1$, $\omega_2$) is neglected to a first approximation, which is supported by the results of Fig.~\ref{fig:effectCos2}(b). The latter indeed show that standard deviation $s(\snn/\Sigma)$, evaluated on various $[abc]$-$[def]$ GB types at fixed GB tilts $\cos^2\theta$ with respect to external tensile loading $\Sigma$, is practically independent of $\cos^2\theta$ (and thus of $\snn$ itself), but slightly dependent on GB type%

To account for loading fluctuations in the estimation of $\snn$ and $\pdf(\snn)$, it is assumed for simplicity that fluctuation normal stress $f_{nn}$ is a random variable with Gaussian distribution $\mathcal{N}(0,s^2(f_{nn}))$, where the standard deviation depends on the external loading and on grain anisotropy, $s(f_{nn})\approx\mathcal{F}(\Sigma,A^u)$. Dependence of $s(f_{nn})$ on internal GB degrees of freedom (\textit{e.g.}, $\theta$, $\Enn$, $\omega_1$, $\omega_2$) is neglected to a first approximation, which is supported by the results of Fig.~\ref{fig:effectCos2}(b). The latter indeed show that standard deviation $s(\snn/\Sigma)$, evaluated on various $[abc]$-$[def]$ GB types at fixed GB tilts $\cos^2\theta$ with respect to external tensile loading $\Sigma$, is practically independent of $\cos^2\theta$ (and thus of $\snn$ itself), but slightly dependent on GB type%
\footnote{While the primary source of fluctuations in Fig.~\ref{fig:effectCos2}(b) is the anisotropic GB neighborhood, the secondary source is a finite range of GB tilt angles, $\delta(\cos\theta)=0.05$, which provides negligible contribution to $s(\snn/\Sigma)$.}.
Model of stress fluctuations is derived in Appendix~\ref{app:gauss}.

% where larger fluctuations are observed on GBs with more (internal) degrees of freedom.

%There is significant effect of $\Delta\omega$ on $s(\snn)$ when evaluated on $[112]$-$[112]$-$\Delta\omega$ GB. However, looking at the amplitudes, stress fluctuations evaluated over GBs with approximately fixed inclinations ($\theta$) are much smaller than stress fluctuations evaluated over all GBs.

Considering that stress fluctuations are independent of stresses themselves, a new update can be proposed as
\ba
\begin{split}
\label{eq:conv}
	\tilde{\sigma}_{nn}^{(k)}&=\snn^{(k)}+f_{nn}^{(k)} , \\
	s^2(\tilde{\sigma}_{nn}^{(k)})&=s^2(\snn^{(k)})+s^2(f_{nn}^{(k)}) , \\
	\pdf(\tilde{\sigma}_{nn}^{(k)})&=\left( \pdf\star\ \mathcal{N}(0,s^2(f_{nn}^{(k)})\right)(\snn^{(k)}) ,
\end{split}
\ea
where symbol $\star$ denotes a convolution.

\begin{figure}
	\includegraphics[width=0.9\columnwidth]{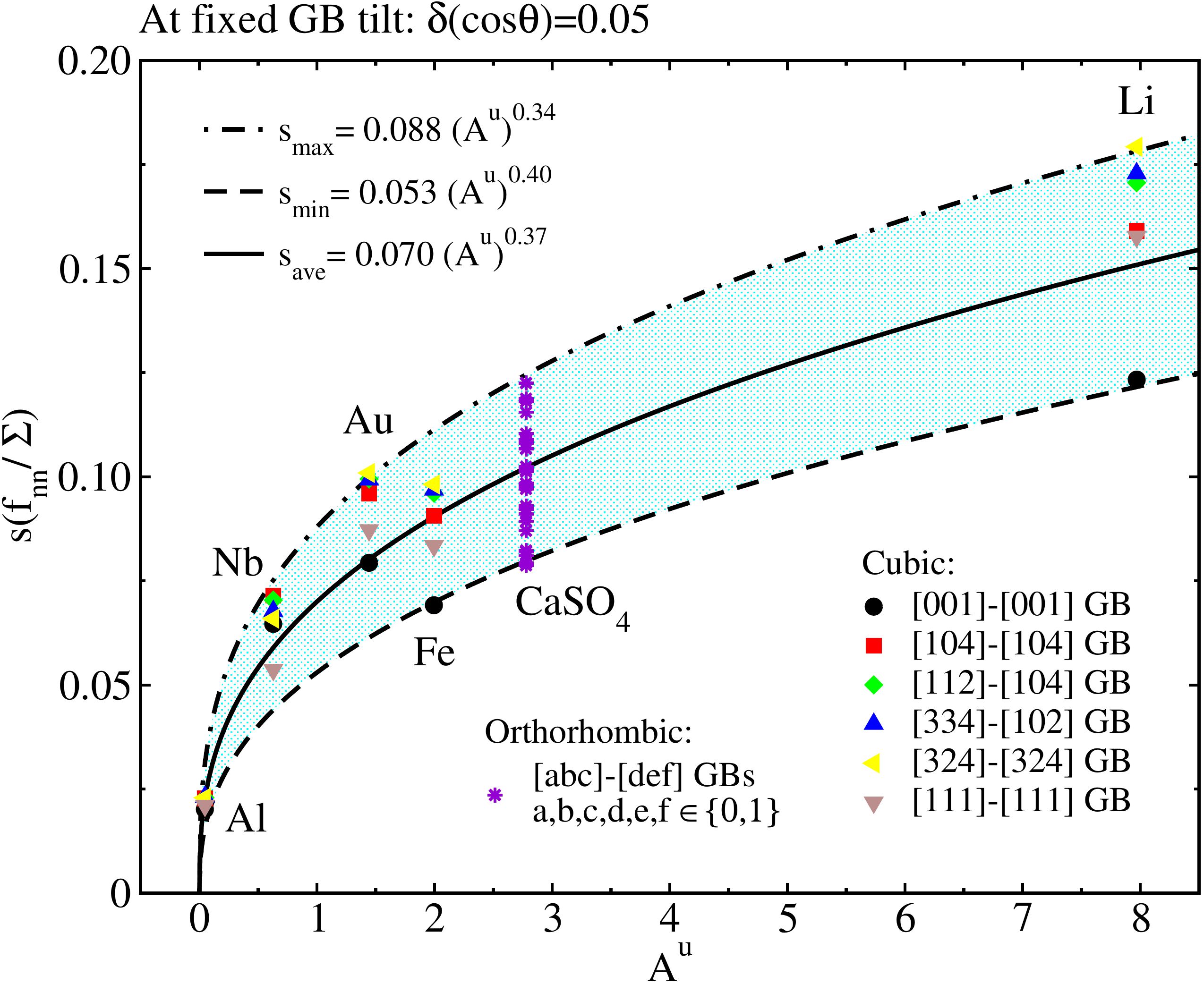}
	\caption{Standard deviation of GB\TM{-stress} (normal) fluctuations $s(f_{nn}/\Sigma)$, evaluated numerically on different GB types with fixed GB tilt (and later averaged over different GB tilts, see Fig.~\ref{fig:effectCos2}(b)) and for different materials ($A^u$) under tensile loading $\Sigma$. A shaded region represents the proposed empirical domain of fluctuations. A generalization to arbitrary loading can be done by substituting $\Sigma$ with $\Sigma_{\text{mis}}$ on the vertical axis (see Appendix~\ref{app:gauss}).}
	\label{fig:stdG}
\end{figure}

To identify standard deviation $s(f_{nn}^{(k)})$ for tensile $\Sigma$, the results from Fig.~\ref{fig:effectCos2}(b) can be averaged over different GB tilts and shown in Fig.~\ref{fig:stdG} for different materials as a function of $A^u$. Obtained standard deviation $s(f_{nn}/\Sigma)$ is set to be a measure of local stress fluctuations $f_{nn}$. As expected, $s(f_{nn}/\Sigma)$ increases with $A^u$ following a simple empirical law $s(f_{nn}/\Sigma)=(0.070\pm0.018) \left(A^u\right)^{0.37\mp 0.03}$. The $\pm$ sign denotes a finite width of $s(f_{nn}/\Sigma)$ domain, which is attributed to GB internal degrees of freedom.

The empirical fit is generalized further to arbitrary loading using the familiar normalization for the second statistical moment (see Appendix~\ref{app:gauss} for more detail),
\be
\label{eq:conv2}
	s(f_{nn})=\Sigma_{\text{mis}} (0.070\pm0.018) \left(A^u\right)^{0.37\mp 0.03}.
\ee
The above relation applies not only to cubic but also to non-cubic materials%
\footnote{It is also interesting to note that a hydrostatic loading $\Sigma$ provides no GB stress fluctuations even in the case of anisotropic grains.}.
For example, tensile fluctuations evaluated in calcium sulfate (CaSO$_4$), with orthorhombic lattice symmetry and $A^u=2.78$, fit accurately within the proposed domain in Fig.~\ref{fig:stdG}. 

%In the following, a simple assumption about the shape of tensor $f$ is given by accounting for the rotational symmetry of the GB models. Loading fluctuations $f$ are prescribed to be of hydrostatic form, $f_{ij}=f_0 \delta_{ij}$, due to the isotropy of grain shapes and random texture, which updates the final solution simply as: snn + fluc. 

%We expect that tensor $f$ can be eventually reduced to a scalar variable. The amplitude of $f$ will 

% ==========================================================================
\section{\label{verifi}Verification of upgraded models}
\label{sec:4}
% ==========================================================================

\subsection{Cubic materials}

\begin{figure*}
	\includegraphics[width=0.6\columnwidth]{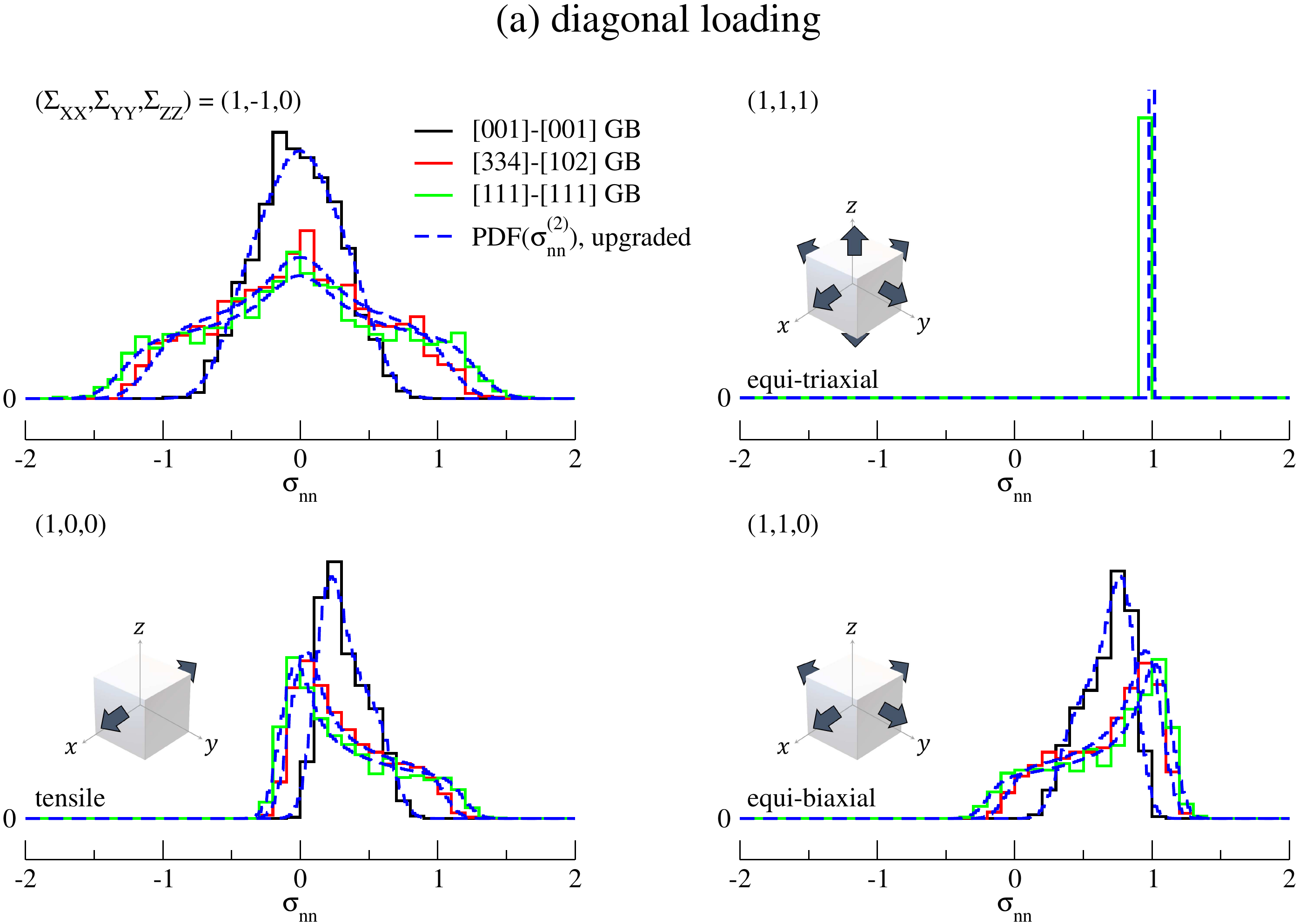}
	\includegraphics[width=0.6\columnwidth]{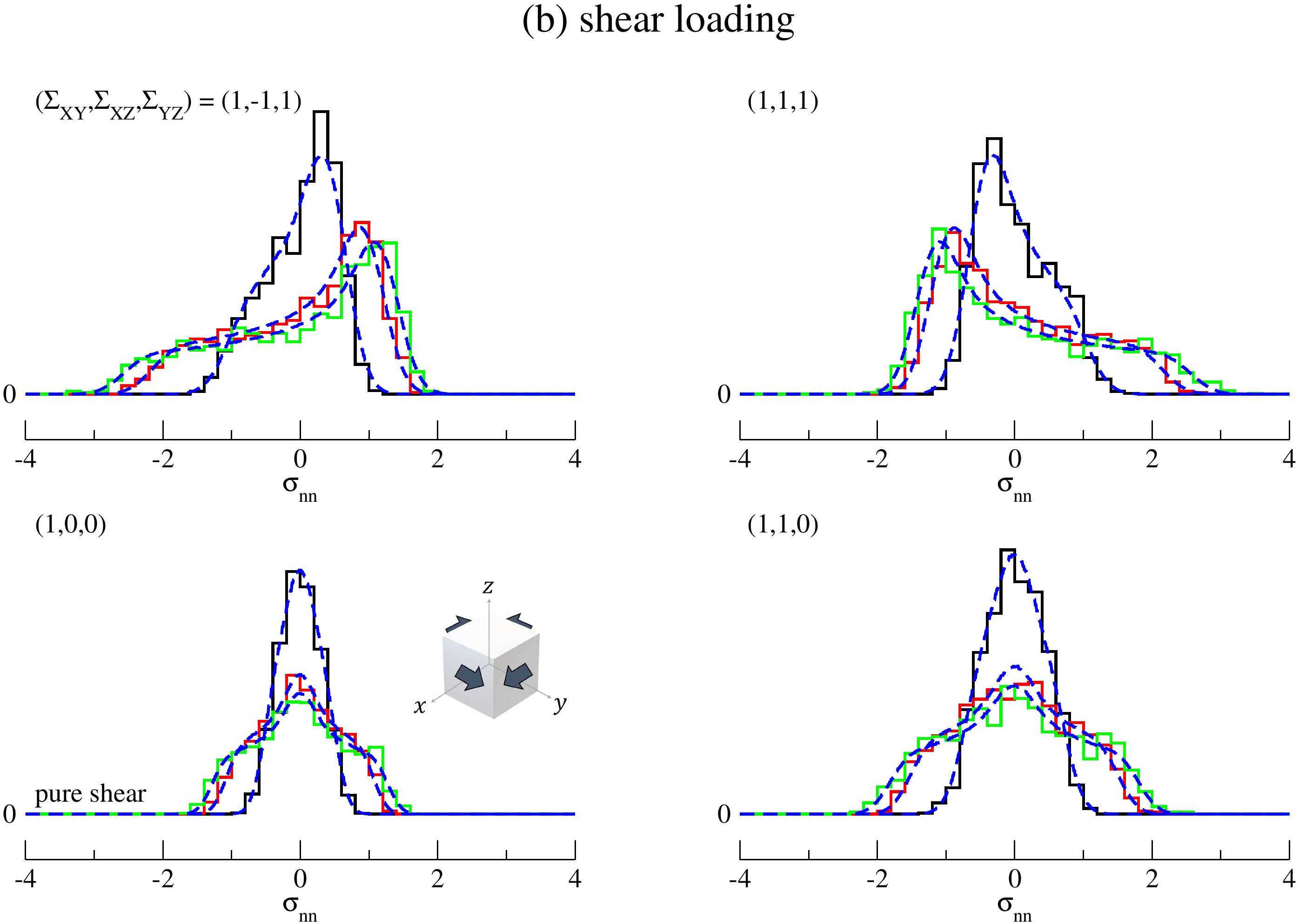}
	\includegraphics[width=0.6\columnwidth]{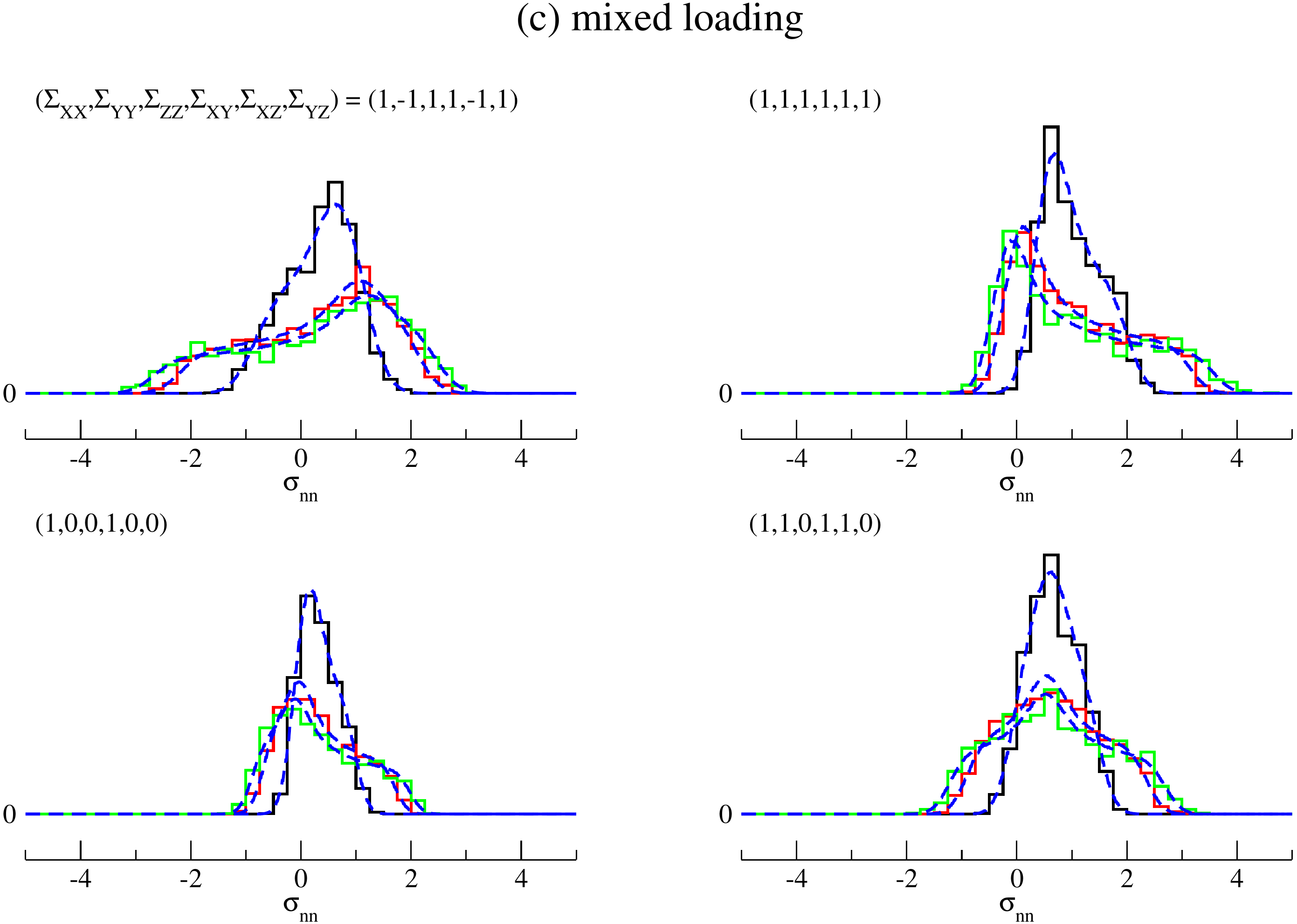}
	\caption{Statistical stress distributions $\pdf(\snn)$ evaluated on three different GB types in Fe for $12$ different macroscopic loadings (grouped into purely diagonal, purely shear and mixed loadings $\mathbf{\Sigma}$). An excellent agreement is shown between simulation results (solid lines) and upgraded model predictions (dashed lines) for all the cases; cf.~Eq.~\eqref{eq:cubic2}.}
	\label{fig:effectLoad}
\end{figure*}

Statistical response $\pdf(\tilde{\sigma}_{nn}^{(2)})$ of the upgraded cubic GB model (using $L_n= 2$ and $E_3+\delta E_3= 1$ in Eq.~\eqref{eq:cubic}),
\ba
\begin{split}
\label{eq:cubic2}
	\tilde{\sigma}_{nn}^{(2)}&=\frac{2}{(\Enn+\delta\Enn)^{-1}+1}\Sigma_{zz} + \\
	&+\frac{1}{2}\left(1-\frac{2}{(\Enn+\delta\Enn)^{-1}+1}\right)\left(\Sigma_{xx}+\Sigma_{yy}\right) + \\
	&+ f_{nn} ,
\end{split}
\ea
where $\delta\Enn$ is estimated by Eqs.~\eqref{eq:fit}, \eqref{eq:fit2} and $s(f_{nn})$ by Eq.~\eqref{eq:conv2}, is verified in Fig.~\ref{fig:effectLoad} for polycrystalline Fe under different macroscopic loadings $\Sigma$. The predicted $\pdf(\tilde{\sigma}_{nn}^{(2)})$ distributions are calculated numerically using Monte Carlo sampling of the two%
\footnote{Since $\snn^{(k)}=A^{(k)} \Sigma_{zz}+B^{(k)} (\Sigma_{xx}+\Sigma_{yy})$ for any $k$, the third Euler angle $\phi$ drops out from the $\snn^{(k)}$ expression.}
Euler angles $(\theta,\psi)$, which are used to evaluate $\Sigma_{xx}$, $\Sigma_{yy}$ and $\Sigma_{zz}$ defined in Eq.~\eqref{eq:sigGB}. An excellent agreement with simulation results is demonstrated, confirming the accuracy of the proposed model for $\textit{arbitrary}$ GB type, $\textit{arbitrary}$ (cubic) material and $\textit{arbitrary}$ macroscopic loading conditions. 

\begin{figure}
	\includegraphics[width=0.9\columnwidth]{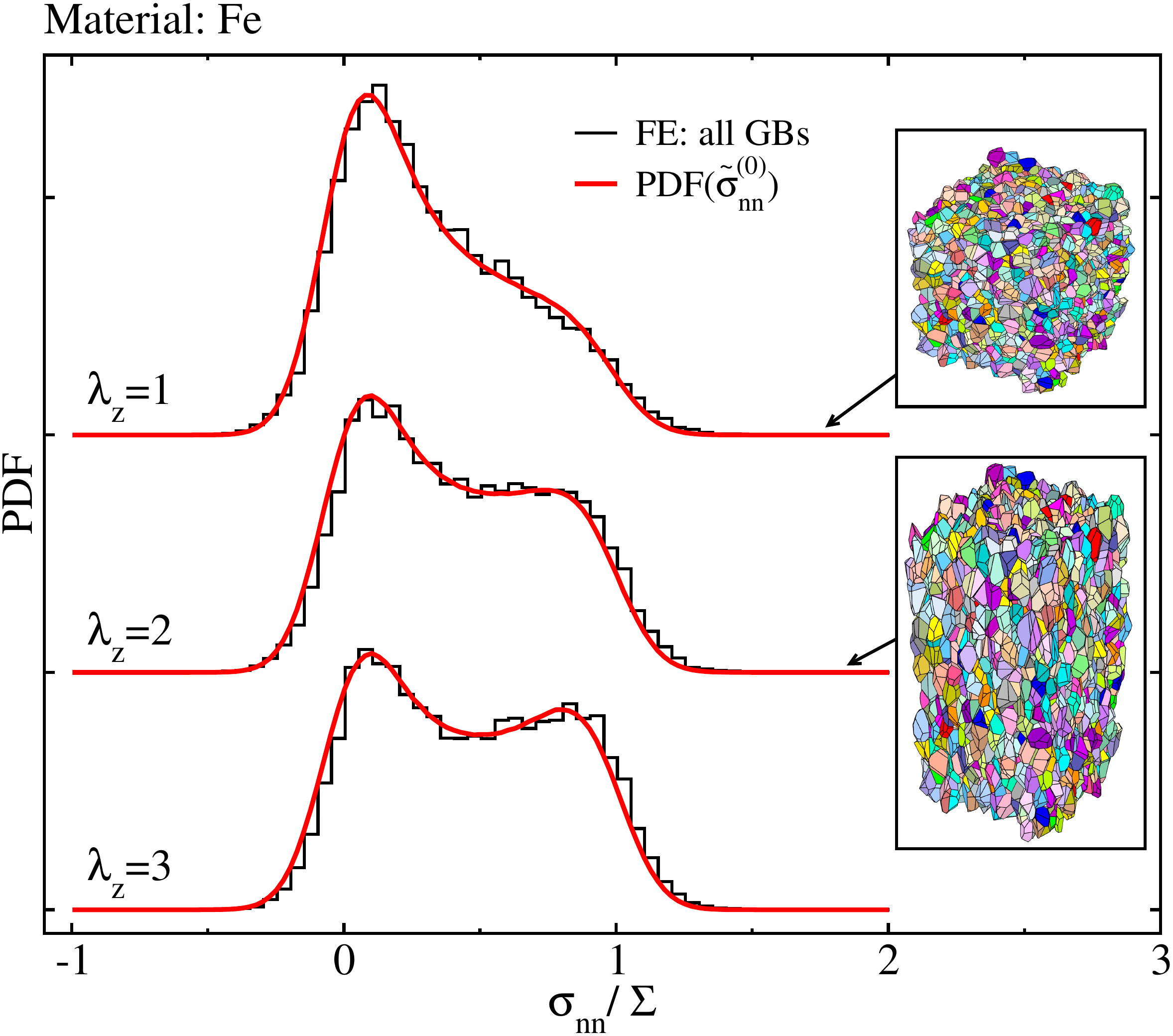}
	\caption{Stress distributions $\pdf(\snn/\Sigma)$ evaluated on all (random) GBs in Fe with grains elongated along the $Z$-axis (elongation factor $\lambda_z$) for tensile loading $\Sigma$ along the $X$-axis. An excellent agreement is shown between simulation results (black) and model predictions (red) for all three cases; cf.~Eqs.~\eqref{eq:elong}--\eqref{eq:elong2}.}
	\label{fig:elongated}
\end{figure}

In Fig.~\ref{fig:elongated} a comparison is shown for a polycrystalline Fe with elongated grains to verify the applicability of the derived models in materials with non-zero morphological texture (but with zero crystallographic texture). In this comparison, the $\pdf$ response is calculated on all GBs (random type) using the following simple relation (see Appendix~\ref{app:random})
\be
\label{eq:elong}
	\pdf_{\text{rnd}}(\tilde{\sigma}_{nn}^{(k)})\approx\pdf(\tilde{\sigma}_{nn}^{(0)}).
\ee
The response of random GBs is calculated using the convolution of the isotropic solution $\pdf(\snn^{(0)})$ and Gaussian distribution $\mathcal{N}(0,s^2(f_{nn})$ with $s(f_{nn})$ from Eq.~\eqref{eq:conv2}%
\footnote{Since the $\pdf$ of the FE model is calculated on all GBs of an aggregate, including those with smallest GB areas, the finite-size effects (due to poor meshing) result in wider $\pdf$ distributions. For this reason, a $\sim$40$\%$ larger $s(f_{nn})$ is used in Fig.~(\ref{fig:elongated}) to fit accurately the FE results.}.
The distributions are calculated numerically using Monte Carlo sampling of the two Euler angles $(\theta,\psi)$ with the following distribution functions (see Appendix~\ref{app:elongated})
\ba
\begin{split}
\label{eq:elong2}
	f(\cos\theta)&=\frac{\lambda_z}{2}\left(\frac{1}{1+(\lambda_z^2-1)\cos^2\theta}\right)^{3/2} , \\
%	f(\phi)&=&\frac{1}{2\pi}\nonumber\\
	f(\psi)&=\frac{1}{2\pi} ,
\end{split}
\ea
for $-1\le\cos\theta\le1$ and $0\le\psi\le2\pi$, with a scaling factor $\lambda_z>0$ accounting for grain elongation along the $Z$-axis ($\lambda_z=1$ denoting no scaling).

Again, an excellent agreement with simulation results is demonstrated in Fig.~\ref{fig:elongated}, which confirms the accuracy of the proposed model when applied to materials with $\textit{arbitrary}$ morphological texture. 

\begin{figure*}
	\includegraphics[width=0.8\columnwidth]{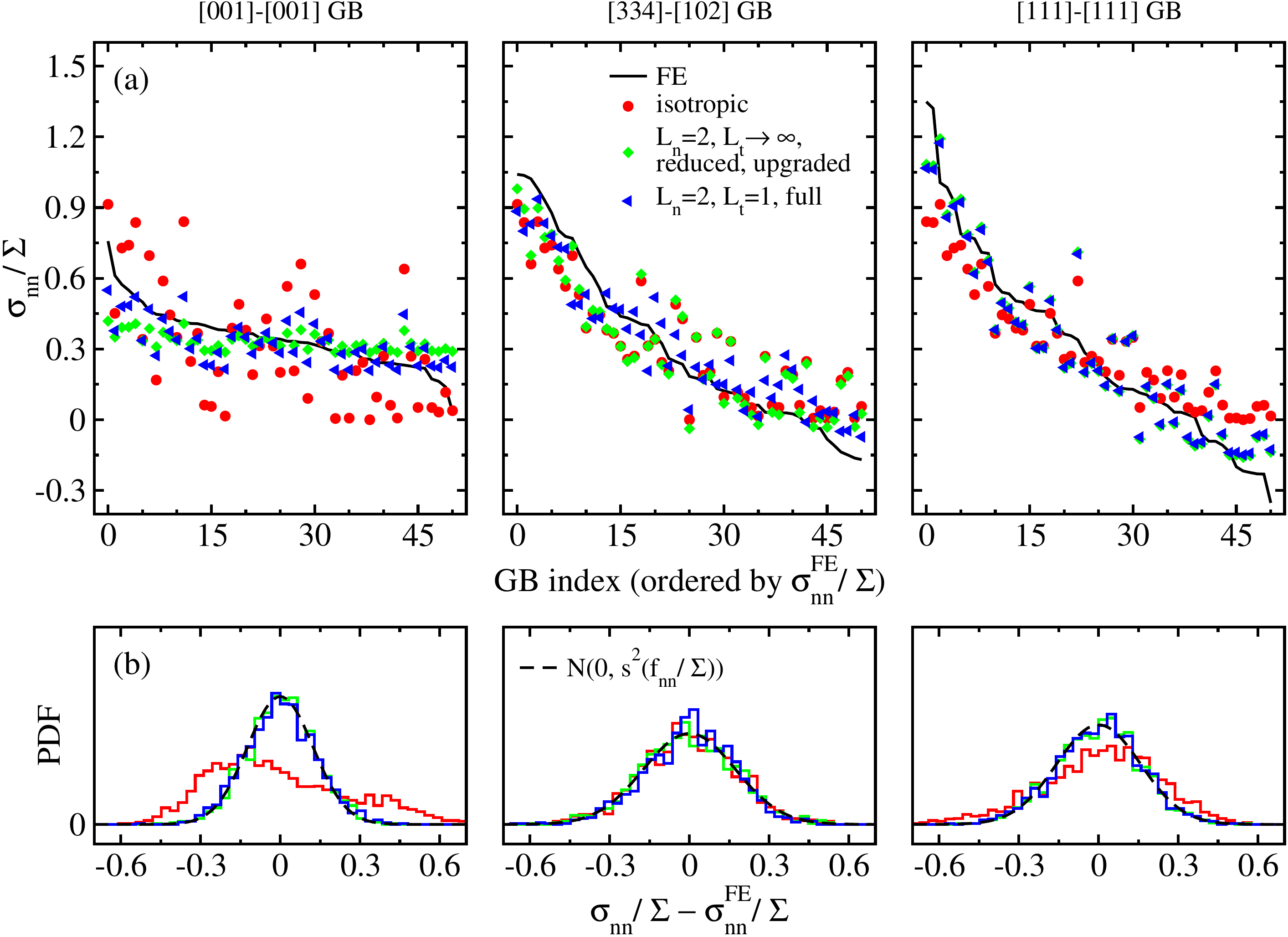}
	\caption{(a) Local $\snn/\Sigma$ stress response in Li under macroscopic tensile loading $\Sigma$. A comparison is shown between FE simulation (lines) and the \TM{results of three GB models (shapes). The $50$ largest GBs in a $4000$-grain aggregate (see Fig.~\ref{fig:geom}) are shown, to which a specific GB type ($[001]$-$[001]$, $[334]$-$[102]$ or $[111]$-$[111]$) is assigned in each panel. The GBs are sorted (indexed) in descending order with respect to FE results for stresses. (b) Probability distributions ($\pdf$) of discrepancy between the model prediction and FE result for GB stress, demonstrating how accurate the three GB models are (locally). All $1631$ special GBs are considered in the $\pdf$ (as opposed to $50$ shown in (a)).} Gaussian distributions with standard deviations $s(f_{nn}/\Sigma)$ from Fig.~\ref{fig:stdG} are added for comparison (no fitting has been applied).}
	\label{fig:local}
\end{figure*}

In Fig.~\ref{fig:local}, the accuracy of GB models is furthermore tested on a local GB scale using FE simulations of a polycrystalline Li under tensile loading $\Sigma$ as a reference. In particular, three models (of increasing complexity) are compared: (i) the isotropic model $\snn^{(0)}$, (ii) the reduced and upgraded version of the $L_n$-chain ($L_n=2$, $L_t\to\infty$) model $\snn^{(2)}$ and (iii) the full version of the $L_n$-$L_t$-chain ($L_n=2$, $L_t=1$) model $\snn^{(3)}$. The accuracy of the models is tested locally by comparing $\snn^{(k)}$ values with FE results $\snn^{\text{FE}}$ evaluated on individual GBs of particular type (three GB types are tested in total). 

According to Fig.~\ref{fig:local}, both $\snn^{(2)}$ and $\snn^{(3)}$ models are comparable in accuracy, outperforming the simplest $\snn^{(0)}$ model on softer $[001]$-$[001]$ and stiffer $[111]$-$[111]$ GBs. The uncertainties (deviations from true FE values) in the $\snn^{(2)}$ model (and also $\snn^{(3)}$ model) are of Gaussian type with zero mean and standard deviation (exactly!) equal to $s(f_{nn})$ from Fig.~(\ref{fig:stdG}) (see dashed lines in Fig.~\ref{fig:local}(b)). This confirms the validity (and consistency) of the $\snn^{(2)}$ model, which is shown to be accurate up to unknown loading fluctuations $f_{nn}$ (which are substantial in Li). The latter are therefore the only%
\footnote{In case of an invalid GB model, standard deviation of local stress errors would be larger than that of loading stress fluctuations, $s(\snn^{(2)}-\snn^{\text{FE}})>s(f_{nn})$.}
source of local stress uncertainties (errors), $\snn^{(2)}-\snn^{\text{FE}}\approx f_{nn}$, suggesting that $\tilde{\sigma}_{nn}^{(2)}\approx\snn^{\text{FE}}$.

\subsection{Non-cubic materials}

\begin{figure}
	\includegraphics[width=0.9\columnwidth]{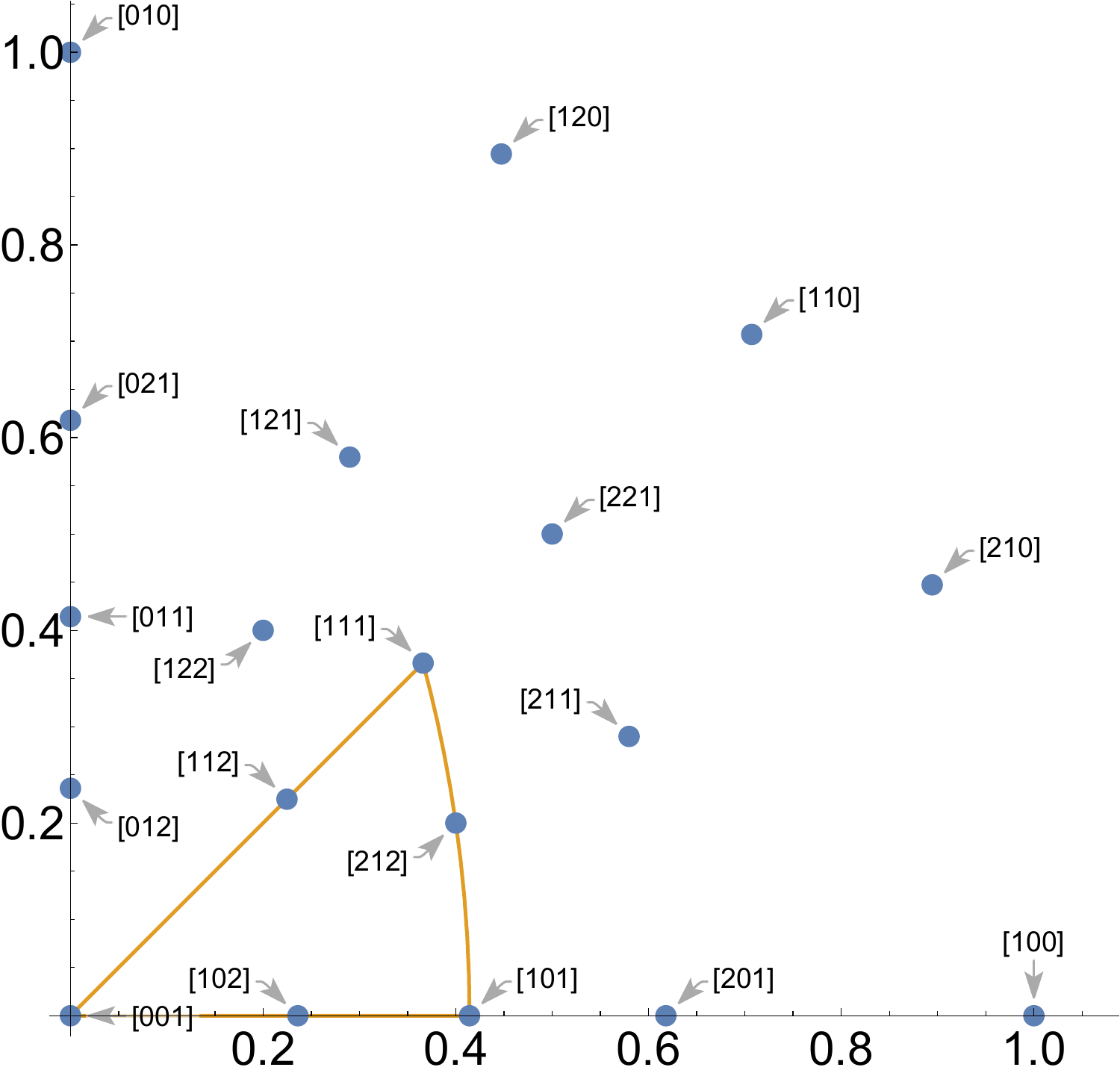}
	\caption{\TM{$19$ representative directions $[abc]$, from which GB normals in either GB grain were selected for orthorhombic material (CaSO$_4$). They correspond to $190$ different GB types $[abc]$-$[def]$, considered in our numerical studies.} The standard stereographic triangle is shown for reference.}
	\label{fig:stereo}
\end{figure}
\begin{figure}
	\includegraphics[width=0.9\columnwidth]{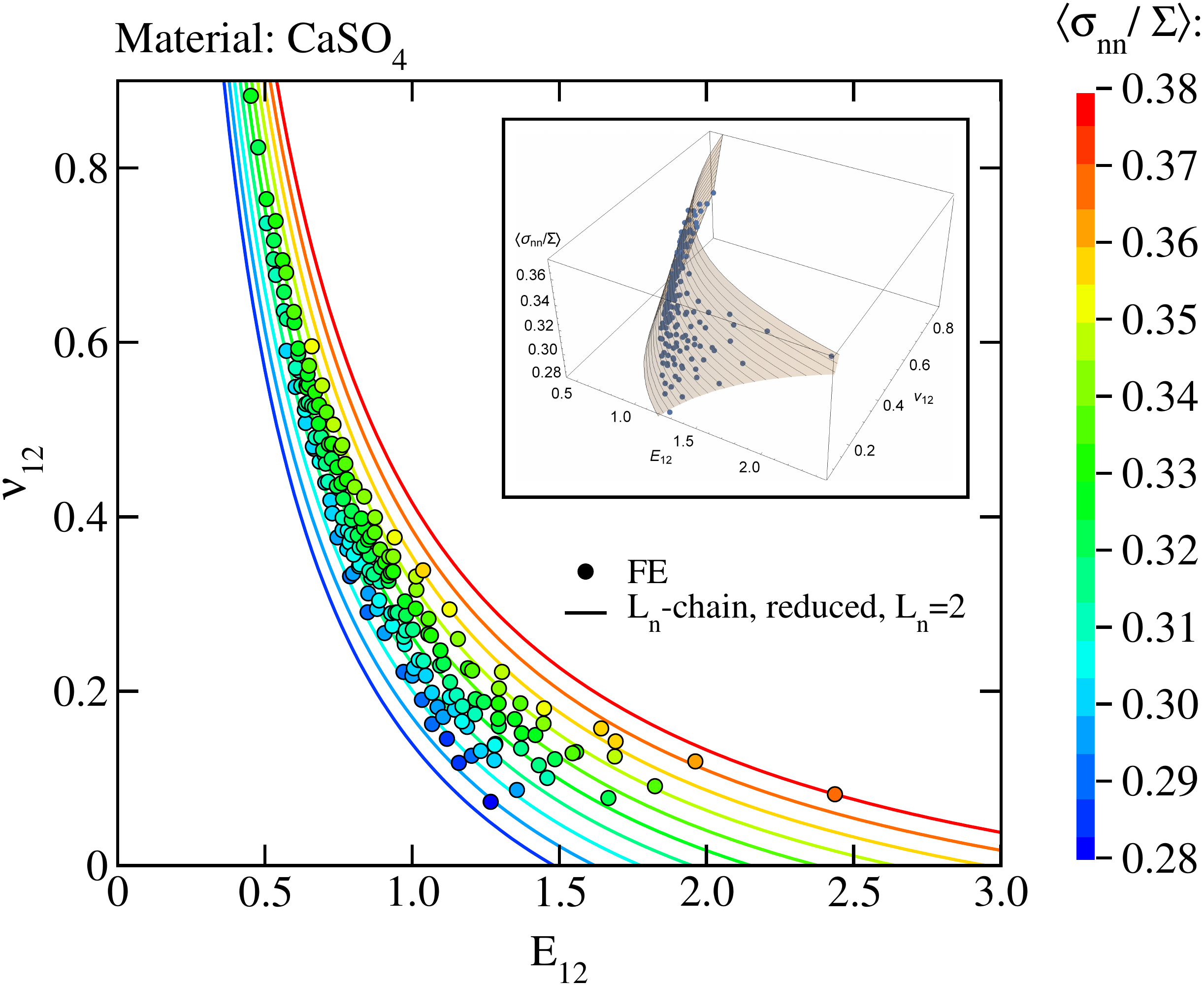}
	\caption{\TM{Simulation results (circles) for $\ave{\snn/\Sigma}$ in CaSO$_4$ under tensile loading $\Sigma$, evaluated on $190$ $[abc]$-$[def]$ GB types, constructed from the $19$ selected directions shown in Fig.~\ref{fig:stereo}. For comparison, a smooth prediction (iso-lines) is shown for $\ave{\snn^{(2)}/\Sigma}$ of the $L_n=2$ reduced model; cf.~Eq.~\eqref{eq:chainpdf}.} The inset shows the same results in 3D plot.}
	\label{fig:ortho}
\end{figure}

To provide accurate stress distributions for non-cubic materials, the evaluation of $\delta\Enn$ and $\delta\nu_{12}$ would need to be derived (see Eqs.~\eqref{eq:dE12} and~\eqref{eq:dnu12}) to account for variable axial strain constraint and 3D effects missing in the $L_n$-chain model. The procedure for that should follow the one described for cubic materials in Sec.~\ref{sec:3Deffects}. However, this is left for future analyses.

In Fig.~\ref{fig:ortho} the simulation results for average stress $\ave{\snn/\Sigma}$ are presented which are evaluated on 190 $[abc]$-$[def]$ GBs obtained as combinations of 19 planes defined in Fig.~\ref{fig:stereo} for the orthorhombic material CaSO$_4$ under tensile loading $\Sigma$. For comparison, a smooth prediction of $\ave{\snn^{(2)}/\Sigma}$ from Eq.~\eqref{eq:chainpdf} is shown for arbitrarily chosen $L_n=2$. A good qualitative agreement is demonstrated (without fine-tuning of $L_n$), implying that, to a good approximation, only two parameters, $\Enn$ and $\nu_{12}$, are needed to characterize the response of a general GB, in agreement with the prediction of the GB model.

% ==========================================================================
\section{\label{discussion}Discussion}
\label{sec:5}
% ==========================================================================

%\begin{itemize}
%	\item We can mention one practical implication. For example, we can consider the following (general) case: 1. Solve stresses in an arbitrary component under complex loading assuming isotropic material (using e.g. Abaqus). 2. Use the resulting 3D stress field as an input to our modeling PDFs. 3. Knowing the strengths $\sigma_c$ for different GB types, we would be able to quickly provide the probability of $\snn>\sigma_c$ at any position in the complex component.

%	\item What can be done in short time: semi-analytical solutions have been provided for cubic lattices (solved in general!), for non-cubic lattices it has been only indicated how to evaluate $\delta\Enn$ and $\delta\nu_{12}$.

%	\item Extended applicability: although static elastic loads have been used, I think it is safe to extend the validity of the results also (i) to dynamic loading (e.g. fatigue), (ii) to estimate trans-granular normal stresses given the plane normal $(abc)$ and using the results for $(abc)$-$(abc)$-$0^{\circ}$ GBs.

%	\item Possible extensions: only elastic grains (what about plasticity?), only bulk GBs (what about surface GBs?), zero-texture materials (what about finite texture?).
%\end{itemize}

The derived $\pdf(\snn)$ distributions \TM{are not only very accurate, as demonstrated for various scenarios (see Figs.~\ref{fig:effectLoad} and~\ref{fig:elongated}), but also relatively undemanding in computational sense. If they are produced numerically, using Monte Carlo sampling for GB-normal directions, %while the corresponding $\snn$ are provided by the model, 
the results} can be immediately used for several practical applications. \TM{For instance, we could predict} the GB-damage initiation in complex geometries, using the probabilistic approach. \TM{If the GB strength $\sigma_c$ of each GB type was known (or measured), and stress field $\tou{\Sigma}(\tou{r})$ in the investigated component at least roughly estimated (\textit{e.g.}, in FE simulations, using %a very coarse mesh and isotropic 
homogeneous material)}, one can immediately obtain the probability of finding an overloaded GB of a specific type at \textit{arbitrary} location $\tou{r}$ in the component, $P(\tou{r})=\int_{\sigma_c}^{\infty}\pdf(\snn)d\snn$, using $\tou{\Sigma}(\tou{r})$ as an input for external loading to produce $\pdf(\snn)$. If that probability exceeded the threshold value, $P(\tou{r})>P_f$, a macroscopic-size crack may develop at $\tou{r}$, which might result in a catastrophic failure of the component. With such approach, potentially dangerous regions, susceptible to intergranular cracking, can be quickly identified for any component and its loading. A more detailed analysis of such an application will be presented in a separate publication~\cite{elshawish2022draft}.
	
In all the examples presented so far, static elastic loads have been assumed in expressions for $\snn$ and $\pdf(\snn)$. The procedure can be generalized also to dynamic stresses, provided that stress amplitudes remain in the elastic domain and inertia effects are negligible. In this respect, $\pdf(\snn)$ spectra can be used to predict even the initiation of GB-fatigue cracks~\cite{koyama2015}. Following the above procedure for static load and assuming time-dependent evolution of GB strength (due to the build-up of strain localization~\cite{koyama2015}), the probability $P(\tou{r},t)=\int_{\sigma_c(t)}^{\infty}\pdf(\snn)d\snn$ becomes time dependent too. The \TM{measurement} data can then be used, for example, to estimate \TM{how $P_f$ and GB-strength evolution $\sigma_c(t)$ change with the number of loading cycles}.
		
Although the semi-analytical $\snn$ expression, derived for cubic crystal lattices, provides accurate $\pdf(\snn)$ distributions for a wide range of situations, it relies not only on analytical, but also on empirical considerations (estimation of $\delta\Enn$ and $s(f_{nn})$). \TM{A (quasi) symmetry of $\delta\Enn(\Enn)$ curves was observed for $L_n=2$ and all investigated materials. The origin of this feature is not yet understood, it might even be only accidental. Nonetheless, it can be very useful, since it allows us to make the search of the fitting function significantly simpler (Fig.~\ref{fig:dE12cubic}). Due to that, it is important to gain a better understanding of this (quasi) symmetry for cubic lattices (and possibly even non-cubic lattices) in the future.} 
	
In a similar way, the effect of \TM{%different GB neighborhoods due to the anisotropic properties of 
more distant grains} has not been modeled explicitly. Instead it was conveniently packed into an empirical fit of $s(f_{nn})$, which represents the amplitude of GB-stress fluctuations (Fig.~\ref{fig:stdG}). The fact that these fluctuations are more or less independent of stresses, makes the fitting function $s(f_{nn})$ relatively simple and, most importantly, the calculation of $\pdf(\snn)$ very accurate (by applying Gaussian broadening with known \TM{width} $s(f_{nn})$). However, the accuracy on a local GB scale is limited by the same $s(f_{nn})$ (representing the uncertainty \TM{of model predictions}), and can be substantial in highly anisotropic materials. A possible improvement would necessarily include an exact modeling of the \TM{more distant} grains (\TM{whose structure should probably be considered in a similar level of detail as the two GB grains}). \TM{Unfortunately, this} would probably result in very cumbersome and impractical solutions.\\
	
The \TM{derivation of GB models was based on two major ideas}: the \TM{perturbative approach} and the Saint Venant's principle. A possible alternative approach could follow one of the well-known methods, used for calculation of the effective elastic constants of polycrystals from single-crystal and structure properties. For example, in the self-consistent method invented by Kr\"oner~\cite{kroner58}, an effective stress-strain relation is derived, taking into account the boundary conditions for stresses and strains at the GBs, \TM{which are only statistically correct}. Analytical results are given for macroscopically isotropic polycrystals, composed of crystal grains with cubic symmetry~\cite{hershey}, and also for a general lattice symmetry~\cite{kroner58}. Replicating such approach, the established relation between the local (single-grain) and macroscopic quantities would need to be modified to account for bi-crystal instead of single-crystal local quantities. While this might be worth trying, it has (at least) one significant shortcoming, common to all multi-scale techniques. It fails to reproduce additional degrees of freedom on a local scale (that would manifest themselves in stress fluctuations and thus in wider $\pdf(\snn)$ distributions), given there are fewer degrees of freedom on a macroscopic scale. Therefore, additional improvements would be needed (as it was done here) to obtain accurate $\pdf(\snn)$. A detailed analysis \TM{along these lines} is left for future work.

% ==========================================================================
\section{Conclusions}
\label{sec:6}
% ==========================================================================

In this study, a perturbative \TM{model} of grain-boundary-normal stresses has been derived for an arbitrary grain-boundary type \TM{within} a general polycrystalline material, composed of randomly shaped elastic continuum grains with arbitrary lattice symmetry, and under a general uniform external loading. The \TM{constructed} perturbative models have been solved under reasonable assumptions, \TM{needed to obtain compact, yet still} accurate analytical and semi-analytical expressions for local grain-boundary-normal stresses and the corresponding statistical distributions. The strategy \TM{for} deriving the models \TM{was based on two central} concepts. Using the perturbation principle, the \TM{complexity of the model is gradually increased in each successive step, allowing us to first solve and understand simpler variants of the model}. Following the Saint Venant's principle, \TM{anisotropic elastic properties of the two grains closest to grain boundary have been considered in full}, while the effect of \TM{more distant} grains has been modeled in much smaller detail, using average quantities such as elastic grain anisotropy or bulk isotropic stiffness parameter.

The following conclusions have been reached from the solutions of derived perturbative models:
\begin{itemize}
	\item The general $k$-th order solution for the local grain-boundary-normal stress is of the following form: $\tilde{\sigma}_{nn}^{(k)}=A^{(k)}\Sigma_{zz}+B^{(k)}\left(\Sigma_{xx}+\Sigma_{yy}\right)+f_{nn}^{(k)}$, where $A^{(k)}$ and $B^{(k)}$ are the analytic functions of grain-boundary type and elastic material properties, $\Sigma_{ii}$ is a \TM{diagonal} component of the external loading tensor $\mathbf{\Sigma}$, \TM{expressed in a local grain-boundary system}, and $f_{nn}^{(k)}$ is a random variable, representing loading fluctuations.
	\item To a good approximation ($k=2$), the response on a chosen grain boundary can be characterized by just two parameters: $\Enn$ measures the average stiffness of grain-boundary neighborhood along the normal direction, while $\nu_{12}$ is an effective Poisson's \TM{ratio}, measuring the average ratio of \TM{transverse and} axial responses in both adjacent grains.
	\item For an arbitrary lattice symmetry, $A^{(2)}$ and $B^{(2)}$ are simple functions, $\mathcal{F}(\Enn,\nu_{12},\ave{E},\ave{\nu},L_n)$, where $\ave{E}$ and $\ave{\nu}$ denote average elastic bulk properties, and $L_n\ge 0$ is a modeling parameter accounting for the \TM{amount of buffer grains}. Also higher order solutions ($k>2$) have been obtained, but with resulting expressions too cumbersome to be useful in practice.
	\item To account for 3D effects and realistic boundary conditions, a model upgrade has been proposed by assuming $\Enn\to\Enn+\delta\Enn$ and $\nu_{12}\to\nu_{12}+\delta\nu_{12}$ in the expressions for $A^{(2)}$ and $B^{(2)}$, with $\delta\Enn$ and $\delta\nu_{12}$ \TM{obtained from fitting the results of numerical simulations}. A simple empirical relation for $\delta\Enn$ (and $\delta\nu_{12}$) has been derived for materials with cubic crystal lattices.
	\item To account for realistic stresses acting on a grain-boundary model, the external loading has been dressed by fluctuations, $\mathbf{\Sigma}\to\mathbf{\Sigma}+\mathbf{f}$. To a good approximation, the resulting fluctuations of grain-boundary-normal stresses ($f_{nn}$), have been found to be independent of stresses. \TM{Their distribution is Gaussian}, with standard deviation of the form $s(f_{nn})\approx\Sigma_{\text{mis}}\mathcal{F}(A^u)$, where $\mathcal{F}(A^u)$ is an \TM{empirical function, that increases with the value of} universal elastic anisotropy index $A^u$.
	\item A comparison with finite element simulations has demonstrated that the derived semi-analytical expression for a local $\tilde{\sigma}_{nn}^{(2)}$ is accurate only up to unknown stress fluctuations, \TM{\textit{i.e.}, the uncertainty of model prediction is $s(f_{nn})$}. However, the corresponding statistical distributions, $\pdf(\tilde{\sigma}_{nn}^{(2)})$, have been shown to be very accurate. Indeed, an \textit{excellent agreement} with the simulation results has been found for arbitrary grain-boundary types in a general elastic untextured polycrystalline material%
	\footnote{Materials with cubic lattice symmetry have been chosen in this article for demonstration purposes only.}
	under arbitrary uniform loading.
	\item From the application point of view, a reliable tool has been derived for quick and accurate calculation of grain-boundary-normal-stress distributions. \TM{We expect its results should prove extremely} useful for the probabilistic modeling of grain-boundary-damage initiation such as IGSCC.
\end{itemize}
%

% ==========================================================================
\section*{Acknowledgments}
\label{sec:7}
% ==========================================================================
We gratefully acknowledge financial support provided by Slovenian Research Agency (grant P2-0026). We also thank J\'er\'emy Hure for useful discussions and comments that \TM{helped us to improve} the manuscript.

%\newpage

% ==========================================================================
\appendix
% ==========================================================================

% ==========================================================================
\section{Finite element aggregate model}
% ==========================================================================
\label{app:fem}

Polycrystalline aggregate models are generated upon Voronoi tessellations~\cite{neper} with periodic microstructures in all three spatial directions%
\footnote{Periodic boundary conditions imply \TM{the absence of} free surfaces. Quantities derived in the model therefore correspond to bulk \TM{grains}.
}~\cite{elshawish2019}.
Finite element meshes are generated with quadratic tetrahedral elements to preserve the geometry of the grains. An example of the model with $4000$ grains, used throughout this study, is shown in Fig.~\ref{fig:geom}(a). A general uniform loading $\mathbf{\Sigma}^{\TM{\text{lab}}}$ is applied to the aggregate, where $\Sigma_{ij}=\ave{\sigma_{ij}}$ \TM{for averages taken over the entire volume of the aggregate model}. Since grains are assumed \TM{ideally} elastic, a unit loading can be applied, using a small strain approximation.

It is important to note that the analysis of each GB type requires a dedicated aggregate model. To \TM{isolate} the effect of \TM{selected} GB type, the same grain topology and finite element mesh (see Fig.~\ref{fig:geom}(a)) are used in all of them. 

Due to topological constraints, the same GB character \TM{cannot be assigned to all GBs}. In practice, a \TM{chosen} $[abc]$-$[def]$-$\Delta\omega$ GB type can be \TM{imposed on at most} $\sim 17$\% of GBs% 
\footnote{Fraction $\sim 17$\% denotes a ratio between the area of GBs \TM{of a selected GB type} and the area of all GBs in the model. To improve statistical evaluations, GBs \TM{in the first category (special GBs)} are selected from the largest available GBs in the aggregate. In a given aggregate with $4000$ grains, the total number of GBs is $31154$ and the number of special GBs is $1631$.}
in a given aggregate, with remaining GBs \TM{belonging to a} random type (\textit{i.e.}, \TM{the two grains adjacent to the GB are assigned random orientations}). 

Also, it has been verified that the aggregate size ($4000$ grains) and finite element mesh density ($\sim 5$ million total elements) are sufficiently large to \TM{produce} negligible finite size effects.

The constitutive equations of the generalized Hooke's law are solved with finite element solver Abaqus~\cite{abaqus} in a small strain approximation. Numerically calculated stress fields $\tou{\sigma}_i$ are then used to obtain a single $\snn(k)$ value \TM{for each} GB $k$ of a given type $[abc]$-$[def]$-$\Delta\omega$. In short, $\snn(k)$ is calculated by projecting stress $\tou{\sigma}_i$ onto a GB normal $\tou{n}^{(k)}$ and averaging over all finite elements $i$ touching the GB $k$; $\snn(k)=\sum_i A_i^{(k)} \tou{n}^{(k)}\cdot\tou{\sigma}_i\cdot\tou{n}^{(k)} / \sum_i A_i^{(k)}$, where $A_i^{(k)}$ is the area of GB-element facet touching the GB, and $\tou{\sigma}_i=1/3\sum_{j=1}^{3}\tou{\sigma}_{i,j}$ is Cauchy stress averaged over three Gauss points $j$ of element $i$, that are located closest to the GB plane. Besides local stresses, the first two statistical moments, the mean \TM{value} and standard deviation of $\pdf(\snn)$, are calculated as $\ave{\snn}=\sum_k A_k \snn(k)/ \sum_k A_k$ and $s(\snn)=\sqrt{\ave{\sigma_{nn}^2}-\ave{\sigma_{nn}}^2}$, respectively, for $A_k$ denoting the area of GB $k$. The summation $k$ is performed over all special GBs of a given type.%
\footnote{The response of random GBs is calculated \TM{in an aggregate with randomly oriented grains and summation index $k$ running over all GBs.}}
%

% ==========================================================================
\section{Analytic expressions for grains with cubic lattice symmetry}
% ==========================================================================
\label{app:epsnn}

In this section, analytic expressions for cubic lattice symmetry are given for completeness. It is assumed that rotation $\mathbf{R}^{\text{cry}}$, defined in Eq.~\eqref{eq:Rloc}, \TM{is} used for expressing the \TM{crystallographic properties of the grain} in \TM{a local} GB coordinate system, whose GB-plane \TM{normal (local $z$-axis) is oriented along} the $[h k l]$ \TM{direction of the grain (crystallographic)} coordinate system and with $\omega$ denoting the twist angle \TM{about} the GB normal. 

Strain component in the direction of GB normal can be expressed (for a cubic grain) as
\begin{widetext}
\ba
\begin{split}
	\epsilon_{zz}&= \TM{\sum_{k,l=1}^3 s_{33kl}^{\text{GB}} \, \sigma_{kl} =} \hspace{-.5cm} \sum_{k,l,m,n,o,p=1}^3 \hspace{-.6cm}  R_{3m}^{\text{cry}} R_{3n}^{\text{cry}} R_{ko}^{\text{cry}} R_{lp}^{\text{cry}} s_{mnop}^{\text{cry}} \, \TM{\sigma_{kl}} \\
	&=\TM{\sigma_{xx}} \left(\frac{S_0 \left(h k \left(2 h k \left(h^2+k^2+l^2\right)\sin ^2\omega+l \left(k^2-h^2\right)   \sqrt{h^2+k^2+l^2}\sin 2 \omega\right)+2 l^2 \left(h^2 k^2+h^4+k^4\right) \cos ^2\omega \right)}{\left(h^2+k^2\right)
		\left(h^2+k^2+l^2\right)^2}+S_{12}\right) + \\
	&+\TM{\sigma_{yy}} \left(\frac{S_0 \left(l \left(h k (h-k) (h+k)  \sqrt{h^2+k^2+l^2}\sin 2 \omega+2 l \left(h^2 k^2+h^4+k^4\right) \sin ^2\omega \right)+2 h^2 k^2  \left(h^2+k^2+l^2\right)\cos ^2\omega\right)}{\left(h^2+k^2\right)
		\left(h^2+k^2+l^2\right)^2}+S_{12}\right) + \\
	&+\TM{\sigma_{zz}} \left(\frac{S_0 \left(h^4+k^4+l^4\right)}{\left(h^2+k^2+l^2\right)^2}+S_{12}+\frac{S_{44}}{2}\right) + \\
	&+\TM{\sigma_{xy}}\frac{2 S_0 \left( \left(h^2 k^2 \left(h^2+k^2\right)-l^2 \left(h^4+k^4\right)\right)\sin 2 \omega+h k l \left(k^2-h^2\right)  \sqrt{h^2+k^2+l^2}\cos 2 \omega\right)}{\left(h^2+k^2\right) \left(h^2+k^2+l^2\right)^2} + \\
	&+\TM{\sigma_{xz}}\frac{2 S_0 \left(h k \left(k^2-h^2\right)  \sqrt{h^2+k^2+l^2}\sin \omega+l  \left(-l^2 \left(h^2+k^2\right)+h^4+k^4\right)\cos \omega\right)}{\sqrt{h^2+k^2} \left(h^2+k^2+l^2\right)^2} + \\
	&+\TM{\sigma_{yz}}\frac{2 S_0 \left(l  \left(l^2 \left(h^2+k^2\right)-h^4-k^4\right)\sin \omega+h k \left(k^2-h^2\right)  \sqrt{h^2+k^2+l^2}\cos \omega \right)}{\sqrt{h^2+k^2} \left(h^2+k^2+l^2\right)^2} ,
\end{split}	
\ea
\end{widetext}
where $S_{11}:=s_{1111}^{\text{cry}}$, $S_{12}:=s_{1122}^{\text{cry}}$ and $S_{44}:=\TM{4} \, s_{\TM{2323}}^{\text{cry}}$ are the components of compliance tensor of a grain in Voigt notation and $S_0:=S_{11}-S_{12}-S_{44}/2$. Expressions for \TM{the other two diagonal components of strain tensor}, $\epsilon_{xx}$ and $\epsilon_{yy}$, are derived analogously, but are omitted here for brevity.

The effective GB-stiffness parameter $\Enn$, measuring the average stiffness of GB neighborhood along the GB-normal direction, takes the following form (for cubic grains)
\ba
\label{eq:E12cubic}
	\Enn&=&\frac{2\ave{E}^{-1}}{s^{\text{GB,abc}}_{3333}+s^{\text{GB,def}}_{3333}}\\
	&=&\frac{\ave{E}^{-1}}{S_{11}-S_0\left(\frac{(a b)^2+(a c)^2+(b c)^2}{\left(a^2+b^2+c^2\right)^2}+\frac{(d e)^2+(d f)^2+(e f)^2}{\left(d^2+e^2+f^2\right)^2}\right)}\nonumber.
\ea
The effective GB Poisson's \TM{ratio} $\nu_{12}$, measuring the average ratio of \TM{transverse and} axial responses \TM{(strains)} in both GB grains, takes the following form (for cubic grains)
\ba
	\nu_{12}&=&-\frac{\ave{E}}{4}\left(s^{\text{GB,abc}}_{3311}+s^{\text{GB,abc}}_{3322}+s^{\text{GB,def}}_{3311}+s^{\text{GB,def}}_{3322}\right)\nonumber\\
	&=&\ave{\nu}+\frac{1}{2}(\Enn^{-1}-1),
\ea
\TM{where $\ave{\nu} = \tfrac{1}{2} \left (1-\ave{E} (S_{11}+2 S_{12})\right )$.}

% ==========================================================================
\section{General grain boundary model solution for cubic crystal lattices}
% ==========================================================================
\label{app:cubic}

The highest-order (reduced) solution $\snn^{(3)}$, derived in Sec.~\ref{sec:extended} for the most general grain-lattice symmetry and external loading, simplifies enormously for cubic lattice symmetry, where
	\TM{
	\ba
	\begin{split}
	s_{1111}^{\text{cry}}&=s_{2222}^{\text{cry}}=s_{3333}^{\text{cry}} := S_{11} , \\
	s_{1122}^{\text{cry}}&=s_{1133}^{\text{cry}}=s_{2233}^{\text{cry}} := S_{12} , %\\
	%s_{2323}^{\text{cry}}&=s_{1313}^{\text{cry}}=s_{1212}^{\text{cry}} , \\
	%s_{ijkl}^{\text{cry}}&=0 , \quad \text{otherwise}
	\end{split}
	\ea
    }
	and
	\be
	\ave{\nu}=\frac{1}{2}\left(1-\TM{(S_{11}+2 S_{12})}\ave{E}\right),
	\ee
	which in turn implies
	\ba
	\begin{split}
	s_{tl}^{hkl} &= -s_{ll}^{hkl} + \TM{(S_{11}+2 S_{12})} , \\
	s_{tt}^{hkl} &= \frac{1}{2} \left(s_{ll}^{hkl} + \TM{(S_{11}+2 S_{12})}\right) ,
	\end{split}
	\ea
	meaning $A^{(3)}_{\text{cub}}$ and $B^{(3)}_{\text{cub}}$ are only functions of Young`s moduli $E_{abc}$ and $E_{def}$ along the GB-normal direction \TM{in both grains},
	\begin{widetext}
		\ba
		\begin{split}
		A^{(3)}_{\text{cub}} &= \frac{(1 + (2 L_t + s') E_{12}) (L_n + 2 s' + 4 (L_t + s') E_{12} + (2 L_n L_t + L_n s' - 2 s'^2) E_{12}) + (L_n + 2 s') \Delta_{12}}{(1 + (2 L_t + s') E_{12}) (L_n + 2 s' + 4 (L_t + s') + (2 L_n L_t + L_n s' - 2 s'^2) E_{12}) + (L_n + 2 s' + 4 (L_t + s')) \Delta_{12}} , \\
		%
		%	B^{(3)}_{\text{cub}} &=& \frac{(1 + (2 L_t + s') E_{12}) (2 (L_t + s') - 2 (L_t + s') E_{12}) + 2 (L_t + s') \Delta_{12}}{(1 + (2 L_t + s') E_{12}) (L_n + 2 s' + 4 (L_t + s') + (2 L_n L_t + L_n s' - 2 s'^2) E_{12}) + (L_n + 2 s' + 4 (L_t + s')) \Delta_{12}}.\nonumber\\
		%
		B^{(3)}_{\text{cub}} &= \frac{1}{2}\left(1-A^{(3)}_{\text{cub}}\right).
		\end{split}
		\ea
	\end{widetext}
	All the used quantities are dimensionless, with $s'$ denoting
	\be
	s':=\ave{E} \TM{(S_{11}+2 S_{12})}=1-2\ave{\nu}.
	\ee
	$E_{abc}$ and $E_{def}$ appear only in combinations $E_{12}$ and $\Delta_{12}$, where
	\vspace{-.1cm}
	\ba
	\begin{split}
	\Delta_{12}&:=\frac{4 E_{abc}^{-1} E_{def}^{-1}}{(E_{abc}^{-1} + E_{def}^{-1})^2} - 1 \\
	&=\left(E_{12} \ave{E}\right)^2 E_{abc}^{-1} E_{def}^{-1} - 1
	\end{split}
	\ea
	is a sort of geometric mean of inverse Young's moduli of both grains, measuring the deviation from a single-grain scenario (\textit{i.e.}, the $[abc]$-$[abc]$ GB type), \TM{in which} $\Delta_{12}$ vanishes. If the effective GB stiffness $E_{12}$ is a measure of the average stiffness of the $[abc]$-$[def]$ GB neighborhood along the GB-normal direction, then $\Delta_{12}$ represents an additional, orthogonal degree of freedom, that can break the $\sigma_{nn}$- and $s(\sigma_{nn})$-degeneracies of GB types with the same $E_{12}$ (but different values of $\Delta_{12}$).
	
	The \TM{GB-normal} stress and the corresponding first two statistical moments thus become
	\begin{widetext}
		\ba
		\label{eq:std_dev_cubic}
		\begin{split}
		\snn^{(3)} &= A^{(3)}_{\text{cub}}\Sigma_{zz} + B^{(3)}_{\text{cub}} (\Sigma_{xx}+\Sigma_{yy}) , \\
		\ave{\snn^{(3)}}&=\frac{1}{3}\operatorname{tr}(\mathbf{\Sigma}^{\text{lab}}) , \\
		s(\snn^{(3)})&= \frac{2\Sigma_{\text{mis}}^{\text{lab}}}{3\sqrt{5}}\left\vert \frac{(1 + (2 L_t + s') E_{12}) (L_n - 2 L_t + 6 (L_t + s') E_{12} + (2 L_n L_t + L_n s' - 2 s'^2) E_{12}) + (L_n - 2 L_t) \Delta_{12}}{(1 + (2 L_t + s') E_{12}) (L_n + 2 s' + 4 (L_t + s') + (2 L_n L_t + L_n s' - 2 s'^2) E_{12}) + (L_n + 2 s' + 4 (L_t + s')) \Delta_{12}} \right\vert .
		\end{split}
		\ea
	\end{widetext}
	The relevance of $E_{12}$ parameter determining the $s(\sigma_{nn})$ for materials with cubic lattice symmetry was correctly identified already in Ref.~\cite{elshawish2021}. The new parameter $\Delta_{12}$ in Eq.~\eqref{eq:std_dev_cubic} \TM{represents} a higher-order correction, which can, at least qualitatively, explain the (small) spread of $s(\sigma_{nn})$ values, observed numerically \TM{for GB types with} the same $E_{12}$ (see Fig. 13(a) in Ref.~\cite{elshawish2021}).

% ==========================================================================
\section{Material elastic properties}
% ==========================================================================
\label{app:mat}

\TM{Elastic constants of single crystals with cubic symmetry, together with their aggregate properties, are listed in Table~\ref{tab:au} for several representative materials.} The materials are ordered according to their universal elastic anisotropy index $A^u$~\cite{ranganathan}, where $A^u=0$ corresponds to an isotropic crystal.

\begin{table}[b]
	\caption{\label{tab:au}
		Elastic constants \TM{$C_{ij}$ (in Voigt notation)} of single crystals \TM{with cubic symmetry}~\cite{bower} \TM{and their} aggregate properties. $C_{ij}$ and $\ave{E}$ are in units of GPa. Fe is assumed in gamma phase.}
	\begin{ruledtabular}
		\begin{tabular}{lllllll}
			Crystal & $C_{11}$ & $C_{12}$ & $C_{44}$ & $\ave{E}$ & $\ave{\nu}$ & $A^u$ \\
			\colrule
			Al         & 107.3    & 60.9    &  28.3 & 70.41 & 0.346 & 0.05 \\
			%			Nb         & 240.2    & 125.6   &  28.2 & & & 0.63 \\
			%			Au         & 192.9    & 163.8   &  41.5 & & & 1.44 \\
			Fe         & 197.5    & 125.0   &  122.0 & 195.2 & 0.282 & 2.00 \\
			Li         & 13.5     & 11.44   &  8.78  & 10.94 & 0.350 & 7.97 \\
		\end{tabular}
	\end{ruledtabular}
\end{table}

In Table~\ref{tab:orto} the elastic constants of a single crystal with orthorhombic symmetry (CaSO$_4$) and its aggregate properties are gathered.

\begin{table*}[b]
	\caption{\label{tab:orto}
		Elastic constants of a single crystal \TM{with orthorhombic symmetry} (CaSO$_4$)~\cite{simmonswang} \TM{and its} aggregate properties. $C_{ij}$ and $\ave{E}$ are in units of GPa.}
	\begin{ruledtabular}
		\begin{tabular}{lllllllllllll}
			Crystal & $C_{11}$ & $C_{22}$ & $C_{33}$ & $C_{12}$ & $C_{13}$ & $C_{23}$ & $C_{44}$ & $C_{55}$ & $C_{66}$ & $\ave{E}$ & $\ave{\nu}$ & $A^u$ \\
			\colrule
			CaSO$_4$  & 93.82 & 185.5 & 111.8 & 16.51 & 15.20 & 31.73 & 32.47 & 26.53 &	9.26 & 71.77 & 0.282 & 2.78
		\end{tabular}
	\end{ruledtabular}
\end{table*}
%

% ==========================================================================
\section{Model of stress fluctuations}
% ==========================================================================
\label{app:gauss}

\begin{figure}
	\includegraphics[width=0.9\columnwidth]{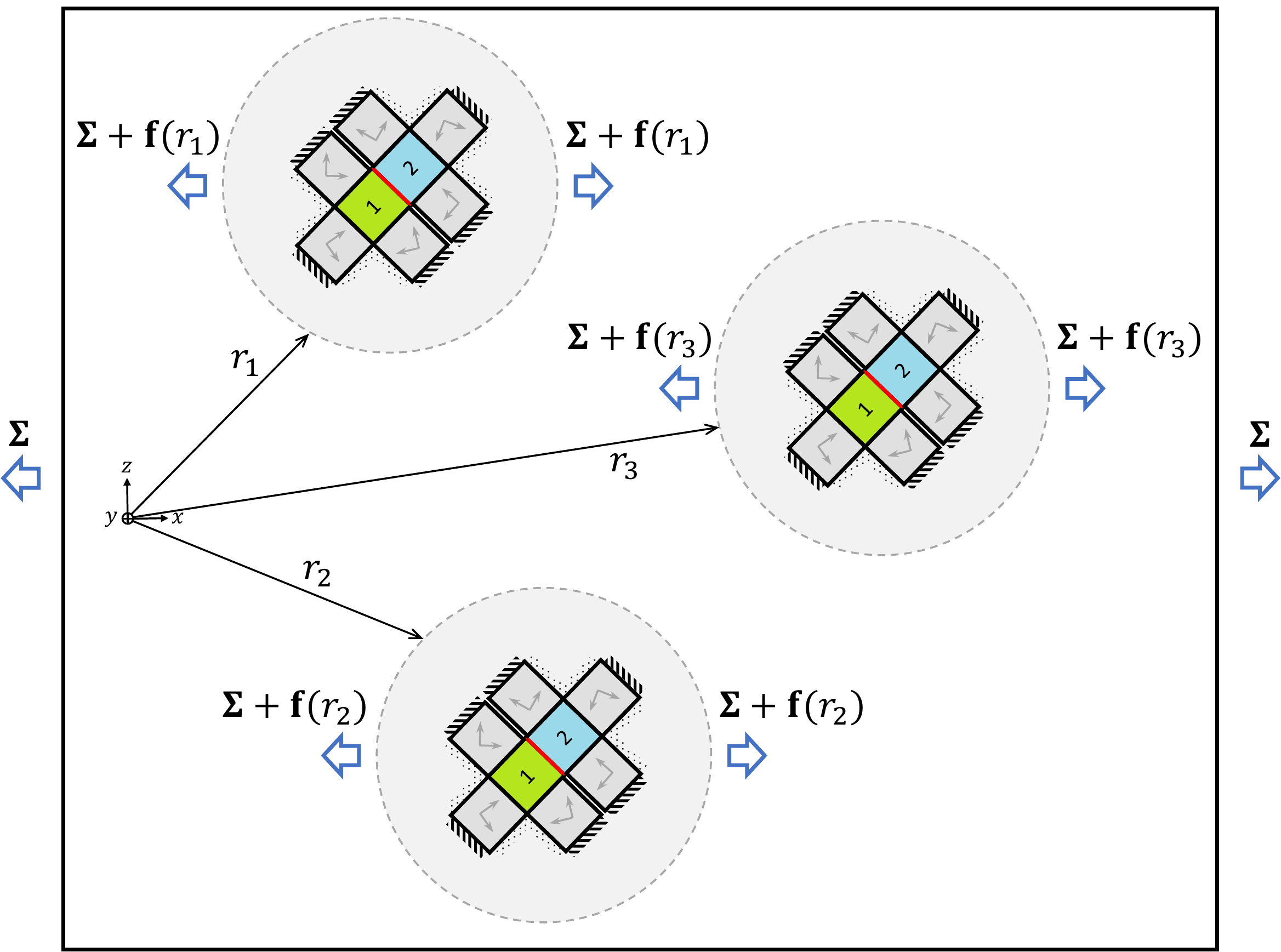}
	\caption{The fluctuation stress tensor $\mathbf{f}(r)$ at position $r$ is defined as the difference between the actual loading stress $\mathbf{\Sigma}(r)$, acting on a given GB neighborhood %(with a fixed GB type and fixed GB normal) 
	at position $r$, and the external loading stress $\mathbf{\Sigma}$, \TM{\textit{i.e.},} $\mathbf{f}(r)=\mathbf{\Sigma}(r)-\mathbf{\Sigma}$.}
	\label{fig:fluc}
\end{figure}

The stress applied to the GB model (see, \textit{e.g.}, Fig.~\ref{fig:chains}) is assumed to be equal to the external stress, modified by fluctuations, $\mathbf{\Sigma}+\mathbf{f}(r)$, where $\mathbf{f}(r)$ is the fluctuation stress tensor at position $r$. In a large aggregate, where crystallographic orientations of the grains are \TM{uncorrelated with} their shapes (assuming zero morphological texture), the \TM{average} fluctuation stress tensor should \TM{go towards} zero, \TM{\textit{i.e.},}
\be
	\ave{f_{ij}(r)}_r=\ave{(\mathbf{R} \, \mathbf{f}(r) \, \mathbf{R}^T)_{ij}}_r= 0,
\ee
when averaged over all GBs of a \TM{chosen} type \TM{(and thus having a specific value of $\Enn$, $\nu_{12}$, ...)}
%
%\footnote{With fixed $\Enn$, $\nu_{12}$, ...}
%
and for a fixed GB-normal direction $n$ (see Fig.~\ref{fig:fluc}). \TM{This should be true in any} coordinate system ($\mathbf{R}$ denotes an arbitrary rotation matrix).

Since fluctuations are induced by external loading $\mathbf{\Sigma}$ and strain incompatibility of the grains (which correlates with the elastic anisotropy index $A^u$), it seems reasonable to assume that the corresponding standard deviation $s$ depends on $\mathbf{\Sigma}$ and $A^u$, possibly on GB model intrinsic parameters (\textit{e.g.}, $\Enn$, $\nu_{12}$, ...), but not on the global aggregate (or external loading) rotation $\mathbf{R}$,
\be
\label{eq:d2}
	s(f_{ij}(r))=s((\mathbf{R} \, \mathbf{f}(r) \, \mathbf{R}^T)_{ij})=\mathcal{F}(\mathbf{\Sigma},A^u,\Enn,...).
\ee
In addition, the rotational invariance of $s$ suggests that $\mathbf{\Sigma}$ dependence in Eq.~\eqref{eq:d2} can be expressed solely in terms of $\mathbf{\Sigma}$ invariants
\be
\label{eq:d3}
	\mathcal{F}(\mathbf{\Sigma})\to\mathcal{F}\left(\operatorname{tr}(\mathbf{\Sigma}),\operatorname{det}(\mathbf{\Sigma}),\operatorname{tr}(\mathbf{\Sigma}_{\text{dev}}^2)\right).
\ee
In the limit where the external loading is of hydrostatic form, $\mathbf{\Sigma}=\mathbf{\Sigma}_{\text{hyd}}$ (and $\mathbf{\Sigma}_{\text{dev}}=0$), a trivial solution is obtained with no stress fluctuations, $\mathbf{f}=0$, implying that $\mathbf{\Sigma}_{\text{dev}}$ is the only relevant loading contribution in Eq.~\eqref{eq:d3}. To account for this limit, the following fluctuation stress tensor is finally proposed at position $r$,
\be
\label{eq:fluc1}
	\mathbf{f}(r)= \eta(r) \, \mathbf{R}(r) \, \mathbf{\Sigma}_{\text{dev}} \, \mathbf{R}(r)^T ,
\ee
where
\be
	\mathbf{R}(r)=\mathbf{R}\left(\alpha_1(r),\alpha_2(r),\alpha_3(r)\right)
\ee
is a random rotation matrix with corresponding Euler angles $\alpha_j(r)$, and $\eta(r)$ is a random number with the assumed Gaussian distribution $\mathcal{N}(0,s^2(\eta))$ with $s(\eta)=\mathcal{F}(A^u,\Enn,...)$%
\footnote{For demonstrating purposes, a simple scalar multiplication $\eta$ is used for rescaling, instead of a more general matrix multiplication.}.

In Eq.~\eqref{eq:fluc1}, the fluctuation stress tensor is modeled as the deviatoric part of the external loading%
\footnote{The hydrostatic part of the external loading is invariant to grain orientations and thus unable to produce strain incompatibility between the grains, which is the source of stress fluctuations.}
rescaled and rotated by a random amount to account for uncertainty of far-away grains that blur the external loading.

Using $\mathbf{f}(r)$ defined in Eq.~\eqref{eq:fluc1} and a general expression $\snn^{(k)}=A^{(k)} \Sigma_{zz}+B^{(k)} (\Sigma_{xx}+\Sigma_{yy})$, the GB-normal-stress fluctuations evaluate to (using notation $f_{nn}^{(k)}=\Delta\snn^{(k)}$)
\ba
\begin{split}
	f_{nn}^{(k)}(r)&=A^{(k)} f_{zz}(r)+B^{(k)} \left(f_{xx}(r)+f_{yy}(r)\right) \\
	&=\left(A^{(k)}-B^{(k)}\right) f_{zz}(r) ,
\end{split}
\ea
where the $\operatorname{tr}(\mathbf{f}(r))=0$ property of Eq.~\eqref{eq:fluc1} has been used. The first two statistical moments follow as
\begin{widetext}
\ba
\begin{split}
	\ave{f_{nn}^{(k)}(r)}_r&=\left(A^{(k)}-B^{(k)}\right)\int\eta\pdf(\eta)d\eta\iiint \Sigma_{\text{dev},zz}\pdf(\alpha_1,\alpha_2,\alpha_3)\ d\alpha_1 d\alpha_2 d\alpha_3 \\
	&=0 ,
\end{split}
\ea
and
\ba
\begin{split}
	s(f_{nn}^{(k)}(r))&=\sqrt{\ave{\left(f_{nn}^{(k)}(r)\right)^2}_r} \\
	&=\left|A^{(k)}-B^{(k)}\right|\sqrt{\int\eta^2\pdf(\eta)d\eta\iiint \Sigma_{\text{dev},zz}^2\pdf(\alpha_1,\alpha_2,\alpha_3)\ d\alpha_1 d\alpha_2 d\alpha_3} \\ 
	&=\frac{2 \Sigma_{\text{mis}}}{3\sqrt{5}} \left|A^{(k)}-B^{(k)}\right| s(\eta),
\end{split}
\ea
\end{widetext}
with $\pdf(\eta)$ corresponding to Gaussian distribution $\mathcal{N}(0,s^2(\eta))$ and $\pdf(\alpha_1,\alpha_2,\alpha_3)$ to random orientation distribution. For crystal lattices with cubic symmetry, $\left|A^{(k)}-B^{(k)}\right|$ simplifies to $\left|3A^{(k)}-1\right|/2$.

In overall, the derived expression for $s$ suggests that the loading contribution to GB-normal-stress fluctuations can be trivially decoupled as
\be
	s(f_{nn}^{(k)}(r))=\Sigma_{\text{mis}}\ \mathcal{F}(A^u,\Enn,...).
\ee
%

% ==========================================================================
\section{Macroscopic response of random grain boundaries}
% ==========================================================================
\label{app:random}

The upgraded models for local \TM{stresses} $\tilde{\sigma}_{nn}^{(k)}$ and \TM{their macroscopic manifestation} $\pdf(\tilde{\sigma}_{nn}^{(k)})$ \TM{are typically used for GBs of a certain GB type, corresponding to fixed values of $\Enn$ and $\nu_{12}$ (together with $\delta\Enn$ and $\delta\nu_{12}$).} The response of random GBs can therefore be estimated by integration over all GB types, \TM{hence taking into account all GBs in a given aggregate},
\be
	\pdf_{\text{rnd}}(\tilde{\sigma}_{nn}^{(k)})=\iint w(\Enn,\nu_{12}) \pdf(\tilde{\sigma}_{nn}^{(k)}) d\Enn d\nu_{12},
\ee
where $w(\Enn,\nu_{12})$ represents the distribution function of GB types in an aggregate and $\pdf(\tilde{\sigma}_{nn}^{(k)})$ the macroscopic response of a specific GB type; cf.~Eq.~$\eqref{eq:conv}$.

For aggregates with zero crystallographic texture, the response of random GBs can also be obtained \TM{more elegantly. There, the average $\snn$ of all GBs with the same GB normal $n$ (but arbitrary $\Enn$ and $\nu_{12}$)} should be equal to the external loading $\mathbf{\Sigma}$, projected onto that GB plane, \textit{i.e.}, $\ave{\snn}=\snn^{(0)}=\Sigma_{zz}$. This is true because crystallographic orientations of grains are \TM{not correlated with orientations of} GB planes, \TM{hence randomly distributed grain orientations} are expected along every GB-normal direction $n$, providing an average (bulk) response. Thus, 
\be
	\pdf_{\text{rnd}}(\tilde{\sigma}_{nn}^{(k)})\approx\pdf(\tilde{\sigma}_{nn}^{(0)}) ,
\ee
or, equivalently,
\be
	\pdf_{\text{rnd}}(\tilde{\sigma}_{nn}^{(k)})\approx\left.\pdf(\tilde{\sigma}_{nn}^{(k\ge1)})\right\vert_{
%		\substack{
%			\Enn+d\Enn:= 1\\
%			E_3+dE_3:= 1\\
%			\nu_{12}+d\nu_{12}:=\ave{\nu}
%		}
		\scriptsize\begin{aligned}
			\Enn+d\Enn&= 1\\[-4pt]
			E_3+dE_3&= 1\\[-4pt]
			\nu_{12}+d\nu_{12}&=\ave{\nu}
		\end{aligned}
	}.
\ee
In practice, the \TM{(approximate) $\pdf$ of stress response in} a random aggregate can be obtained simply by convoluting the isotropic solution $\pdf(\snn^{(0)})$ with the Gaussian distribution $\mathcal{N}(0,s^2(f_{nn})$, taking $s(f_{nn})$ from Eq.~\eqref{eq:conv2}.

% ==========================================================================
\section{Grain-boundary-normal distribution in aggregates with elongated grains}
% ==========================================================================
\label{app:elongated}

In aggregates with zero morphological texture (\textit{i.e.}, with no preferred direction for grain shapes), the GB normals, $n=(\sin\theta\cos\psi,\sin\theta\sin\psi,\cos\theta)$, are uniformly distributed on a sphere, with corresponding distribution functions for the two angles
\ba
\begin{split}
	f(\cos\theta)&=\frac{1}{2} \, ; \quad (-1\le\cos\theta\le1) , \\
	f(\psi)&=\frac{1}{2\pi} \, ; \quad (0\le\psi\le2\pi) .
\end{split}	 
\ea
To generate an aggregate with grains elongated along the $z$-axis, a simple geometrical scaling can be applied to the initially isotropic aggregate
\be
	x\to x,\quad y\to y,\quad z\to \lambda_z z \, ; \quad (\lambda_z>0).
\ee
As a result of such transformation, the two distribution functions become
\ba
\begin{split}
	f(\cos\theta)&=\frac{\lambda_z}{2}\left(\frac{1}{1+(\lambda_z^2-1)\cos^2\theta}\right)^{3/2} \hspace{-.4cm} ; & \ (-1\le\cos\theta\le1) , \\
	f(\psi)&=\frac{1}{2\pi} \, ; & \ (0\le\psi\le2\pi) .
\end{split}	 
\ea
%
%\TM{where their domains remain the same}; $-1\le\cos\theta\le1$ and $0\le\psi\le2\pi$. 
The derived distributions are used to predict the \TM{stress response} $\pdf(\snn^{(k)})$ of any GB type within the elongated aggregate%
\footnote{\TM{In aggregates with non-zero morphological texture the first two statistical moments do not scale anymore with $\operatorname{tr}(\mathbf{\Sigma}^{\text{lab}})$ and $\Sigma^{\text{lab}}_{\text{mis}}$, respectively.}}
(see Sec.~\ref{verifi}).

\bibliography{spebib2}

\begin{thebibliography}{41}%
\makeatletter
\providecommand \@ifxundefined [1]{%
 \@ifx{#1\undefined}
}%
\providecommand \@ifnum [1]{%
 \ifnum #1\expandafter \@firstoftwo
 \else \expandafter \@secondoftwo
 \fi
}%
\providecommand \@ifx [1]{%
 \ifx #1\expandafter \@firstoftwo
 \else \expandafter \@secondoftwo
 \fi
}%
\providecommand \natexlab [1]{#1}%
\providecommand \enquote  [1]{``#1''}%
\providecommand \bibnamefont  [1]{#1}%
\providecommand \bibfnamefont [1]{#1}%
\providecommand \citenamefont [1]{#1}%
\providecommand \href@noop [0]{\@secondoftwo}%
\providecommand \href [0]{\begingroup \@sanitize@url \@href}%
\providecommand \@href[1]{\@@startlink{#1}\@@href}%
\providecommand \@@href[1]{\endgroup#1\@@endlink}%
\providecommand \@sanitize@url [0]{\catcode `\\12\catcode `\$12\catcode
  `\&12\catcode `\#12\catcode `\^12\catcode `\_12\catcode `\%12\relax}%
\providecommand \@@startlink[1]{}%
\providecommand \@@endlink[0]{}%
\providecommand \url  [0]{\begingroup\@sanitize@url \@url }%
\providecommand \@url [1]{\endgroup\@href {#1}{\urlprefix }}%
\providecommand \urlprefix  [0]{URL }%
\providecommand \Eprint [0]{\href }%
\providecommand \doibase [0]{https://doi.org/}%
\providecommand \selectlanguage [0]{\@gobble}%
\providecommand \bibinfo  [0]{\@secondoftwo}%
\providecommand \bibfield  [0]{\@secondoftwo}%
\providecommand \translation [1]{[#1]}%
\providecommand \BibitemOpen [0]{}%
\providecommand \bibitemStop [0]{}%
\providecommand \bibitemNoStop [0]{.\EOS\space}%
\providecommand \EOS [0]{\spacefactor3000\relax}%
\providecommand \BibitemShut  [1]{\csname bibitem#1\endcsname}%
\let\auto@bib@innerbib\@empty
%</preamble>
\bibitem [{\citenamefont {Nishioka}\ \emph {et~al.}(2008)\citenamefont
  {Nishioka}, \citenamefont {Fukuya}, \citenamefont {Fujii},\ and\
  \citenamefont {Torimaru}}]{nishioka2008}%
  \BibitemOpen
  \bibfield  {author} {\bibinfo {author} {\bibfnamefont {H.}~\bibnamefont
  {Nishioka}}, \bibinfo {author} {\bibfnamefont {K.}~\bibnamefont {Fukuya}},
  \bibinfo {author} {\bibfnamefont {K.}~\bibnamefont {Fujii}},\ and\ \bibinfo
  {author} {\bibfnamefont {T.}~\bibnamefont {Torimaru}},\ }\bibfield  {title}
  {\bibinfo {title} {{IASCC} initiation in highly irradiated stainless steels
  under uniaxial constant load conditions},\ }\href@noop {} {\bibfield
  {journal} {\bibinfo  {journal} {J. Nuc. Sci. and Tech.}\ }\textbf {\bibinfo
  {volume} {45}},\ \bibinfo {pages} {1072} (\bibinfo {year}
  {2008})}\BibitemShut {NoStop}%
\bibitem [{\citenamefont {Le~Millier}\ \emph {et~al.}(2013)\citenamefont
  {Le~Millier}, \citenamefont {Calonne}, \citenamefont {Cr\'epin},
  \citenamefont {Duhamel}, \citenamefont {Fournier}, \citenamefont {Gaslain},
  \citenamefont {H\'eripr\'e}, \citenamefont {Toader}, \citenamefont
  {Vidalenc},\ and\ \citenamefont {Was}}]{lemillier}%
  \BibitemOpen
  \bibfield  {author} {\bibinfo {author} {\bibfnamefont {M.}~\bibnamefont
  {Le~Millier}}, \bibinfo {author} {\bibfnamefont {O.}~\bibnamefont {Calonne}},
  \bibinfo {author} {\bibfnamefont {J.}~\bibnamefont {Cr\'epin}}, \bibinfo
  {author} {\bibfnamefont {C.}~\bibnamefont {Duhamel}}, \bibinfo {author}
  {\bibfnamefont {L.}~\bibnamefont {Fournier}}, \bibinfo {author}
  {\bibfnamefont {F.}~\bibnamefont {Gaslain}}, \bibinfo {author} {\bibfnamefont
  {E.}~\bibnamefont {H\'eripr\'e}}, \bibinfo {author} {\bibfnamefont
  {O.}~\bibnamefont {Toader}}, \bibinfo {author} {\bibfnamefont
  {Y.}~\bibnamefont {Vidalenc}},\ and\ \bibinfo {author} {\bibfnamefont
  {G.}~\bibnamefont {Was}},\ }\bibfield  {title} {\bibinfo {title} {Influence
  of strain localization on {IASCC} of proton irradiated {304L} stainless steel
  in simulated {PWR} primary water},\ }in\ \href@noop {} {\emph {\bibinfo
  {booktitle} {16th International Conference on Environmental Degragation of
  Materials in Nuclear Power Systems - Water Reactors}}}\ (\bibinfo {year}
  {2013})\BibitemShut {NoStop}%
\bibitem [{\citenamefont {Stephenson}\ and\ \citenamefont
  {Was}(2014)}]{stephenson2014}%
  \BibitemOpen
  \bibfield  {author} {\bibinfo {author} {\bibfnamefont {K.}~\bibnamefont
  {Stephenson}}\ and\ \bibinfo {author} {\bibfnamefont {G.}~\bibnamefont
  {Was}},\ }\bibfield  {title} {\bibinfo {title} {Crack initiation behavior of
  neutron irradiated model and commercial stainless steels in high temperature
  water},\ }\href@noop {} {\bibfield  {journal} {\bibinfo  {journal} {J. Nuc.
  Mat.}\ }\textbf {\bibinfo {volume} {444}},\ \bibinfo {pages} {331} (\bibinfo
  {year} {2014})}\BibitemShut {NoStop}%
\bibitem [{\citenamefont {Gupta}\ \emph {et~al.}(2016)\citenamefont {Gupta},
  \citenamefont {Hure}, \citenamefont {Tanguy}, \citenamefont {Laffont},
  \citenamefont {Lafont},\ and\ \citenamefont {Andrieu}}]{gupta}%
  \BibitemOpen
  \bibfield  {author} {\bibinfo {author} {\bibfnamefont {J.}~\bibnamefont
  {Gupta}}, \bibinfo {author} {\bibfnamefont {J.}~\bibnamefont {Hure}},
  \bibinfo {author} {\bibfnamefont {B.}~\bibnamefont {Tanguy}}, \bibinfo
  {author} {\bibfnamefont {L.}~\bibnamefont {Laffont}}, \bibinfo {author}
  {\bibfnamefont {M.}~\bibnamefont {Lafont}},\ and\ \bibinfo {author}
  {\bibfnamefont {E.}~\bibnamefont {Andrieu}},\ }\bibfield  {title} {\bibinfo
  {title} {Evaluation of stress corrosion cracking of irradiated 304 stainless
  steel in {PWR} environment using heavy ion irradiation},\ }\href@noop {}
  {\bibfield  {journal} {\bibinfo  {journal} {J. Nuc. Mat.}\ }\textbf {\bibinfo
  {volume} {476}},\ \bibinfo {pages} {82} (\bibinfo {year} {2016})}\BibitemShut
  {NoStop}%
\bibitem [{\citenamefont {Fujii}\ \emph {et~al.}(2019)\citenamefont {Fujii},
  \citenamefont {Tohgo}, \citenamefont {Mori}, \citenamefont {Miura},\ and\
  \citenamefont {Shimamura}}]{fujii2019}%
  \BibitemOpen
  \bibfield  {author} {\bibinfo {author} {\bibfnamefont {T.}~\bibnamefont
  {Fujii}}, \bibinfo {author} {\bibfnamefont {K.}~\bibnamefont {Tohgo}},
  \bibinfo {author} {\bibfnamefont {Y.}~\bibnamefont {Mori}}, \bibinfo {author}
  {\bibfnamefont {Y.}~\bibnamefont {Miura}},\ and\ \bibinfo {author}
  {\bibfnamefont {Y.}~\bibnamefont {Shimamura}},\ }\bibfield  {title} {\bibinfo
  {title} {Crystallographic and mechanical investigation of intergranular
  stress corrosion crack initiation in austenitic stainless steel},\
  }\href@noop {} {\bibfield  {journal} {\bibinfo  {journal} {Mat. Sci. Eng. A}\
  }\textbf {\bibinfo {volume} {751}},\ \bibinfo {pages} {160} (\bibinfo {year}
  {2019})}\BibitemShut {NoStop}%
\bibitem [{\citenamefont {Cox}(1970)}]{cox}%
  \BibitemOpen
  \bibfield  {author} {\bibinfo {author} {\bibfnamefont {B.}~\bibnamefont
  {Cox}},\ }\href@noop {} {\emph {\bibinfo {title} {Environmentally induced
  cracking of zirconium alloys}}},\ \bibinfo {type} {Tech. Rep.}\ (\bibinfo
  {institution} {{A}tomic {E}nergy of {C}anada {L}imited},\ \bibinfo {year}
  {1970})\BibitemShut {NoStop}%
\bibitem [{\citenamefont {Cox}(1990)}]{cox1990}%
  \BibitemOpen
  \bibfield  {author} {\bibinfo {author} {\bibfnamefont {B.}~\bibnamefont
  {Cox}},\ }\bibfield  {title} {\bibinfo {title} {Envrionmentally-induced
  cracking of zirconium alloys - {A} review},\ }\href@noop {} {\bibfield
  {journal} {\bibinfo  {journal} {J. Nuc. Mat}\ }\textbf {\bibinfo {volume}
  {170}},\ \bibinfo {pages} {1} (\bibinfo {year} {1990})}\BibitemShut {NoStop}%
\bibitem [{\citenamefont {Van~Rooyen}(1975)}]{rooyen1975}%
  \BibitemOpen
  \bibfield  {author} {\bibinfo {author} {\bibfnamefont {D.}~\bibnamefont
  {Van~Rooyen}},\ }\bibfield  {title} {\bibinfo {title} {Review of the stress
  corrosion cracking of inconel 600},\ }\href@noop {} {\bibfield  {journal}
  {\bibinfo  {journal} {Corrosion}\ }\textbf {\bibinfo {volume} {31}},\
  \bibinfo {pages} {327} (\bibinfo {year} {1975})}\BibitemShut {NoStop}%
\bibitem [{\citenamefont {Shen}\ and\ \citenamefont
  {Shewmon}(1990)}]{shen1990}%
  \BibitemOpen
  \bibfield  {author} {\bibinfo {author} {\bibfnamefont {C.}~\bibnamefont
  {Shen}}\ and\ \bibinfo {author} {\bibfnamefont {P.}~\bibnamefont {Shewmon}},\
  }\bibfield  {title} {\bibinfo {title} {A mechanism for hydrogen-induced
  intergranular stress corrosion cracking in alloy 600},\ }\href@noop {}
  {\bibfield  {journal} {\bibinfo  {journal} {Metall. Trans. A.}\ }\textbf
  {\bibinfo {volume} {21A}},\ \bibinfo {pages} {1261} (\bibinfo {year}
  {1990})}\BibitemShut {NoStop}%
\bibitem [{\citenamefont {Panter}\ \emph {et~al.}(2006)\citenamefont {Panter},
  \citenamefont {Viguier}, \citenamefont {Clou\'e}, \citenamefont {Foucault},
  \citenamefont {Combrade},\ and\ \citenamefont {Andrieu}}]{panter2006}%
  \BibitemOpen
  \bibfield  {author} {\bibinfo {author} {\bibfnamefont {J.}~\bibnamefont
  {Panter}}, \bibinfo {author} {\bibfnamefont {B.}~\bibnamefont {Viguier}},
  \bibinfo {author} {\bibfnamefont {J.}~\bibnamefont {Clou\'e}}, \bibinfo
  {author} {\bibfnamefont {M.}~\bibnamefont {Foucault}}, \bibinfo {author}
  {\bibfnamefont {P.}~\bibnamefont {Combrade}},\ and\ \bibinfo {author}
  {\bibfnamefont {E.}~\bibnamefont {Andrieu}},\ }\bibfield  {title} {\bibinfo
  {title} {Influence of oxide films on primary water stress corrosion cracking
  initiation of alloy 600},\ }\href@noop {} {\bibfield  {journal} {\bibinfo
  {journal} {J. Nuc. Mat.}\ }\textbf {\bibinfo {volume} {348}},\ \bibinfo
  {pages} {213} (\bibinfo {year} {2006})}\BibitemShut {NoStop}%
\bibitem [{\citenamefont {{IAEA}}(2011)}]{IASCC_IAEA}%
  \BibitemOpen
  \bibfield  {author} {\bibinfo {author} {\bibnamefont {{IAEA}}},\ }\href@noop
  {} {\emph {\bibinfo {title} {Stress corrosion cracking in light water
  reactors: Good practices and lessons learned}}},\ \bibinfo {type}
  {{NP-T}-3.13}\ (\bibinfo  {institution} {{IAEA} Nuclear Energy Series},\
  \bibinfo {year} {2011})\BibitemShut {NoStop}%
\bibitem [{\citenamefont {Speidel}(1975)}]{speidel}%
  \BibitemOpen
  \bibfield  {author} {\bibinfo {author} {\bibfnamefont {M.}~\bibnamefont
  {Speidel}},\ }\bibfield  {title} {\bibinfo {title} {Stress corrosion cracking
  of aluminum alloys},\ }\href@noop {} {\bibfield  {journal} {\bibinfo
  {journal} {Metallurgical and Materials Transactions A}\ }\textbf {\bibinfo
  {volume} {6A}},\ \bibinfo {pages} {631} (\bibinfo {year} {1975})}\BibitemShut
  {NoStop}%
\bibitem [{\citenamefont {Burleigh}(1991)}]{burleigh}%
  \BibitemOpen
  \bibfield  {author} {\bibinfo {author} {\bibfnamefont {T.}~\bibnamefont
  {Burleigh}},\ }\bibfield  {title} {\bibinfo {title} {The postulated
  mechanisms for stress corrosion cracking of aluminum alloys: A review of the
  literature 1980-1989},\ }\href@noop {} {\bibfield  {journal} {\bibinfo
  {journal} {Corrosion}\ }\textbf {\bibinfo {volume} {47}},\ \bibinfo {pages}
  {89} (\bibinfo {year} {1991})}\BibitemShut {NoStop}%
\bibitem [{\citenamefont {Wang}\ and\ \citenamefont {Atrens}(1996)}]{wang}%
  \BibitemOpen
  \bibfield  {author} {\bibinfo {author} {\bibfnamefont {Z.}~\bibnamefont
  {Wang}}\ and\ \bibinfo {author} {\bibfnamefont {A.}~\bibnamefont {Atrens}},\
  }\bibfield  {title} {\bibinfo {title} {Initiation of stress corrosion
  cracking for pipeline steels in a carbonate-bicarbonate solution},\
  }\href@noop {} {\bibfield  {journal} {\bibinfo  {journal} {Metallurgical and
  Materials Transactions A}\ }\textbf {\bibinfo {volume} {27A}},\ \bibinfo
  {pages} {2686} (\bibinfo {year} {1996})}\BibitemShut {NoStop}%
\bibitem [{\citenamefont {Arafin}\ and\ \citenamefont
  {Szpunar}(2009)}]{arafin}%
  \BibitemOpen
  \bibfield  {author} {\bibinfo {author} {\bibfnamefont {M.}~\bibnamefont
  {Arafin}}\ and\ \bibinfo {author} {\bibfnamefont {J.}~\bibnamefont
  {Szpunar}},\ }\bibfield  {title} {\bibinfo {title} {A new understanding of
  intergranular stress corrosion cracking resistance of pipeline steel through
  grain boundary character and crystallographic texture studies},\ }\href@noop
  {} {\bibfield  {journal} {\bibinfo  {journal} {Corrosion Science}\ }\textbf
  {\bibinfo {volume} {51}},\ \bibinfo {pages} {119} (\bibinfo {year}
  {2009})}\BibitemShut {NoStop}%
\bibitem [{\citenamefont {Rahimi}\ and\ \citenamefont
  {Marrow}(2011)}]{rahimi2011}%
  \BibitemOpen
  \bibfield  {author} {\bibinfo {author} {\bibfnamefont {S.}~\bibnamefont
  {Rahimi}}\ and\ \bibinfo {author} {\bibfnamefont {T.~J.}\ \bibnamefont
  {Marrow}},\ }\bibfield  {title} {\bibinfo {title} {Effects of orientation,
  stress and exposure time on short intergranular stress corrosion crack
  behaviour in sensitised type 304 austenitic stainless steel},\ }\href@noop {}
  {\bibfield  {journal} {\bibinfo  {journal} {Fatigue Fract. Eng. Mater.
  Struct.}\ }\textbf {\bibinfo {volume} {35}},\ \bibinfo {pages} {359}
  (\bibinfo {year} {2011})}\BibitemShut {NoStop}%
\bibitem [{\citenamefont {Rahimi}\ \emph {et~al.}(2009)\citenamefont {Rahimi},
  \citenamefont {Engelberg}, \citenamefont {Duff},\ and\ \citenamefont
  {Marrow}}]{rahimi2009}%
  \BibitemOpen
  \bibfield  {author} {\bibinfo {author} {\bibfnamefont {S.}~\bibnamefont
  {Rahimi}}, \bibinfo {author} {\bibfnamefont {D.~L.}\ \bibnamefont
  {Engelberg}}, \bibinfo {author} {\bibfnamefont {J.~A.}\ \bibnamefont
  {Duff}},\ and\ \bibinfo {author} {\bibfnamefont {T.~J.}\ \bibnamefont
  {Marrow}},\ }\bibfield  {title} {\bibinfo {title} {In situ observation of
  intergranular crack nucleation in a grain boundary controlled austenitic
  stainless steel},\ }\href@noop {} {\bibfield  {journal} {\bibinfo  {journal}
  {Journal of Microscopy}\ }\textbf {\bibinfo {volume} {233}},\ \bibinfo
  {pages} {423} (\bibinfo {year} {2009})}\BibitemShut {NoStop}%
\bibitem [{\citenamefont {Liu}\ \emph {et~al.}(2019)\citenamefont {Liu},
  \citenamefont {Xia}, \citenamefont {Bai}, \citenamefont {Zhou}, \citenamefont
  {Lu},\ and\ \citenamefont {Shoji}}]{liu2019}%
  \BibitemOpen
  \bibfield  {author} {\bibinfo {author} {\bibfnamefont {T.}~\bibnamefont
  {Liu}}, \bibinfo {author} {\bibfnamefont {S.}~\bibnamefont {Xia}}, \bibinfo
  {author} {\bibfnamefont {Q.}~\bibnamefont {Bai}}, \bibinfo {author}
  {\bibfnamefont {B.}~\bibnamefont {Zhou}}, \bibinfo {author} {\bibfnamefont
  {Y.}~\bibnamefont {Lu}},\ and\ \bibinfo {author} {\bibfnamefont
  {T.}~\bibnamefont {Shoji}},\ }\bibfield  {title} {\bibinfo {title}
  {Evaluation of grain boundary network and improvement of intergranular
  cracking resistance in 316{L} stainless steel after grain boundary
  engineering},\ }\href@noop {} {\bibfield  {journal} {\bibinfo  {journal}
  {Materials}\ }\textbf {\bibinfo {volume} {12}},\ \bibinfo {pages} {242}
  (\bibinfo {year} {2019})}\BibitemShut {NoStop}%
\bibitem [{\citenamefont {Diard}\ \emph {et~al.}(2002)\citenamefont {Diard},
  \citenamefont {Leclercq}, \citenamefont {Rousselier},\ and\ \citenamefont
  {Cailletaud}}]{diard2002}%
  \BibitemOpen
  \bibfield  {author} {\bibinfo {author} {\bibfnamefont {O.}~\bibnamefont
  {Diard}}, \bibinfo {author} {\bibfnamefont {S.}~\bibnamefont {Leclercq}},
  \bibinfo {author} {\bibfnamefont {G.}~\bibnamefont {Rousselier}},\ and\
  \bibinfo {author} {\bibfnamefont {G.}~\bibnamefont {Cailletaud}},\ }\bibfield
   {title} {\bibinfo {title} {Distribution of normal stress at grain boundaries
  in multicrystals: {A}pplication to an intergranular damamge modeling},\
  }\href@noop {} {\bibfield  {journal} {\bibinfo  {journal} {Comp. Mat. Sci.}\
  }\textbf {\bibinfo {volume} {25}},\ \bibinfo {pages} {73} (\bibinfo {year}
  {2002})}\BibitemShut {NoStop}%
\bibitem [{\citenamefont {Diard}\ \emph {et~al.}(2005)\citenamefont {Diard},
  \citenamefont {Leclercq}, \citenamefont {Rousselier},\ and\ \citenamefont
  {Cailletaud}}]{diard2005}%
  \BibitemOpen
  \bibfield  {author} {\bibinfo {author} {\bibfnamefont {O.}~\bibnamefont
  {Diard}}, \bibinfo {author} {\bibfnamefont {S.}~\bibnamefont {Leclercq}},
  \bibinfo {author} {\bibfnamefont {G.}~\bibnamefont {Rousselier}},\ and\
  \bibinfo {author} {\bibfnamefont {G.}~\bibnamefont {Cailletaud}},\ }\bibfield
   {title} {\bibinfo {title} {Evaluation of finite element based analysis of
  {3D} multicrystalline aggregates plasticity. {A}pplication to crystal
  plasticity model identification and the study of strain fields near grain
  boundaries},\ }\href@noop {} {\bibfield  {journal} {\bibinfo  {journal} {Int.
  J. Plasticity}\ }\textbf {\bibinfo {volume} {21}},\ \bibinfo {pages} {691}
  (\bibinfo {year} {2005})}\BibitemShut {NoStop}%
\bibitem [{\citenamefont {Kanjarla}\ \emph {et~al.}(2010)\citenamefont
  {Kanjarla}, \citenamefont {Van~Houtte},\ and\ \citenamefont
  {Delannay}}]{kanjarla}%
  \BibitemOpen
  \bibfield  {author} {\bibinfo {author} {\bibfnamefont {A.}~\bibnamefont
  {Kanjarla}}, \bibinfo {author} {\bibfnamefont {P.}~\bibnamefont
  {Van~Houtte}},\ and\ \bibinfo {author} {\bibfnamefont {L.}~\bibnamefont
  {Delannay}},\ }\bibfield  {title} {\bibinfo {title} {Assessment of plastic
  heterogeneity in grain interaction models using crystal plasticity finite
  element method},\ }\href@noop {} {\bibfield  {journal} {\bibinfo  {journal}
  {Int. J. Plasticity}\ }\textbf {\bibinfo {volume} {26}},\ \bibinfo {pages}
  {1220} (\bibinfo {year} {2010})}\BibitemShut {NoStop}%
\bibitem [{\citenamefont {Gonzalez}\ \emph {et~al.}(2014)\citenamefont
  {Gonzalez}, \citenamefont {Simonovski}, \citenamefont {Withers},\ and\
  \citenamefont {Quinta~da Fonseca}}]{gonzalez2014}%
  \BibitemOpen
  \bibfield  {author} {\bibinfo {author} {\bibfnamefont {D.}~\bibnamefont
  {Gonzalez}}, \bibinfo {author} {\bibfnamefont {I.}~\bibnamefont
  {Simonovski}}, \bibinfo {author} {\bibfnamefont {P.}~\bibnamefont
  {Withers}},\ and\ \bibinfo {author} {\bibfnamefont {J.}~\bibnamefont
  {Quinta~da Fonseca}},\ }\bibfield  {title} {\bibinfo {title} {Modelling the
  effect of elastic and plastic anisotropies on stresses at grain boundaries},\
  }\href@noop {} {\bibfield  {journal} {\bibinfo  {journal} {Int. J.
  Plasticity}\ }\textbf {\bibinfo {volume} {61}},\ \bibinfo {pages} {49}
  (\bibinfo {year} {2014})}\BibitemShut {NoStop}%
\bibitem [{\citenamefont {Hure}\ \emph {et~al.}(2016)\citenamefont {Hure},
  \citenamefont {El~Shawish}, \citenamefont {Cizelj},\ and\ \citenamefont
  {Tanguy}}]{hure2016}%
  \BibitemOpen
  \bibfield  {author} {\bibinfo {author} {\bibfnamefont {J.}~\bibnamefont
  {Hure}}, \bibinfo {author} {\bibfnamefont {S.}~\bibnamefont {El~Shawish}},
  \bibinfo {author} {\bibfnamefont {L.}~\bibnamefont {Cizelj}},\ and\ \bibinfo
  {author} {\bibfnamefont {B.}~\bibnamefont {Tanguy}},\ }\bibfield  {title}
  {\bibinfo {title} {Intergranular stress distributions in polycrystalline
  aggregates of irradiated stainless steel},\ }\href@noop {} {\bibfield
  {journal} {\bibinfo  {journal} {J. Nuc. Mat.}\ } (\bibinfo {year}
  {2016})}\BibitemShut {NoStop}%
\bibitem [{\citenamefont {El~Shawish}\ and\ \citenamefont
  {Hure}(2018)}]{elshawish2018}%
  \BibitemOpen
  \bibfield  {author} {\bibinfo {author} {\bibfnamefont {S.}~\bibnamefont
  {El~Shawish}}\ and\ \bibinfo {author} {\bibfnamefont {J.}~\bibnamefont
  {Hure}},\ }\bibfield  {title} {\bibinfo {title} {Intergranular normal stress
  distributions in untextured polycrystalline aggregates},\ }\href@noop {}
  {\bibfield  {journal} {\bibinfo  {journal} {Eur. J. Mech. / A Solids}\
  }\textbf {\bibinfo {volume} {72}},\ \bibinfo {pages} {354} (\bibinfo {year}
  {2018})}\BibitemShut {NoStop}%
\bibitem [{\citenamefont {Lebensohn}\ \emph {et~al.}(2012)\citenamefont
  {Lebensohn}, \citenamefont {Kanjarla},\ and\ \citenamefont
  {Eisenlohr}}]{lebensohn2012}%
  \BibitemOpen
  \bibfield  {author} {\bibinfo {author} {\bibfnamefont {R.}~\bibnamefont
  {Lebensohn}}, \bibinfo {author} {\bibfnamefont {A.}~\bibnamefont
  {Kanjarla}},\ and\ \bibinfo {author} {\bibfnamefont {P.}~\bibnamefont
  {Eisenlohr}},\ }\bibfield  {title} {\bibinfo {title} {An elasto-viscoplastic
  formulation based on fast fourier transforms for the prediction of
  micromechanical fields in polycrystalline materials},\ }\href@noop {}
  {\bibfield  {journal} {\bibinfo  {journal} {Int. J. Plasticity}\ }\textbf
  {\bibinfo {volume} {32-33}},\ \bibinfo {pages} {59} (\bibinfo {year}
  {2012})}\BibitemShut {NoStop}%
\bibitem [{\citenamefont {Liang}\ \emph {et~al.}(2020)\citenamefont {Liang},
  \citenamefont {Hure}, \citenamefont {Courcelle}, \citenamefont {El~Shawish},\
  and\ \citenamefont {Tanguy}}]{disen2020}%
  \BibitemOpen
  \bibfield  {author} {\bibinfo {author} {\bibfnamefont {D.}~\bibnamefont
  {Liang}}, \bibinfo {author} {\bibfnamefont {J.}~\bibnamefont {Hure}},
  \bibinfo {author} {\bibfnamefont {A.}~\bibnamefont {Courcelle}}, \bibinfo
  {author} {\bibfnamefont {S.}~\bibnamefont {El~Shawish}},\ and\ \bibinfo
  {author} {\bibfnamefont {B.}~\bibnamefont {Tanguy}},\ }\bibfield  {title}
  {\bibinfo {title} {A micromechanical analysis of intergranular stress
  corrosion cracking of an irradiated austenitic stainless steel},\ }\href@noop
  {} {\bibfield  {journal} {\bibinfo  {journal} {under review}\ } (\bibinfo
  {year} {2020})}\BibitemShut {NoStop}%
\bibitem [{\citenamefont {West}\ and\ \citenamefont {Was}(2011)}]{west2011}%
  \BibitemOpen
  \bibfield  {author} {\bibinfo {author} {\bibfnamefont {E.~A.}\ \bibnamefont
  {West}}\ and\ \bibinfo {author} {\bibfnamefont {G.~S.}\ \bibnamefont {Was}},\
  }\bibfield  {title} {\bibinfo {title} {A model for the normal stress
  dependence of intergranular cracking of irradiated 316{L} stainless steel in
  supercritical water},\ }\href@noop {} {\bibfield  {journal} {\bibinfo
  {journal} {Journal of Nuclear Materials}\ }\textbf {\bibinfo {volume}
  {408}},\ \bibinfo {pages} {142 } (\bibinfo {year} {2011})}\BibitemShut
  {NoStop}%
\bibitem [{\citenamefont {El~Shawish}\ \emph {et~al.}(2021)\citenamefont
  {El~Shawish}, \citenamefont {Mede},\ and\ \citenamefont
  {Hure}}]{elshawish2021}%
  \BibitemOpen
  \bibfield  {author} {\bibinfo {author} {\bibfnamefont {S.}~\bibnamefont
  {El~Shawish}}, \bibinfo {author} {\bibfnamefont {T.}~\bibnamefont {Mede}},\
  and\ \bibinfo {author} {\bibfnamefont {J.}~\bibnamefont {Hure}},\ }\bibfield
  {title} {\bibinfo {title} {A single grain boundary parameter to characterize
  normal stress fluctuations in materials with elastic cubic grains},\
  }\href@noop {} {\bibfield  {journal} {\bibinfo  {journal} {European Journal
  of Mechanics, A/Solids}\ }\textbf {\bibinfo {volume} {89}} (\bibinfo {year}
  {2021})}\BibitemShut {NoStop}%
\bibitem [{\citenamefont {Zener}(1948)}]{zener}%
  \BibitemOpen
  \bibfield  {author} {\bibinfo {author} {\bibfnamefont {C.}~\bibnamefont
  {Zener}},\ }\href@noop {} {\emph {\bibinfo {title} {Elasticity and
  Anelasticity of Metals}}}\ (\bibinfo  {publisher} {University of Chicago},\
  \bibinfo {year} {1948})\BibitemShut {NoStop}%
\bibitem [{\citenamefont {Stratulat}\ \emph {et~al.}(2014)\citenamefont
  {Stratulat}, \citenamefont {Duff},\ and\ \citenamefont
  {Marrow}}]{stratulat2014}%
  \BibitemOpen
  \bibfield  {author} {\bibinfo {author} {\bibfnamefont {A.}~\bibnamefont
  {Stratulat}}, \bibinfo {author} {\bibfnamefont {J.~A.}\ \bibnamefont
  {Duff}},\ and\ \bibinfo {author} {\bibfnamefont {T.~J.}\ \bibnamefont
  {Marrow}},\ }\bibfield  {title} {\bibinfo {title} {Grain boundary structure
  and intergranular stress corrosion crack initiation in high temperature water
  of a thermally sensitised austenitic stainless steel, observed in situ},\
  }\href@noop {} {\bibfield  {journal} {\bibinfo  {journal} {Corrosion
  Science}\ }\textbf {\bibinfo {volume} {85}},\ \bibinfo {pages} {428 }
  (\bibinfo {year} {2014})}\BibitemShut {NoStop}%
\bibitem [{\citenamefont {Zhang}\ \emph {et~al.}(2019)\citenamefont {Zhang},
  \citenamefont {Xia}, \citenamefont {Bai}, \citenamefont {Liu}, \citenamefont
  {Li}, \citenamefont {Zhou}, \citenamefont {Wang},\ and\ \citenamefont
  {Ma}}]{zhang2019}%
  \BibitemOpen
  \bibfield  {author} {\bibinfo {author} {\bibfnamefont {Z.}~\bibnamefont
  {Zhang}}, \bibinfo {author} {\bibfnamefont {S.}~\bibnamefont {Xia}}, \bibinfo
  {author} {\bibfnamefont {Q.}~\bibnamefont {Bai}}, \bibinfo {author}
  {\bibfnamefont {T.}~\bibnamefont {Liu}}, \bibinfo {author} {\bibfnamefont
  {H.}~\bibnamefont {Li}}, \bibinfo {author} {\bibfnamefont {B.}~\bibnamefont
  {Zhou}}, \bibinfo {author} {\bibfnamefont {L.}~\bibnamefont {Wang}},\ and\
  \bibinfo {author} {\bibfnamefont {W.}~\bibnamefont {Ma}},\ }\bibfield
  {title} {\bibinfo {title} {Effects of 3{D} grain boundary geometrical angles
  and the net normal stress on intergranular stress corrosion cracking
  initiation in a 316 stainless steel},\ }\href@noop {} {\bibfield  {journal}
  {\bibinfo  {journal} {Materials Science and Engineering: A}\ }\textbf
  {\bibinfo {volume} {765}},\ \bibinfo {pages} {138277} (\bibinfo {year}
  {2019})}\BibitemShut {NoStop}%
\bibitem [{\citenamefont {Simulia}(2016)}]{abaqus}%
  \BibitemOpen
  \bibfield  {author} {\bibinfo {author} {\bibnamefont {Simulia}},\ }\href@noop
  {} {\emph {\bibinfo {title} {ABAQUS 6.14-2}}} (\bibinfo {year}
  {2016})\BibitemShut {NoStop}%
\bibitem [{\citenamefont {Ranganathan}\ and\ \citenamefont
  {Ostaja-Starzewski}(2008)}]{ranganathan}%
  \BibitemOpen
  \bibfield  {author} {\bibinfo {author} {\bibfnamefont {S.~I.}\ \bibnamefont
  {Ranganathan}}\ and\ \bibinfo {author} {\bibfnamefont {M.}~\bibnamefont
  {Ostaja-Starzewski}},\ }\bibfield  {title} {\bibinfo {title} {Universal
  elastic anisotropy index},\ }\href@noop {} {\bibfield  {journal} {\bibinfo
  {journal} {Phys. Rev. Lett.}\ }\textbf {\bibinfo {volume} {101}},\ \bibinfo
  {pages} {055504} (\bibinfo {year} {2008})}\BibitemShut {NoStop}%
\bibitem [{\citenamefont {El~Shawish}()}]{elshawish2022draft}%
  \BibitemOpen
  \bibfield  {author} {\bibinfo {author} {\bibfnamefont {S.}~\bibnamefont
  {El~Shawish}},\ }\bibfield  {title} {\bibinfo {title} {Probabilistic
  two-scale modeling approach for predicting grain boundary damage initiation
  in complex geometries},\ }\href@noop {} {\bibinfo  {journal} {In
  preparation}\ }\BibitemShut {NoStop}%
\bibitem [{\citenamefont {Koyama}\ \emph {et~al.}(2015)\citenamefont {Koyama},
  \citenamefont {Xi}, \citenamefont {Yoshida}, \citenamefont {Yoshimura},
  \citenamefont {Ushioda},\ and\ \citenamefont {Noguchi}}]{koyama2015}%
  \BibitemOpen
\bibfield  {journal} {  }\bibfield  {author} {\bibinfo {author} {\bibfnamefont
  {M.}~\bibnamefont {Koyama}}, \bibinfo {author} {\bibfnamefont {Z.-J.}\
  \bibnamefont {Xi}}, \bibinfo {author} {\bibfnamefont {Y.}~\bibnamefont
  {Yoshida}}, \bibinfo {author} {\bibfnamefont {N.}~\bibnamefont {Yoshimura}},
  \bibinfo {author} {\bibfnamefont {K.}~\bibnamefont {Ushioda}},\ and\ \bibinfo
  {author} {\bibfnamefont {H.}~\bibnamefont {Noguchi}},\ }\bibfield  {title}
  {\bibinfo {title} {Intergranular fatigue crack initiation and its associated
  small fatigue crack propagation in water-quenched fe–c fully ferritic
  steel},\ }\href@noop {} {\bibfield  {journal} {\bibinfo  {journal} {ISIJ
  International}\ }\textbf {\bibinfo {volume} {55}},\ \bibinfo {pages} {2463}
  (\bibinfo {year} {2015})}\BibitemShut {NoStop}%
\bibitem [{\citenamefont {Kr\"oner}(1958)}]{kroner58}%
  \BibitemOpen
  \bibfield  {author} {\bibinfo {author} {\bibfnamefont {E.}~\bibnamefont
  {Kr\"oner}},\ }\bibfield  {title} {\bibinfo {title} {Berechnung der
  elastischen konstanten des vielkristalls aus den konstanten des
  einskristalls},\ }\href@noop {} {\bibfield  {journal} {\bibinfo  {journal}
  {Z. Physik}\ }\textbf {\bibinfo {volume} {151}},\ \bibinfo {pages} {504}
  (\bibinfo {year} {1958})}\BibitemShut {NoStop}%
\bibitem [{\citenamefont {Hershey}(1954)}]{hershey}%
  \BibitemOpen
  \bibfield  {author} {\bibinfo {author} {\bibfnamefont {A.}~\bibnamefont
  {Hershey}},\ }\bibfield  {title} {\bibinfo {title} {The elasticity of an
  isotropic aggregate of anisotropic cubic crystals},\ }\href@noop {}
  {\bibfield  {journal} {\bibinfo  {journal} {J. Appl. Mech.}\ }\textbf
  {\bibinfo {volume} {21}},\ \bibinfo {pages} {236} (\bibinfo {year}
  {1954})}\BibitemShut {NoStop}%
\bibitem [{\citenamefont {Quey}\ \emph {et~al.}(2011)\citenamefont {Quey},
  \citenamefont {Dawson},\ and\ \citenamefont {Barbe}}]{neper}%
  \BibitemOpen
  \bibfield  {author} {\bibinfo {author} {\bibfnamefont {R.}~\bibnamefont
  {Quey}}, \bibinfo {author} {\bibfnamefont {P.~R.}\ \bibnamefont {Dawson}},\
  and\ \bibinfo {author} {\bibfnamefont {F.}~\bibnamefont {Barbe}},\ }\bibfield
   {title} {\bibinfo {title} {Large-scale 3{D} random polycrystals for the
  finite element method: Generation, meshing and remeshing},\ }\href@noop {}
  {\bibfield  {journal} {\bibinfo  {journal} {Comput. Methods Appl. Mech.
  Eng.}\ }\textbf {\bibinfo {volume} {200}},\ \bibinfo {pages} {1729} (\bibinfo
  {year} {2011})}\BibitemShut {NoStop}%
\bibitem [{\citenamefont {El~Shawish}\ \emph {et~al.}(2020)\citenamefont
  {El~Shawish}, \citenamefont {Vincent}, \citenamefont {Moulinec},
  \citenamefont {Cizelj},\ and\ \citenamefont
  {G\'{e}l\'{e}bart}}]{elshawish2019}%
  \BibitemOpen
  \bibfield  {author} {\bibinfo {author} {\bibfnamefont {S.}~\bibnamefont
  {El~Shawish}}, \bibinfo {author} {\bibfnamefont {P.~G.}\ \bibnamefont
  {Vincent}}, \bibinfo {author} {\bibfnamefont {H.}~\bibnamefont {Moulinec}},
  \bibinfo {author} {\bibfnamefont {L.}~\bibnamefont {Cizelj}},\ and\ \bibinfo
  {author} {\bibfnamefont {L.}~\bibnamefont {G\'{e}l\'{e}bart}},\ }\bibfield
  {title} {\bibinfo {title} {Full-field polycrystal plasticity simulations of
  neutron-irradiated austenitic stainless steel: A comparison between fe and
  fft-based approaches},\ }\href@noop {} {\bibfield  {journal} {\bibinfo
  {journal} {J. Nuc. Mat.}\ }\textbf {\bibinfo {volume} {529}},\ \bibinfo
  {pages} {151927} (\bibinfo {year} {2020})}\BibitemShut {NoStop}%
\bibitem [{\citenamefont {Bower}(2010)}]{bower}%
  \BibitemOpen
  \bibfield  {author} {\bibinfo {author} {\bibfnamefont {A.~F.}\ \bibnamefont
  {Bower}},\ }\href@noop {} {\emph {\bibinfo {title} {Applied Mechanics of
  Solids}}}\ (\bibinfo  {publisher} {Taylor \& Francis Group},\ \bibinfo {year}
  {2010})\BibitemShut {NoStop}%
\bibitem [{\citenamefont {Simmons}\ and\ \citenamefont
  {Wang}(1971)}]{simmonswang}%
  \BibitemOpen
  \bibfield  {author} {\bibinfo {author} {\bibfnamefont {G.}~\bibnamefont
  {Simmons}}\ and\ \bibinfo {author} {\bibfnamefont {H.}~\bibnamefont {Wang}},\
  }\href@noop {} {\emph {\bibinfo {title} {Single Crystal Elastic Constants and
  Calculated Aggregate Properties. A Handbook}}}\ (\bibinfo  {publisher} {The
  {MIT} Press},\ \bibinfo {year} {1971})\BibitemShut {NoStop}%
\end{thebibliography}%
% ==========================================================================

\end{document}